\documentclass{jfm}
\usepackage{graphicx}
\usepackage{tikz}
\usetikzlibrary{positioning}
\usepackage{subcaption}
\usepackage{pgfplots}
\usepackage{epstopdf, epsfig}
\usepackage{float} 

\usepackage{xcolor}
\usepackage[font=small]{caption}
\usepackage[labelformat=empty, position=top]{subcaption}
\usepackage[export]{adjustbox}
\usepackage{anyfontsize}
\usepackage{amsmath}
\usepackage[labelsep=period]{caption}
\usepackage{upgreek}
\usepackage{esvect}
\usepackage[hidelinks,colorlinks=true,linkcolor=blue,citecolor=blue]{hyperref}  

\usepackage{chngcntr}
\newcounter{appendixfigure}
\renewcommand{\theappendixfigure}{A\arabic{appendixfigure}}

\numberwithin{equation}{section}

\shorttitle{Turbulent emulsions in Rayleigh–Bénard convection}
\shortauthor{A. Moradi Bilondi, N. Scapin, L. Brandt, and P. Mirbod}

\title{Turbulent convection in emulsions: the Rayleigh–Bénard configuration}

\author{Abbas Moradi Bilondi\aff{{\color{blue}1}}, 
        Nicolò Scapin\aff{{\color{blue}2},{\color{blue}3}}, 
        Luca Brandt\aff{{\color{blue}3},{\color{blue}4}}, \and 
        Parisa Mirbod\aff{{\color{blue}1}}{\color{blue}\corresp{\email{{\color{blue}pmirbod@uic.edu}}}}
}

\affiliation{
\aff{1} Department of Mechanical and Industrial Engineering, 842 W. Taylor Street, \\ University of Illinois at Chicago, Chicago, IL 60607, USA
\aff{2} Department of Mechanical and Aerospace Engineering, \\ Princeton University, Princeton, NJ 08544
\aff{3} FLOW, Department of Engineering Mechanics, KTH, Stockholm, Sweden 
\aff{4} Department of Energy and Process Engineering, \\ Norwegian University of Science and Technology (NTNU), Trondheim, Norway
}

\begin{document}

\maketitle

\begin{abstract}
This study explores heat and turbulent modulation in three-dimensional multiphase Rayleigh–Bénard convection using direct numerical simulations. Two immiscible fluids with identical reference density undergo systematic variations in dispersed-phase volume fractions, $0.0 \leq \Upphi \leq 0.5$, and ratios of dynamic viscosity, $\lambda_{\mu}$, and thermal diffusivity, $\lambda_{\alpha}$, within the range $[0.1-10]$. The Rayleigh, Prandtl, Weber, and Froude numbers are held constant at $10^8$, $4$, $6000$, and $1$, respectively. Initially, when both fluids share the same properties, a 10\% Nusselt number increase is observed at the highest volume fractions. In this case, despite a reduction in turbulent kinetic energy, droplets enhance energy transfer to smaller scales, smaller than those of single-phase flow, promoting local mixing. By varying viscosity ratios, while maintaining a constant Rayleigh number based on the average mixture properties, the global heat transfer rises by approximately 25\% at $\Upphi=0.2$ and $\lambda_{\mu}=10$. This is attributed to increased small-scale mixing and turbulence in the less viscous carrier phase. In addition, a dispersed phase with higher thermal diffusivity results in a 50\% reduction in the Nusselt number compared to the single-phase counterpart, owing to faster heat conduction and reduced droplet presence near walls. The study also addresses droplet-size distributions, confirming two distinct ranges dominated by coalescence and breakup with different scaling laws.
\end{abstract}


\section{Introduction} \label{introduction}

Thermal convection, which involves fluid motion induced by temperature gradients, is a ubiquitous and vital phenomenon in nature, with far-reaching implications for diverse fields of study, including fundamental sciences, technology, and environmental flows. In the geophysical and astrophysical context, thermal convection plays a pivotal role in shaping the dynamics of the oceans, atmosphere, and interior of celestial bodies such as stars and planets \citep{busse1978non, busse1989fundamentals,wyngaard1992atmospheric,mapes1993cloud,maxworthy1994unsteady,atkinson1996mesoscale,getling1998rayleigh,marshall1999open,thorpe2004recent,garcia2013atmospheric,young2017forward,schumacher2020colloquium,hanson2020solar}. In oceanic flows, thermohaline convection drives the deep-ocean circulations \citep{marshall1999open, rahmstorf2000thermohaline}, whereas in the atmosphere, an accurate estimate of thermally driven convection is crucial for weather predictions and climate calculations \citep{hartmann2001tropical}. Moreover, thermal convection operates in both the Earth's outer core and mantle \citep{mckenzie1974convection,cardin1994chaotic,christensen1995effects,zhong2000role,finlay2011flow,guervilly2019turbulent}. 

Emulsions (multiphase flows composed of two immiscible liquid phases with similar densities) play an important role in many contexts, from several industries to oil spills in oceans, where e.g. the distribution of oil droplets is crucial for assessing environmental damage \citep{li1998relationship,french2004oil,gopalan2010turbulent}. Heat transfer is also important in these flows; therefore, herein we focus on exploring turbulent Rayleigh–Bénard convection in liquid-liquid emulsions using Direct Numerical Simulations (DNS), which allow for a detailed analysis of fluid dynamics at the smallest scales. 
The Rayleigh–Bénard (RB) convection, the buoyancy-driven flow arising from heating a fluid from below and cooling it from above, represents the most common configuration for turbulent thermal convection \citep{ahlers2009turbulent,ahlers2009heat,lohse2010small,chilla2012new,shishkina2021rayleigh}. This has been extensively investigated for diverse geometries and various scenarios. \par
In thermally driven turbulent flows, such as turbulent RB convection, the thermal plumes serve as the primary carriers of heat \citep{ahlers2009turbulent,ahlers2009heat,lohse2010small,chilla2012new,shishkina2021rayleigh}. Therefore, researchers have explored various approaches to enhance the overall transport of mass, momentum, and heat within the flow by manipulating these coherent thermal plumes \citep{holmes2012turbulence,graham2021exact},  
Among several examples, we mention here surface roughness and grooved walls to facilitate the detachment of plumes from the boundary layers \citep{shen1996turbulent,du1998enhanced,du2000turbulent,roche2001observation,qiu2005experimental,stringano2006turbulent,shishkina2011modelling,tisserand2011comparison,salort2014thermal,wei2014heat,wagner2015heat,goluskin2016bounds,zhu2017roughness,jiang2018controlling,zhang2018surface}, altering the wettability of the walls \citep{liu2022enhancing}, implementing a slippery surface \citep{huang2022heat}, applying geometry modifications \citep{huang2013confinement,chong2017confined}, the combination of inclination of the convection cell and confined geometries \citep{zwirner2018confined,zwirner2020influence} and the insertion of vertical partition walls into the convection cell \citep{bao2015enhanced}. Also, altering the fluid properties \citep{roche2002prandtl,silano2010numerical}, implementing pulsed heating power on the lower plate \citep{jin2008experimental}, adding polymer additives \citep{ahlers2010effect,benzi2010effect}, incorporating shear \citep{wang2020vibration,blass2020flow}, rotating the convection cell \citep{king2009boundary,zhong2009prandtl}, have all been explored aiming to enhance heat transport. An alternative strategy to enhance the global heat transfer in thermal flows entails the introduction of a secondary phase. These encompass two immiscible fluids \citep{ahlers2009heat,lohse2010small,chilla2012new}, two immiscible fluid layers in the non-turbulent \citep{nataf1988responsible,prakash1994convection,busse2009homologous,diwakar2014stability} and turbulent regime \citep{xie2013dynamics,yoshida2016numerical,liu2021two,liu2022heat}, liquid-liquid emulsions \citep{pelusi2021rayleigh,liu2022enhancing}, the generation of turbulence through the injection of air bubbles  \citep{deckwer1980mechanism,sekoguchi1980forced,sato1981momentum,tokuhiro1994natural,deen2013direct,dabiri2015heat,gvozdic2018experimental,ng2020non} or through the formation of vapor bubbles via boiling~\citep{oresta2009heat,zhong2009enhanced,schmidt2011modification,biferale2012convection,lakkaraju2013heat,guzman2016vapour,guzman2016heat,wang2019self}, bubbles attached to the plate to mimic the boiling \citep{liu2022turbulent}, the inclusion of small particles \citep{oresta2013effects,park2018rayleigh}, as well as non-colloidal suspensions experiencing both laminar \citep{kang2021onset} and turbulent flows \citep{chang2020direct,demou2022turbulent}.

In the case of emulsions, the breakup (or coalescence) of the fluid interface is expected to play a vital role in the heat transfer mechanism of RB convection flows. Previous studies have extensively investigated breakup and coalescence in turbulence \citep{deane2002scale,villermaux2007fragmentation,rosti2019droplets,soligo2019breakage,mukherjee2019droplet,wang2019self,villermaux2020fragmentation,crialesi2022modulation,de2019effect}. However, the presence of walls significantly impacts the behavior of emulsion droplets, leading to the formation of clusters and complex structures \citep{scarbolo2015coalescence}.
The classical Kolmogorov–Hinze theory \citep{kolmogorov1949breakage,hinze1955fundamentals} explains fluid breakup based on the balance between surface tension and inertial forces. However, in turbulent RB convection flows, buoyancy becomes a significant factor that can impact the breakup criteria \citep{liu2021two}, altering the heat transfer mechanism. This aspect remains an area of research that has not been comprehensively studied, and the present study aims to investigate it in detail. The objective of this work is, therefore, to fill this knowledge gap by comprehensively studying the behavior of emulsions and their influence on heat transfer modulation in turbulent Rayleigh-Bénard convection flows. We aim to gain a deeper understanding of emulsion dynamics near walls and their impact on heat transfer, which can provide valuable insights into the underlying mechanisms that govern the behavior of emulsions in turbulent RB convection flows. These insights hold significance for optimizing and designing industrial processes that involve multiphase turbulent thermal convection. 
%

In particular, we investigate the influence of key parameters, including the concentration of the dispersed phase and the dynamic viscosity and thermal diffusivity ratios between the two fluids, on the dynamics of the emulsions and their impact on overall heat transfer in the system. This research also addresses the knowledge gap on how interactions of dispersed droplets and their concentration affect the temperature and velocity fields of both carrier and dispersed phases, the internal energy budget, and turbulent kinetic energy, leading to changes in global heat transfer. The study conducted by \cite{demou2022turbulent} focused on the impact of different volume fractions of rigid particles on heat transfer modulations in turbulent Rayleigh-Bénard (RB) convection. Their findings showed that adding up to 25\% particle volume fraction to the single-phase flow enhanced the Nusselt number. However, surpassing this threshold led to a decrease in the heat transfer rate, attributed to the dense layering of particles near the wall. Building upon the insights of \cite{demou2022turbulent}, we investigate a similar flow condition with a deformable secondary phase, represented by liquid-liquid emulsions instead of suspensions. This investigation provides a detailed analysis of the effects of the secondary phase deformability on heat transfer modulation, which, to the best of our knowledge, has not been previously explored. 
In this paper, since we did not employ any stabilizing mechanisms, the term "\textit{emulsion}" is not entirely accurate. Nevertheless, we use "\textit{emulsion}" together with "\textit{mixture}" for simplicity and consistency with previous literature ~\citep{mukherjee2019droplet,yi2021global,crialesi2022modulation,crialesi2023intermittency}.
%
The paper is structured as follows. Section \ref{sec:methodology} outlines the equations governing our problem, the employed numerical methods, and a detailed description of our computational setup with the list of our simulations. Section \ref{sec:results} features and analyses the significant findings of our work. This includes the Nusselt number variations, temperature and velocity statistics, a detailed description of heat transfer and turbulent kinetic energy budgets, and the droplet size distribution (DSD) analysis. In the last parts of Section \ref{sec:results}, we evaluate the effects of dynamic viscosity and thermal diffusivity ratios on DSD and heat transfer modulations. Conclusions and a future outlook are provided in Section \ref{sec:conclusion}. 


\section{Methodology}\label{sec:methodology}

\subsection{Governing equations} \label{subsec:governing_equations}
To study emulsions in a turbulent Rayleigh–Bénard convection, we introduce an indicator function $\mathcal{H}$, equal to $1$ in the volume $V_1$ occupied by the disperse phase (fluid $1$) and $0$ in the volume $V_2$ occupied by the carrier phase (fluid $2$). The function $\mathcal{H}$ is governed by the following transport equation:
\begin{equation}
  \dfrac{\partial\mathcal{H}}{\partial\tilde{t}} + \tilde{\mathbf{u}}\cdot\tilde{\nabla}\mathcal{H} = 0\mathbf{,}
  \label{eqn:H_tran}
\end{equation}
where $\tilde{\mathbf{u}}=(\tilde{u},\tilde{v},\tilde{w})$ is the one-fluid velocity field, assumed continuous in the whole domain. Note that, henceforth, the symbol $\tilde{\cdot}$ indicates a dimensionless scalar or vectorial quantity. The transport of $\mathcal{H}$ is coupled with the incompressibility constraint, the Navier-Stokes equations for a Newtonian fluid, and the transport equation for the temperature field. These read, in dimensionless form, as
\begin{equation}
  \tilde{\bf{\nabla}}  \cdot \tilde{{\bf u}}=0,
  \label{eqn:mass_equation}
\end{equation}
\begin{equation}
\tilde{\rho}\left[\dfrac{\partial \tilde{{\bf u}}}{\partial \tilde{t}} + (\tilde{{\bf u}} \cdot \tilde{{\bf \nabla}}) \tilde{{\bf u}} \right] = -\tilde{{\bf \nabla}} \tilde{p}+\sqrt{\frac{Pr}{Ra}}\tilde{\nabla}\cdot\left[\tilde{\mu}\left(\nabla\tilde{{\bf u}}+\nabla\tilde{{\bf u}}^T\right)\right] + \frac{\tilde{\mathbf{f}}_\sigma}{We} + \dfrac{\tilde{\hat{\rho}}\mathbf{e}_z}{Fr}\mathrm{,}
  \label{eqn:NS_equation}
\end{equation}
\begin{equation}
	\tilde{\rho}\tilde{c}_p\left[\dfrac{\partial \tilde{{\theta}}}{\partial \tilde{t}} + (\tilde{{\bf u}} \cdot \tilde{{\nabla}}) \tilde{{\theta}} \right] = \dfrac{\tilde{\nabla}\cdot(\tilde{k}\tilde{\nabla}\tilde{{\theta}})}{\sqrt{{Pr}{Ra}}}\mathrm{.}
  \label{eqn:energy_equation}
\end{equation}
In the above, $\tilde{p}$ is the hydrodynamic pressure, $\tilde{\theta}$ is the temperature and $\tilde{\mathbf{f}}_\sigma=\tilde{\kappa}\mathbf{n}_\Gamma\tilde{\delta}_\Gamma$ is the surface tension forces with $\tilde{\kappa}$ the interfacial curvature, $\mathbf{n}_\Gamma$ the normal vector and $\tilde{\delta}_\Gamma$ the Dirac-delta function~\citep{scardovelli1999direct}. Note that $\mathbf{e}_z=(0,0,-1)$ is the unit normal vector oriented in the gravity direction. 
Equations~\eqref{eqn:mass_equation},~\eqref{eqn:NS_equation} and~\eqref{eqn:energy_equation} are made dimensionless by introducing a reference length-scale $L_s=H$ with $H$ the cavity height, the reference temperature difference $\Delta T=T_h-T_c$, i.e. the temperature difference between the bottom and top boundary, and a reference velocity taken as the free-fall velocity 
$U_f=\sqrt{\beta_r \mathrm{g} L_s \Delta T}$ where $\beta_r$ is the reference isothermal expansion coefficient
and $\mathrm{g}$ is the module of the gravitational acceleration. 
\par
For each generic thermophysical property, we introduce a reference $\psi_r$, which is chosen in two different ways. In the first choice, $\psi_r$ is taken equal to the average property of the dispersed and continuous phase weighted by the total volume fraction $\Upphi$, i.e. $\psi_r = \psi_1\Upphi+\psi_2(1-\Upphi)$, where $\Upphi = \left(\int_V\mathcal{H}dV\right)/V$ and $V=V_1+V_2$ the total volume of the domain. Following this definition, $\psi_r$ is also used to define the generic dimensionless thermophysical property $\tilde{\psi}$ (density $\tilde{\rho}$, dynamic viscosity $\tilde{\mu}$, thermal conductivity $\tilde{k}$ or specific heat capacity $\tilde{c}_p$) as $\tilde{\psi} = \psi/\psi_r$, where $\psi$ is computed with an arithmetic average, i.e. $\psi = \psi_1\mathcal{H}+\psi_2(1-\mathcal{H})$. Accordingly, $\tilde{\psi}$ can be finally expressed as
\begin{equation}\label{eqn:tildepsi}
  \tilde{\psi} = \dfrac{\psi_1\mathcal{H}+\psi_2(1-\mathcal{H})}{\psi_1\Upphi+\psi_2(1-\Upphi)} = \dfrac{1+(\lambda_\psi-1)\mathcal{H}}{1+(\lambda_\psi-1)\Upphi}\mathrm{,}
\end{equation}
where $\lambda_\psi=\psi_1/\psi_2$ is the property ratio. In the second choice, $\psi_r$ is taken equal to $\psi_2$ and, therefore, $\tilde{\psi}$ in equation~\eqref{eqn:tildepsi} simply reduces to $1+(\lambda_\psi-1)\mathcal{H}$. \par
Regardless of the employed approach to define $\psi_r$, the different dimensionless numbers in equations~\eqref{eqn:NS_equation} and~\eqref{eqn:energy_equation} are expressed as follows. First, we define the Prandtl number, $Pr=\nu_r/\alpha_r$ as the ratio of the reference viscous and thermal diffusivity and the Rayleigh number $Ra=\beta_r \mathrm{g} L_s^3 \Delta T/(\alpha_r\nu_r)$ to characterize the importance of buoyancy forces to the viscous forces. Next, we introduce the Weber number  $We = \rho_rU_f^2L_s/{\sigma}$ as the ratio between the inertial and the surface tension forces with $\sigma$ representing the surface tension coefficient. Finally, we define the Froude number $Fr=U_f^2/(gL_s)$ as the ratio between the inertial and gravity forces. Note that $\hat{\tilde{\rho}}$ in the last term of equation~\eqref{eqn:NS_equation} is the volumetric density field modified to account for the thermal effects in the gravity forces. By assuming that the flow is incompressible within the limits of the Oberbeck–Boussinesq approximation~\citep{oberbeck1879warmeleitung,boussinesq1903theorie,gray1976validity}, $\hat{\rho}$ takes the following form:
\begin{equation}
  \tilde{\hat{\rho}} = \dfrac{1}{\rho_r}\left[\rho_{1} \left( 1 - \beta_1 \Delta T \tilde{\theta} \right) \mathcal{H} + \rho_{2} \left( 1 - \beta_2 \Delta T \tilde{\theta} \right) (1-\mathcal{H})\right]\mathrm{.}
  \label{Oberbeck-Boussinesq approximation}
\end{equation}
An important dimensionless parameter is the Nusselt number $Nu$, i.e. the dimensionless heat flux, which serves as an indicator of the overall heat transfer rate within the Rayleigh–Bénard cell. $Nu$ is defined as follows
\begin{equation}
Nu = \frac{h L_s}{k}=\frac{\text{total heat flux}}{\text{conductive heat flux}}=\frac{-\displaystyle{\left[k \frac{dT}{dz}\right]}_\text{wall}L_s}{k_r \Delta T}\mathrm{.} 
  \label{eq:Nusselt_number}
\end{equation}
Here, $h$ represents the convective heat transfer coefficient of the flow, $k$ is the local thermal conductivity of the emulsion, and $k_r$ is the reference thermal conductivity taken equal to the average thermal conductivity of the emulsion. \par
Throughout this work, we adopt the first approach, where for each change in $\Upphi$ and $\lambda_\psi$ during the simulation campaign, $\psi_r$ is kept equal to the value for $\Upphi=0$, i.e., the single-phase configuration. This approach allows us to investigate turbulence and heat transfer modulation by fixing dimensionless parameters defined using the thermophysical properties of the entire emulsion, rather than those of one of the phases. For the sake of completeness and comparison with this first methodology, we perform two additional simulations using the second definition and report the results in Appendix~\ref{sec:appendix_A}.

\subsection{Numerical methodology} \label{subsec:numerics}
The governing equations \eqref{eqn:mass_equation}-\eqref{eqn:energy_equation} are solved on a uniform Cartesian grid with constant grid spacing in the three directions, $\Delta x = \Delta y = \Delta z$. The grid spacing is defined as $\Delta x = L_x/N_x$, $\Delta y = L_y/N_y$ and $\Delta z = L_z/N_z$ where $L_x$, $L_y$ and $L_z$ are the lengths of the computational domain and $N_x$, $N_y$ and $N_z$ the number of grid cells in the three directions. The so-called one-fluid formulation \citep{prosperetti2009computational} is employed to discretize the governing equations so that only one set of equations valid for both phases is solved over the whole domain. The procedure is as follows. \par
First, the numerical algorithm defines a cell-averaged value of $\mathcal{H}$, which is called volume-of-fluid (VOF) function or volume fraction, 
\begin{equation}
 \phi = \frac{1}{V_c} \int_{V_c}^{}  {\mathcal{H}(\mathrm{\tilde{x}},\tilde{t})dV_c},
 \label{volume fraction}
\end{equation}
where $V_c=\Delta x\Delta y\Delta z$ is the volume of each computational cell. By applying the definition~\eqref{volume fraction} to equation~\eqref{eqn:H_tran}, the advection for $\phi$ reads:
\begin{equation}
 \frac{\partial {\phi}}{\partial \tilde{t}} + \tilde{\nabla} \cdot {(\tilde{{\bf u}}{\mathcal{H}})}={\phi} \tilde{\nabla} \cdot \tilde{{\bf u}}.
  \label{advection of VOF}
\end{equation}
The various VOF methods proposed in the literature differ in the way $\mathcal{H}$ is approximated. In the present work, the Multi-dimensional Tangent Hyperbola Interface Capturing (MTHINC) algorithm developed by \cite{ii2012interface} is employed, in which $\mathcal{H}$ is approximated with a hyperbolic tangent as:
\begin{equation}
 \mathcal{H}(\tilde{X},\tilde{Y},\tilde{Z}) \approx \frac{1}{2} +\Biggl\{  1+\mathrm{tanh}\left[\gamma_{th}\left(\mathcal{S}(\tilde{X},\tilde{Y},\tilde{Z}) + {d}_{th}\right) \right] \Biggl\},
  \label{MTHINC}
\end{equation}
where $(\tilde{X},\tilde{Y},\tilde{Z})$ is a local coordinate system, i.e. $\tilde{X}=(\tilde{x} - 0.5/L_s)/(\Delta \tilde{x})$, $\tilde{Y}=(\tilde{y} - 0.5/L_s)/(\Delta \tilde{y})$ and $\tilde{Z}=(\tilde{z} - 0.5/L_s)/(\Delta \tilde{z})$ with $(\Delta \tilde{x}, \Delta \tilde{y}, \Delta \tilde{z})$ the grid spacing in the scaled directions of $\tilde{x}$, $\tilde{y}$ and $\tilde{z}$. In equation (\ref{MTHINC}), $\gamma_{th}$ is a parameter controlling the sharpness of color function (set equal to $2$ in the present work), $\mathcal{S}(\tilde{X},\tilde{Y},\tilde{Z})$ is the surface function and $d_{th}$ is the normalization parameter. The implementation details of the MTHINC method are reported in \cite{ii2012interface} and \cite{crialesi2023flutas}. Once the color function is known, the local average thermophysical properties of the emulsion (e.g. density, dynamic viscosity, specific heat capacity, and thermal conductivity) are updated using the local volume fraction, i.e. $\psi = \psi_1\phi + \psi_2(1-\phi)$.
\par
Next, the momentum and the temperature equations~\eqref{eqn:NS_equation},~\eqref{eqn:energy_equation} are discretized on a regular Cartesian grid using a staggered arrangement, i.e. all the scalar fields are defined at the cell centers, except for the velocity components, which are defined at the corresponding cell faces~\citep{harlow1965numerical}. All the spatial derivatives are approximated with second-order central schemes, and the equations are advanced with a second-order Adams-Bashforth scheme. The pressure equation is solved with a direct FFT-based Poisson solver to impose exactly the incompressibility constraint~\eqref {eqn:mass_equation}. The simulations are performed using the open-source code FluTAS\footnote{https://github.com/Multiphysics-Flow-Solvers/FluTAS}, \textit{Fluid Transport Accelerated Solver}, which is parallelized using MPI/OpenMP directives in the CPU version and accelerated using OpenACC directives in the GPU version. FluTAS is capable of performing interface-resolved simulations of incompressible multiphase flows, optionally with heat transfer, as shown in several past studies~\citep{rosti2019numerical,rosti2019droplets,de2019effect,kozul2020aerodynamically,de2020numerical,rosti2021shear,cannon2021effect,crialesi2022modulation,scapin2020volume,scapin2022finite,dalla2021interface,scapin2023evaporating,mirbod2023turbulent,crialesi2023interaction}. More details can be found in~\cite{crialesi2023flutas}, where standard benchmarks and scaling tests for the CPU and GPU versions can be found. 

\subsection{Case description} \label{sec:case_description}
\begin{figure}
    \includegraphics[width=\linewidth]{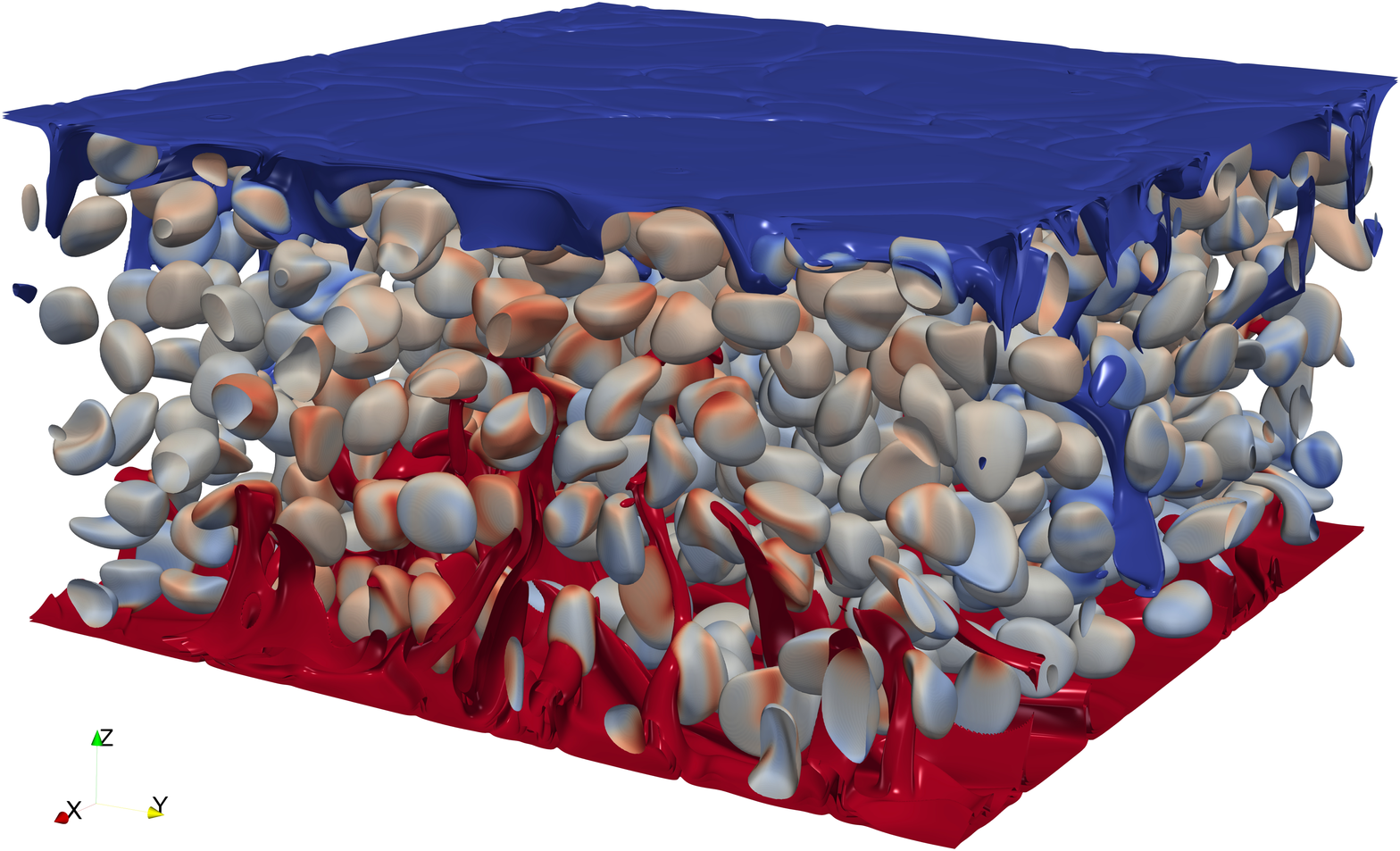} 
    \caption{Schematic of the 3D turbulent RB convection cell with the Cartesian coordinate system. The domain dimensions along the $\tilde{x}$, $\tilde{y}$ and $\tilde{z}$ directions are ($L_{\tilde{x}}$, $L_{\tilde{y}}$, $L_{\tilde{z}}$)=($2$, $2$, $1$). The liquid-liquid emulsions are heated from the bottom wall (depicted in red) and cooled from the top wall (in blue). The figure illustrates instantaneous iso-surfaces of temperature and dispersed droplets a short time ($\Delta \tilde t = 0.4$ time units) after adding droplets ($\Upphi=0.2$) to the single-phase flow.}
    \label{fig:geom}
\end{figure}

Direct numerical simulations are carried out in the three-dimensional Rayleigh-Bénard cell reported in figure \ref{fig:geom}. The cell is periodic along the two horizontal directions, with wall boundary conditions imposed at the bottom and top wall. The temperature of the top and bottom walls is uniform, constant, and equal to $\tilde{\theta}_c = - 0.5$ and $\tilde{\theta}_h = 0.5$. Emulsions are enclosed between two infinitely long plates. 
The figure depicts a sketch of our computational domain with height $L_{\tilde{z}} = 1$ in the scaled wall-normal direction ($\tilde{z}$) and plates' dimensions $L_{\tilde{x}} = L_{\tilde{y}} = 2$ in the periodic streamwise and spanwise directions ($\tilde{x}$ and $\tilde{y}$). In this work, we want to study turbulent emulsions representative of oil-water system at ambient temperature and in a normal gravity condition. Accordingly, we choose $Ra=10^8$, $Pr=4$ and $Fr=1$. The Weber number $We$ is chosen equal to $6000$, which balances the requirements of avoiding excessive unphysical coalescence, which occurs at low $We$, and excessive break-up, which occurs at high $We$. Note that $We$ is a large-scale Weber number, and it is defined based on the height of the cavity and the free-fall velocity. As a preliminary analysis, we also computed a "small-scale" Weber number based on the local droplet radius and local velocity, and we always found it in the order of $10^2$. Furthermore, the two phases share the same densities ($\rho_1=\rho_2=\rho_r$), specific heat capacity ($c_{p,1}=c_{p,2}=c_{p,r}$) and thermal expansion coefficient ($\beta_1=\beta_2=\beta_r$). The viscosity ratio $\lambda_\mu=\mu_d/\mu_c$ and thermal diffusivity ratio $\lambda_\alpha=\alpha_d/\alpha_c$, however, vary in the range of $[0.1-10]$. To investigate the effects of the dispersed phase on heat transfer rate in RB convection, different volume fractions of the dispersed phase are considered, $0.0 \leq \Upphi \leq 0.5$. Table \ref{table:list_of_simulations} summarizes the simulations performed in the present work. \par
\begin{table}
\centering
\setlength{\tabcolsep}{1em}

\begin{tabular}{c c c c c}
\hline\hline
$\text{Case}$ & $\Upphi$ & $\lambda_{\mu}=\mu_d/\mu_c$ & $\lambda_{\alpha}=\alpha_d/\alpha_c$ & $\mu_\text{eff}/\mu_\text{sp}$ \\ 
\hline
$1$ & $0.0$ & $1$ & $1$ & $1$\\
$2$ & $0.2$ & $1$ & $1$ & $1$\\
$3$ & $0.3$ & $1$ & $1$ & $1$\\
$4$ & $0.4$ & $1$& $1$ & $1$\\
$5$ & $0.5$ & $1$& $1$ & $1$\\
$6$ & $0.2$ & $0.1$ & $1$ & $1$\\
$7$ & $0.2$ & $10$ & $1$ & $1$\\
$8$ & $0.5$ & $0.1$ & $1$ & $1$\\
$9$ & $0.2$ & $1$ & $0.1$ & $1$\\
$10$ & $0.2$ & $1$ & $10$ & $1$\\
$11$ & $0.5$ & $1$ & $0.1$ & $1$\\
$12$ & $0.2$ & $0.1$ & $1$ & $0.82$\\
$13$ & $0.2$ & $10$ & $1$ & $2.8$\\
\hline
\end{tabular}

\caption{List of simulations performed in this study.}
\label{table:list_of_simulations}
\end{table}
 To ensure that all spatial scales are appropriately resolved, the grid resolution is determined based on three criteria, as outlined in \cite{shishkina2010boundary}. First, the local mesh size must be smaller than the local Kolmogorov scale, $\eta_K(\textbf{x},t)$, the local Batchelor scale, $\eta_B(\textbf{x},t)$, and the local length scale of $\eta_T(\textbf{x},t)$, where $\eta_K(\textbf{x},t) = [\nu^{3/4}\epsilon(\textbf{x},t)^{-1/4}]$, $\eta_B(\textbf{x},t) = \eta_K(\textbf{x},t)Pr^{-1/2}$ and $\eta_T(\textbf{x},t) = \eta_K(\textbf{x},t)Pr^{-3/4}$. Here, $\epsilon(\textbf{x},t)$ is the local kinetic energy dissipation rates per mass. Therefore, in order to meet this well-established criterion, the number of grid points in the wall-normal direction should satisfy ${N_{\tilde{z}}} \ge {\epsilon_\text{max}(\textbf{x},t)}^{1/4}(Pr/\nu)^{3/4}H$. Thus, for all of our cases ${N_{\tilde{z}}}^\text{min} = 440$. Additionally, the global mesh size should be smaller than the global length scales of Kolmogorov, Batchelor and $\eta_T$. To meet this requirement, the number of grid points in the wall-normal direction should satisfy ${N_{\tilde{z}}} \ge [{Ra}{(Nu_\text{max}-1)}Pr]^{1/4}$. Thus, ${N_{\tilde{z}}}^\text{min} = 352$, considering $Nu_\text{max}=39$ (the maximum $Nu$ achieved in our study). Overall, the minimum number of grid points in the wall-normal direction for our uniform grid is calculated as ${N_{\tilde{z}}}^\text{min} = \text{max}(440, 352) = 440$. Finally, in order to resolve all spatial scales inside the boundary layers, the resolutions within the thermal and kinetic boundary layers (TBL and VBL) should meet the following conditions: ${N_{\tilde{z}}}^\text{TBL} \ge \sqrt{2} (0.482) {Nu}^{1/2}{(0.982)}^{3/2}$ and ${N_{\tilde{z}}}^\text{VBL} \ge \sqrt{2} (0.482) {Nu}^{1/2}{Pr}^{1/3} {(0.982)}^{1/2}$. This translates to a minimum requirement of ${N_{\tilde{z}}}^\text{min, BL} = \text{max}(5, 7) = 7$ grid points within the boundary layers. In this study, we employ a grid size of $1024\times 1024\times 512$ with uniform spacing in the $\tilde{x}$, $\tilde{y}$, and $\tilde{z}$ directions, ensuring a minimum of 10 grid points within the boundary layers. Consequently, all three resolution requirements mentioned above are satisfactorily met.
We first simulated the single-phase case ($ \Upphi = 0$), initiated from $\tilde{{\bf u}} = \tilde{\theta} = 0$, until a statistically stationary state. A random noise equal to $5$ $\%$ of the prescribed temperature difference is superimposed on the initial temperature to promote a faster transition to the turbulent state. The multiphase simulations of different volume fractions $\Upphi$, denoted as cases 2-5, were initialized using the same initial velocity field, that is the statistically steady solution of the single-phase case. Moreover, when investigating the impact of the viscosity ratio (cases 6-8), we started from the statistically stationary solution of cases 2 and 5 and modified the dynamic viscosity of both phases while maintaining the average dynamic viscosity of the emulsion constant. 
A similar procedure is followed for cases 9-11 where we examined the effect of the thermal diffusivity ratio. However, for the last two cases of 12 and 13, similar to cases 6-8, we examined the effects of viscosity ratio, but this time we did this by just varying the value of the dispersed-fluid viscosity while maintaining the carrier-fluid viscosity identical to that of the single-phase scenario. The results regarding cases 12 and 13 are presented in Appendix~\ref{sec:appendix_A}. Once a statistical steady state was reached, all simulations were continued under stationary conditions over a predefined time interval to collect the turbulent statistics. Specifically, at the statistical steady state, simulations were run at a fixed time step for an interval of approximately $\Delta \tilde{t}_{\text{ss}}\approx 500$. During the data collection period, we stored a substantial number of samples,  14000 per case. Note that the time step used to advance the governing equations was always dynamically adjusted according to the Courant-Friedrichs-Lewy (CFL) condition with $\text{CFL} = 0.25$, except during the statistical sampling stage. Here, the time step was held constant and set equal to 90\% of the average time step at $\text{CFL} = 0.25$. 
\begin{figure}
  \centering
  \begin{subfigure}[t]{0.02\textwidth}
  \fontsize{6}{9}
    \textbf{(a)}
  \end{subfigure}
  \begin{subfigure}[t]{0.54\textwidth}
    \includegraphics[width=\linewidth, valign=t]{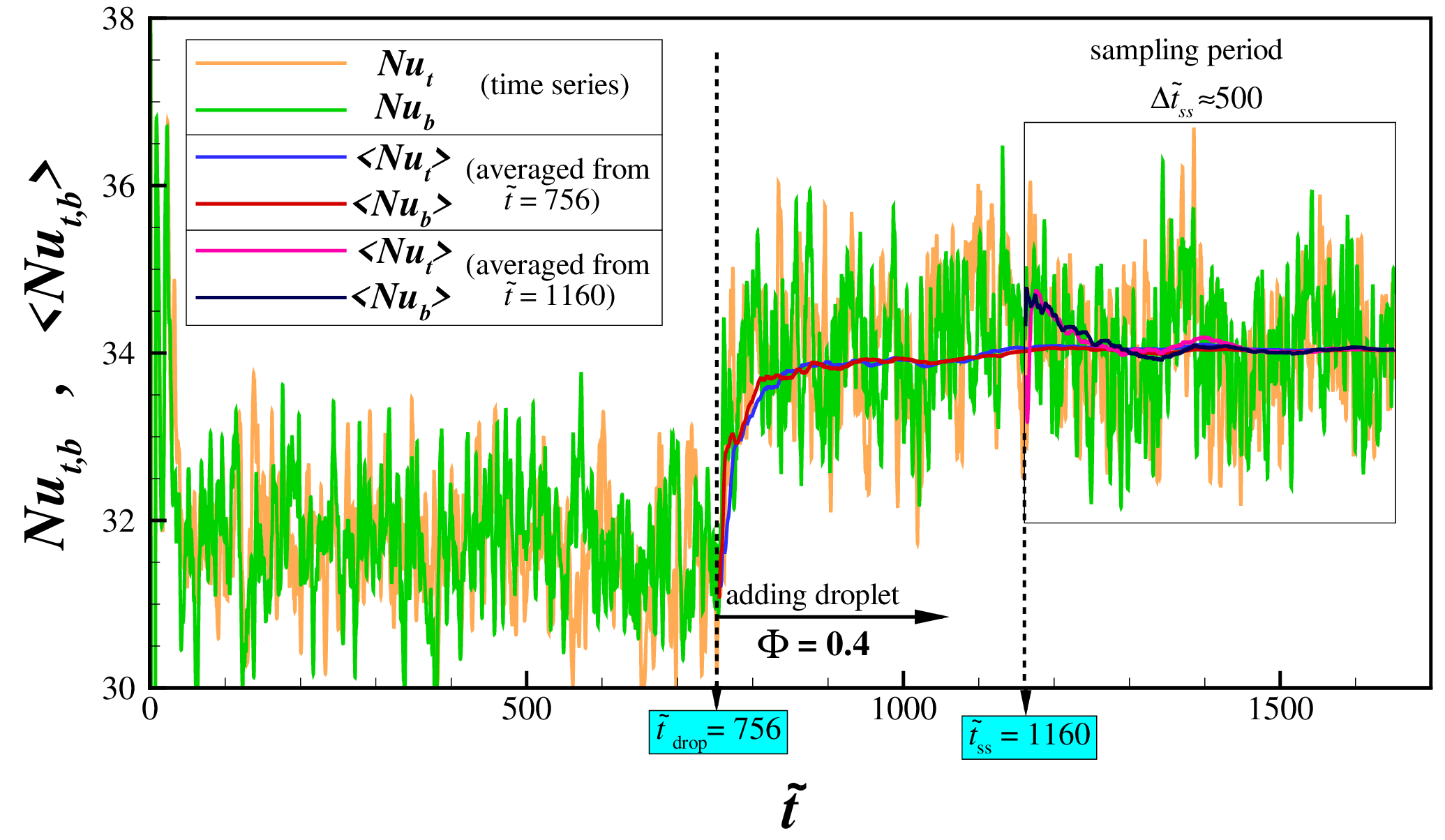}
    \phantomsubcaption\label{fig:nu vs time}
  \end{subfigure}\hfill
  \begin{subfigure}[t]{0.02\textwidth}
  \fontsize{6}{9}
    \textbf{(b)}
  \end{subfigure}
  \begin{subfigure}[t]{0.36\textwidth}
    \includegraphics[width=\linewidth, valign=t]{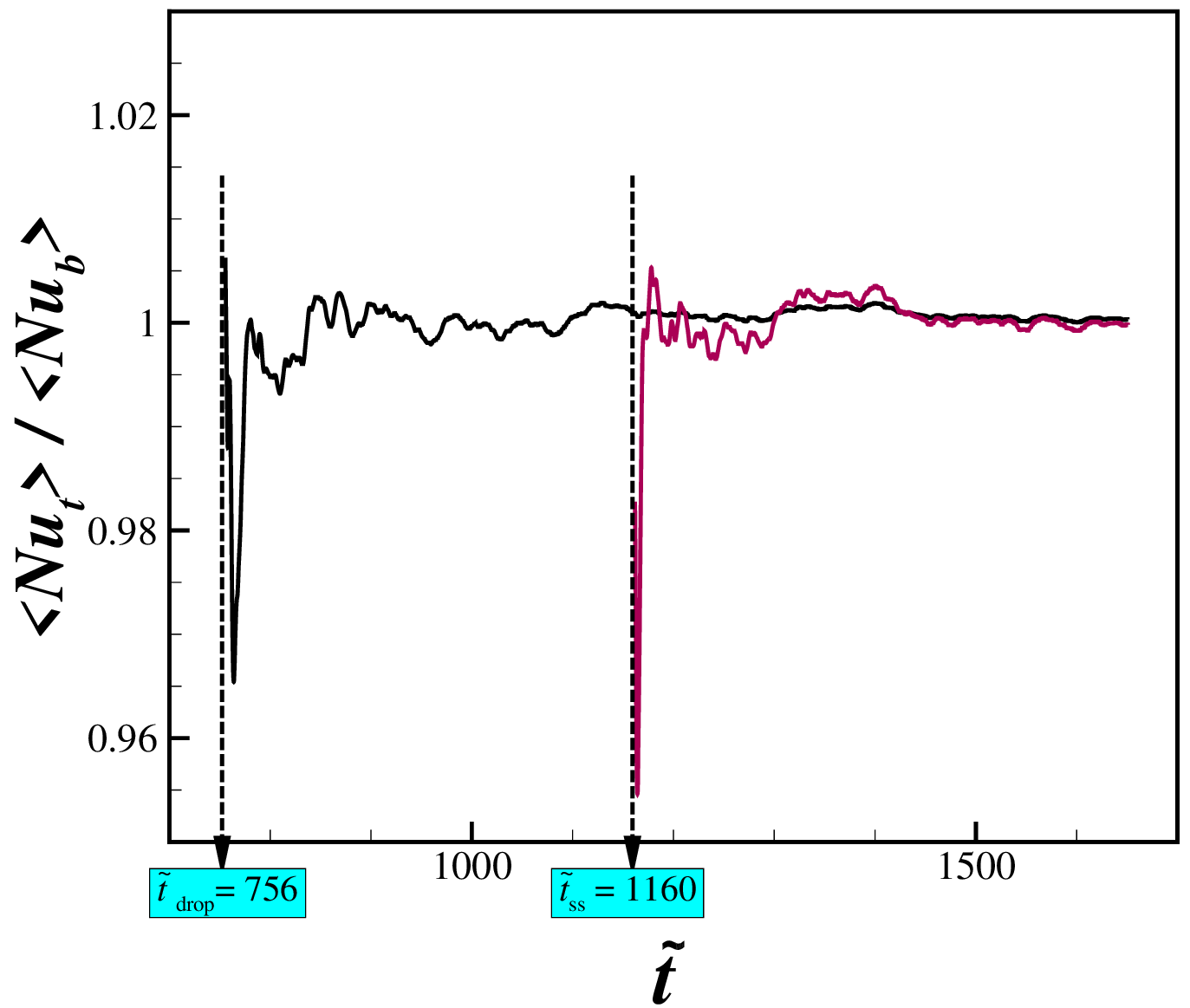}
    \phantomsubcaption\label{fig:nu_ratio_1}
  \end{subfigure}
\caption{(a) Temporal evolution (green and orange solid lines) and temporal average of the Nusselt number on the top and bottom walls for case 4; the time-averaging starts from two different time instants: when adding the droplets, $\tilde{t}=756$ (blue and red solid lines) and $\tilde{t}=1160$ for the final statistical sampling (pink and dark-grey solid lines). (b) The ratio of temporally averaged Nusselt numbers of top and bottom walls.}
  \label{fig:nu_time_evolution_ratio}
\end{figure}
To ensure the convergence of statistics, we systematically computed and averaged both first- and second-order statistics using varying sample sizes. We assessed the differences between these statistics over four distinct time intervals within ${\Delta \tilde{t}_{\text{ss}}}$, i.e.  (a) the first quarter, (b) the first half, (c) the first three quarters and (d) the entire duration of $\Delta \tilde{t}_{\text{ss}}$. The analysis revealed a progressive reduction in the differences between (c) and (d), rendering the difference negligible. For clarity, Figure \ref{fig:nu vs time} illustrates this procedure for case 4, where droplets were added to the single-phase flow at $\tilde{t}=756$, and statistical sampling started at $\tilde{t}=1160$ once the simulation has reached the statistical stationary condition. 
In particular, the temporal evolution of the Nusselt number at the top and bottom walls is monitored starting at ${\tilde{t}_\text{drop}}$, when droplets are added into a statistically stationary single-phase flow, and ${\tilde{t}_\text{ss}}$, when the multiphase flow reaches a statistically stationary condition. The temporally-averaged $Nu$ is computed over the time interval $[{\tilde{t}_\text{start}}, \tilde{t}]$, where $\tilde{t}\ge\tilde{t}_\text{start}$.
Figure \ref{fig:nu_ratio_1} presents the ratio between the time-averaged Nusselt number at the top and bottom walls, which consistently approaches unity once a statistically stationary condition is achieved. 
\section{Results and discussions}\label{sec:results}
\subsection{Flow statistics}\label{subsec:flow_stat}

\begin{figure}
\center
    \includegraphics[width=0.6\linewidth]{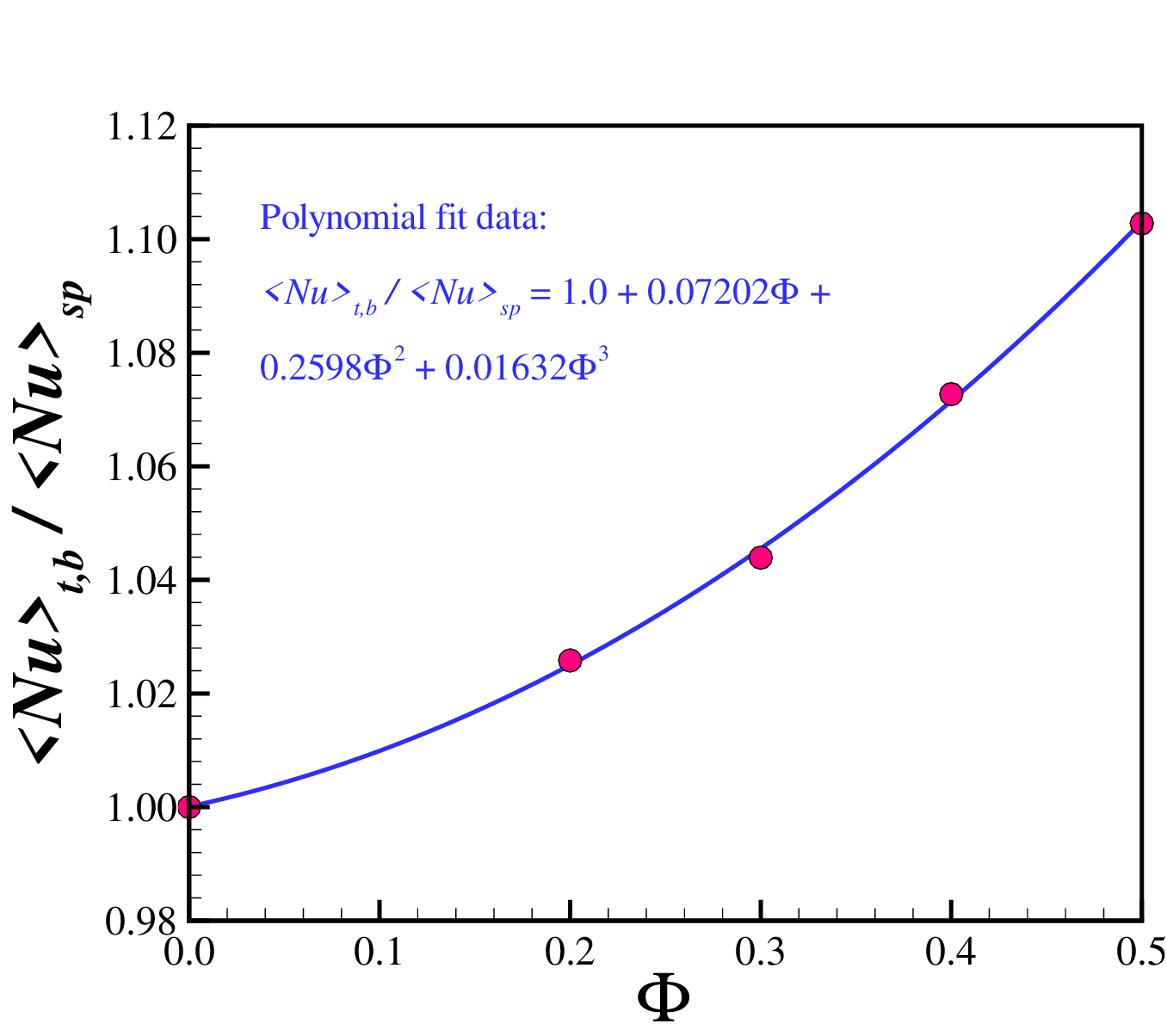} 
    \caption{Nusselt number at different volume fractions of the dispersed phase.} 
    \label{fig:nu_phi} 
\end{figure}
In this section, we present the findings of our investigation, starting with the case of emulsions with property ratios equal to 1 to focus on the modifications induced by the presence of an interface, only. Figure \ref{fig:nu_phi} illustrates the variation of the time-averaged Nusselt number, normalized by the corresponding value in a single-phase configuration, as a function of the droplet volume fraction. Our results indicate that the introduction of droplets into the single-phase flow enhances the Nusselt number, particularly at higher droplet volume fractions. The maximum enhancement observed is $10.2$\% at $\Upphi=0.5$. To provide a more precise assessment of the Nusselt number variation, we fit a third-order polynomial to the simulation data, yielding the following polynomial function: 
\begin{equation}
\dfrac{\left<{Nu}_{t,b}\right>}{\left<{Nu}_{sp}\right>} = 1.0 + 0.07202\Upphi + 0.2598{\Upphi}^2 + 0.01632{\Upphi}^3. 
\end{equation}
Note that as $\Upphi$ approaches zero, $\left<{Nu}_{t,b}\right>$ correctly converges to $\left<{Nu}_{sp}\right >$. Furthermore, with increasing values of $\Upphi$, the linear, quadratic, and cubic terms express an increase of the Nusselt number with $\Upphi$, indicating enhanced heat transfer due to the presence of the dispersed phase. It is interesting to note here that when deformable emulsions are replaced with rigid particles in the same RB convection flow, the Nusselt number exhibits a non-monotonic behavior, as reported in the study by \cite{demou2022turbulent}. In particular, the average heat transfer slightly increases by up to $\Upphi= 30$ \% and then decreases well below the single-phase reference value at $\Upphi=40$ \%. This behavior is attributed to the migration of particles toward the near-wall region, a mechanism absent in the case of emulsions. 
Here, we anticipate that the increase, ${\left<{Nu}_{t,b}\right>}/{\left<{Nu}_{sp}\right>}>1$, and the associated enhanced mixing, is due to the increase of small scale turbulence induced by interfacial stresses, despite 
the decrease of the dispersed-phase concentration in the near-wall region. A comprehensive exploration of the turbulence modulation and of the changes of the diffusion and convection terms at various droplet volume fractions will be provided in detail in section~\ref{subsec:heat_budget}. \par
\begin{figure}
\centering
\begin{subfigure}[t]{0.03\textwidth}
\centering
\fontsize{6}{9}
\textbf{(a)}
\end{subfigure}
\begin{subfigure}[t]{0.45\textwidth}
\includegraphics[width=\linewidth, valign=t]{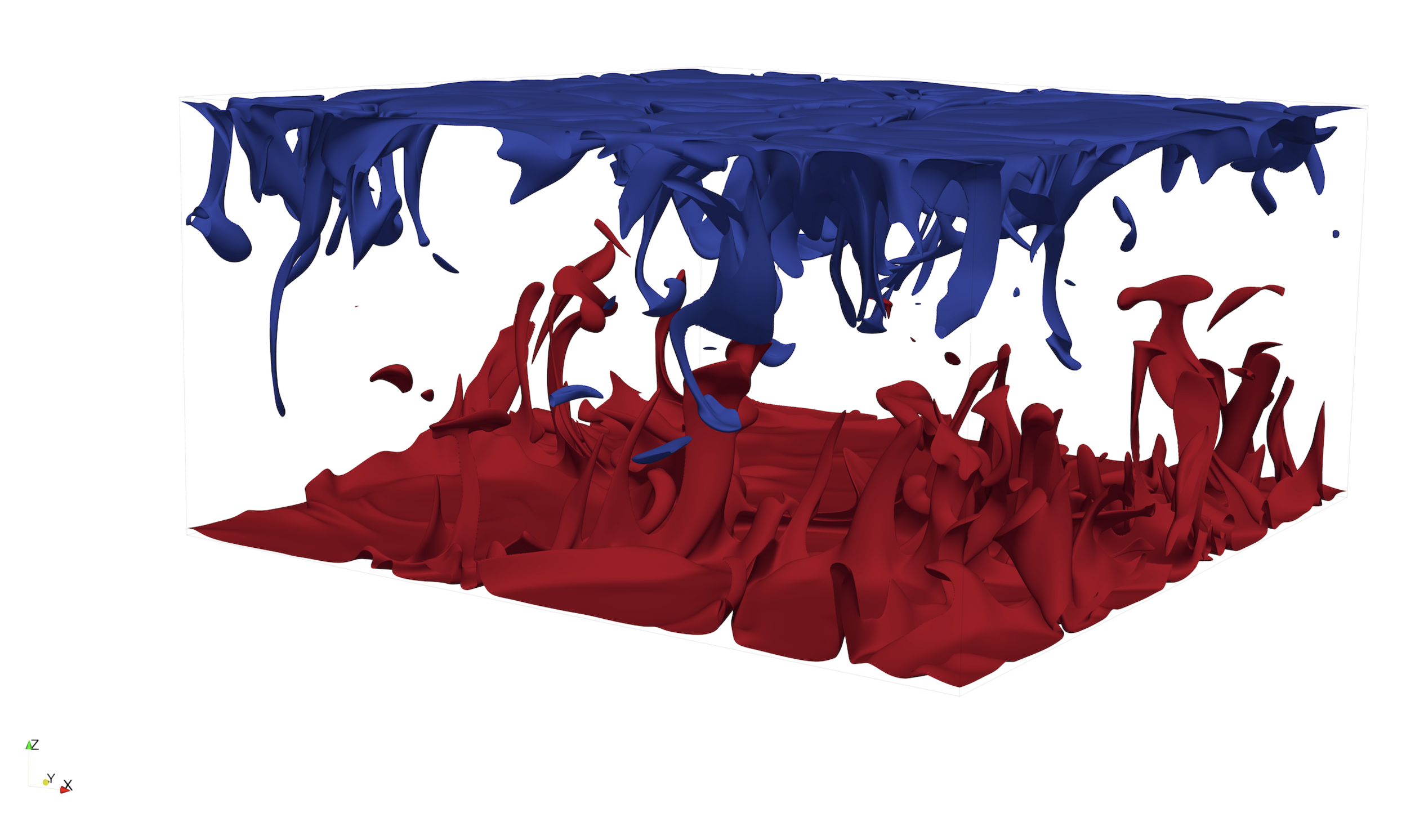}
\phantomsubcaption\label{fig:tmp_iso_surface_00}
\end{subfigure}
\begin{subfigure}[t]{0.03\textwidth}
\centering
\fontsize{6}{9}
\textbf{(b)}
\end{subfigure}
\begin{subfigure}[t]{0.45\textwidth}
\includegraphics[width=\linewidth, valign=t]{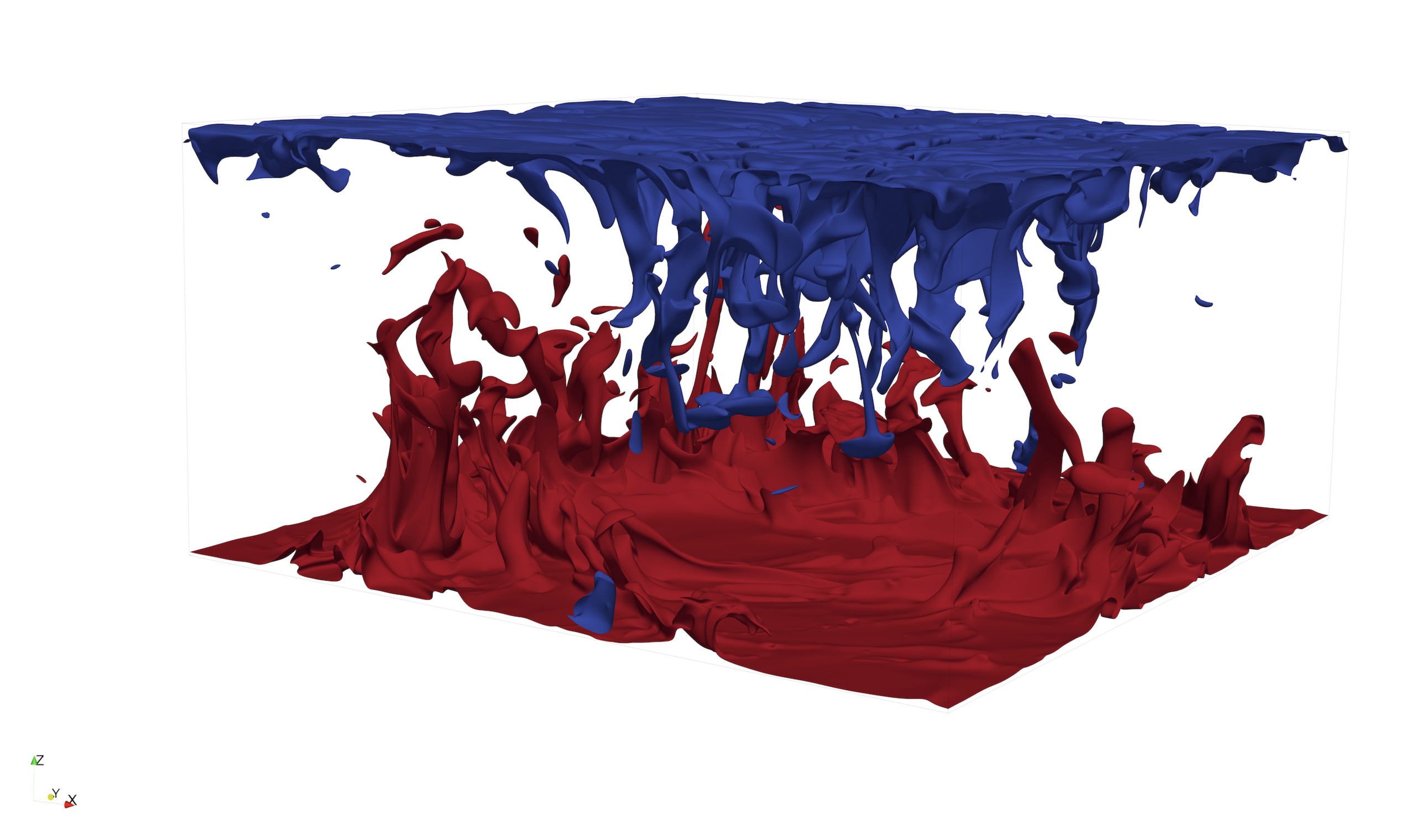}
\phantomsubcaption\label{fig:tmp_iso_surface_20}
\end{subfigure}
\begin{subfigure}[t]{0.03\textwidth}
\centering
\fontsize{6}{9}
\textbf{(c)}
\end{subfigure}
\begin{subfigure}[t]{0.45\textwidth}
\includegraphics[width=\linewidth, valign=t]{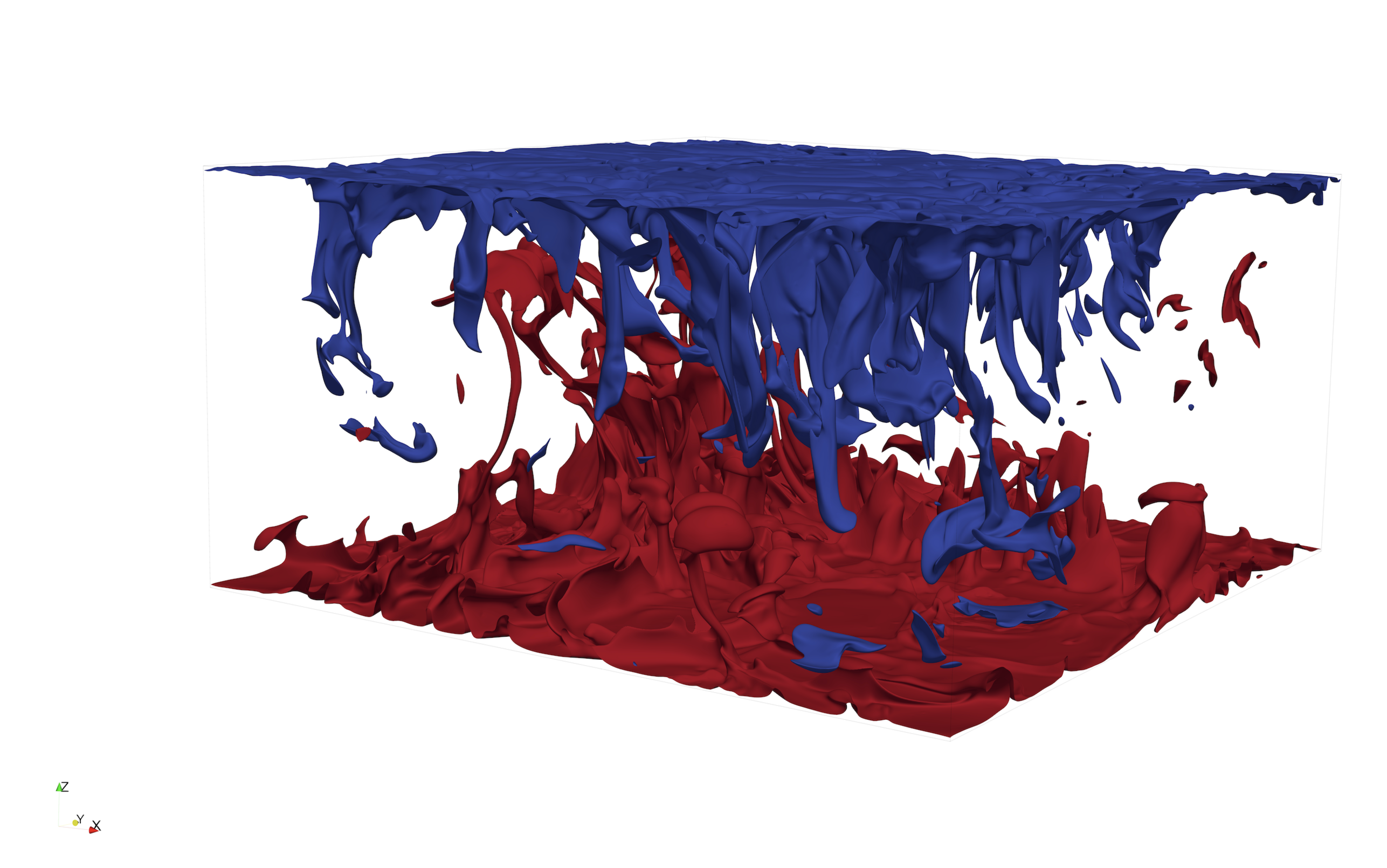}
\phantomsubcaption\label{fig:tmp_iso_surface_40}
\end{subfigure}
\begin{subfigure}[t]{0.03\textwidth}
\centering
\fontsize{6}{9}
\textbf{(d)}
\end{subfigure}
\begin{subfigure}[t]{0.45\textwidth}
\includegraphics[width=\linewidth, valign=t]{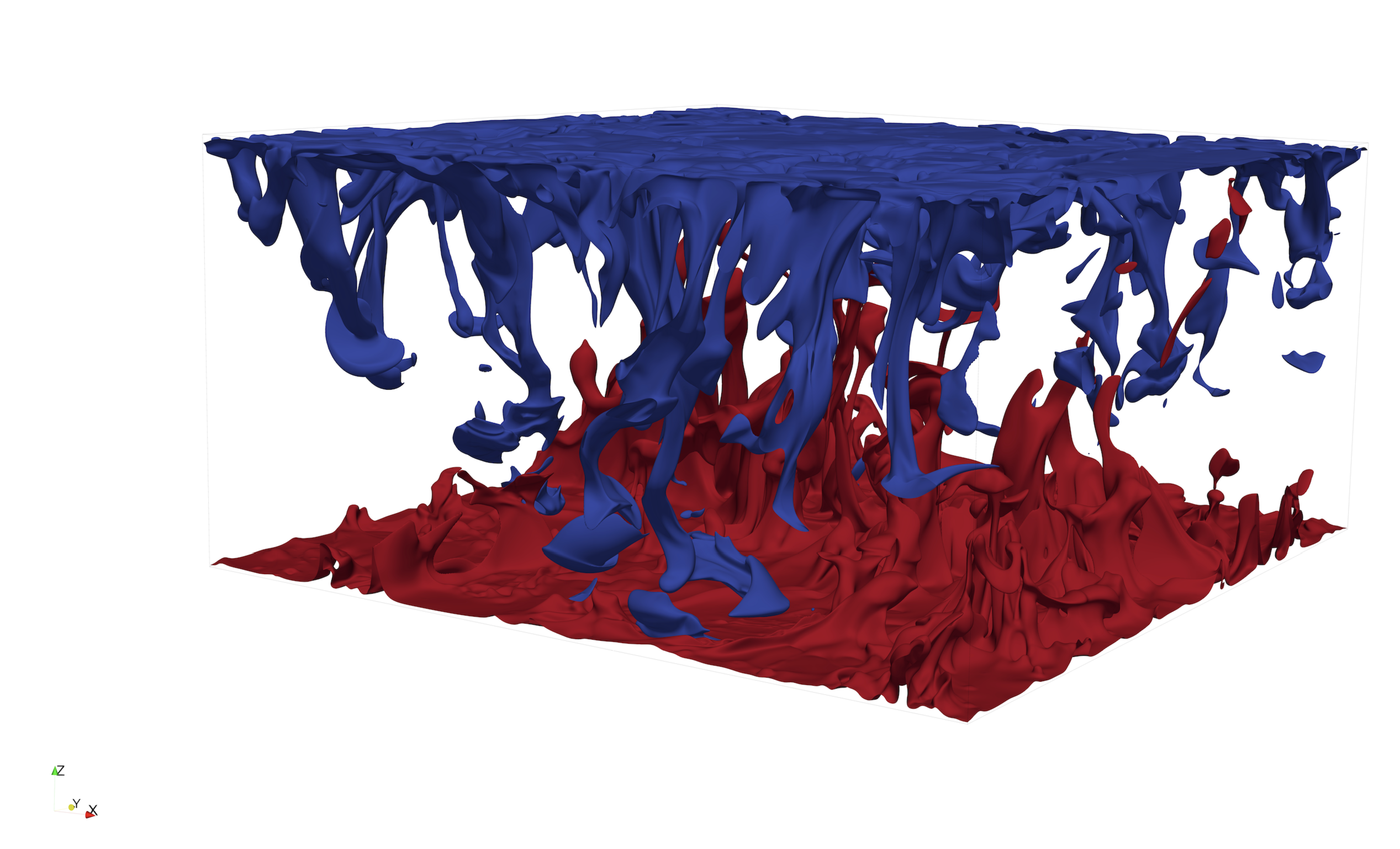}
\phantomsubcaption\label{fig:tmp_iso_surface_50}
\end{subfigure}
\begin{subfigure}[t]{0.03\textwidth}
\centering
\fontsize{6}{9}
\textbf{{\color{red}(e)}}
\end{subfigure}
\begin{subfigure}[t]{0.45\textwidth}
\includegraphics[width=\linewidth, valign=t]{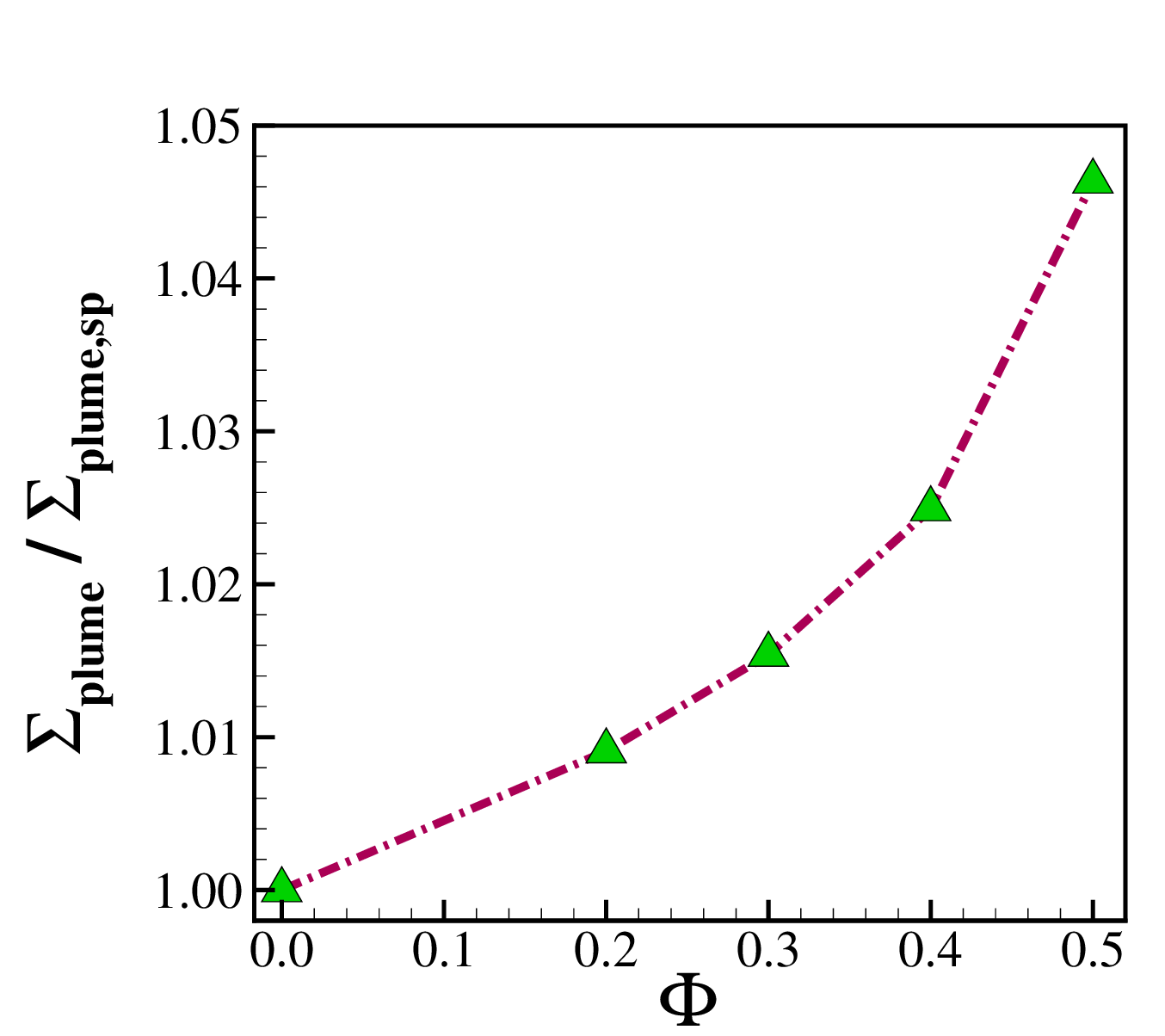}
\phantomsubcaption\label{fig:plume_phi}
\end{subfigure}
\caption{Instantaneous temperature iso-surfaces at $\Tilde{\theta} = \pm 0.1$ (blue color corresponds to $\Tilde{\theta} = - 0.1$ and red color corresponds to $\Tilde{\theta} = 0.1$) and for various dispersed-droplet volume fractions of: (a) $\Upphi= 0$, (b) $\Upphi= 0.2$, (c) $\Upphi= 0.4$ and (d) $\Upphi= 0.5$. (e) Volume fraction of thermal plume for cases 1-5 (Table \ref{table:list_of_simulations}) normalized by the volume fraction of thermal plumes in the single-phase case.}
\label{fig:tmp_3d_isosurfaces}
\end{figure}

Figure \ref{fig:tmp_3d_isosurfaces} offers visual representations of instantaneous temperature iso-surfaces at different dispersed droplet volume fractions. The temperature fields qualitatively corroborate the results presented in Figure \ref{fig:nu_phi}. As the droplet volume fraction increases, we observe an increase in the thermal plumes originating from both the upper and lower plates, accompanied by a reduction of the size of the flow structures.
To quantitatively confirm this conclusion, we conducted a quantitative assessment of thermal plumes, employing the established definition of thermal plumes condition provided by \citet{liu2022enhancing} as:
\begin{equation}
  |\theta - <\theta>| > \sqrt{<(\theta - <\theta>)^2>}
  \label{eq:thermal_plumes}
\end{equation}
where $\theta(\tilde{x}, \Tilde{t})$ being the local temperature and $<>$ the spatial and temporal average. In figure \ref{fig:plume_phi}, we presented the ratio of $\frac{\Sigma_\text{plume}}{\Sigma_\text{plume, sp}}$, which indicates the volume fraction of thermal plume for each case normalized by the volume fraction of thermal plumes in the single-phase case. Evidently, with an increased dispersed-fluid volume fraction, a noticeable rise (around 5\% at $\Upphi=0.5$) in thermal plumes from both the upper and lower plates is observed.
%
\begin{figure}
\centering
\begin{subfigure}[t]{0.03\textwidth}
\centering
\fontsize{6}{9}
\textbf{(a)}
\end{subfigure}
\begin{subfigure}[t]{0.45\textwidth}
\includegraphics[width=\linewidth, valign=t]{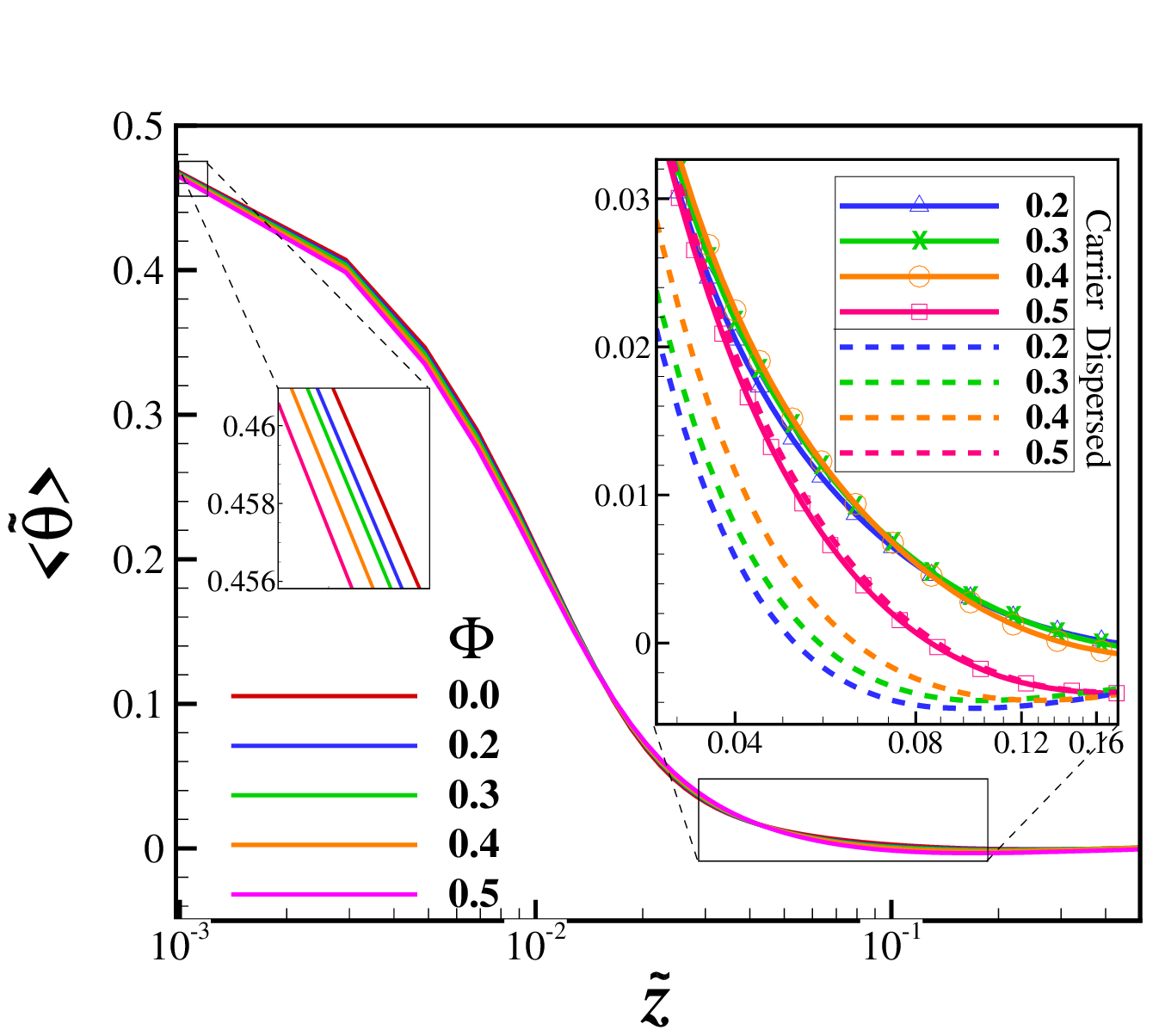}
\phantomsubcaption\label{fig:tmp_mean_tot}
\end{subfigure}
\begin{subfigure}[t]{0.03\textwidth}
\centering
\fontsize{6}{9}
\textbf{(b)}
\end{subfigure}
\begin{subfigure}[t]{0.45\textwidth}
\includegraphics[width=\linewidth, valign=t]{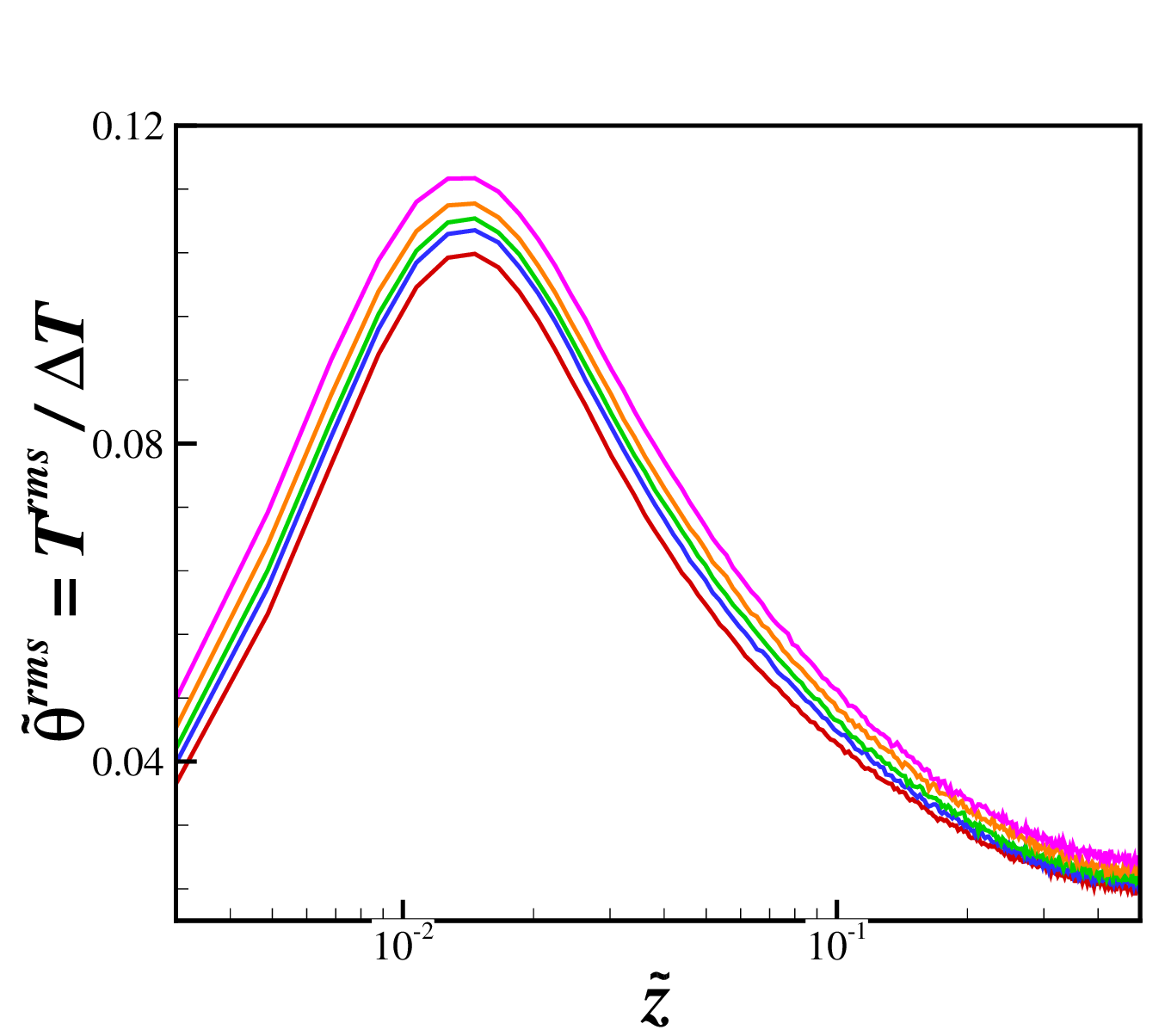}
\phantomsubcaption\label{fig:tmp_rms_tot}
\end{subfigure}
\begin{subfigure}[t]{0.03\textwidth}
\centering
\fontsize{6}{9}
\textbf{(c)}
\end{subfigure}
\begin{subfigure}[t]{0.45\textwidth}
\includegraphics[width=\linewidth, valign=t]{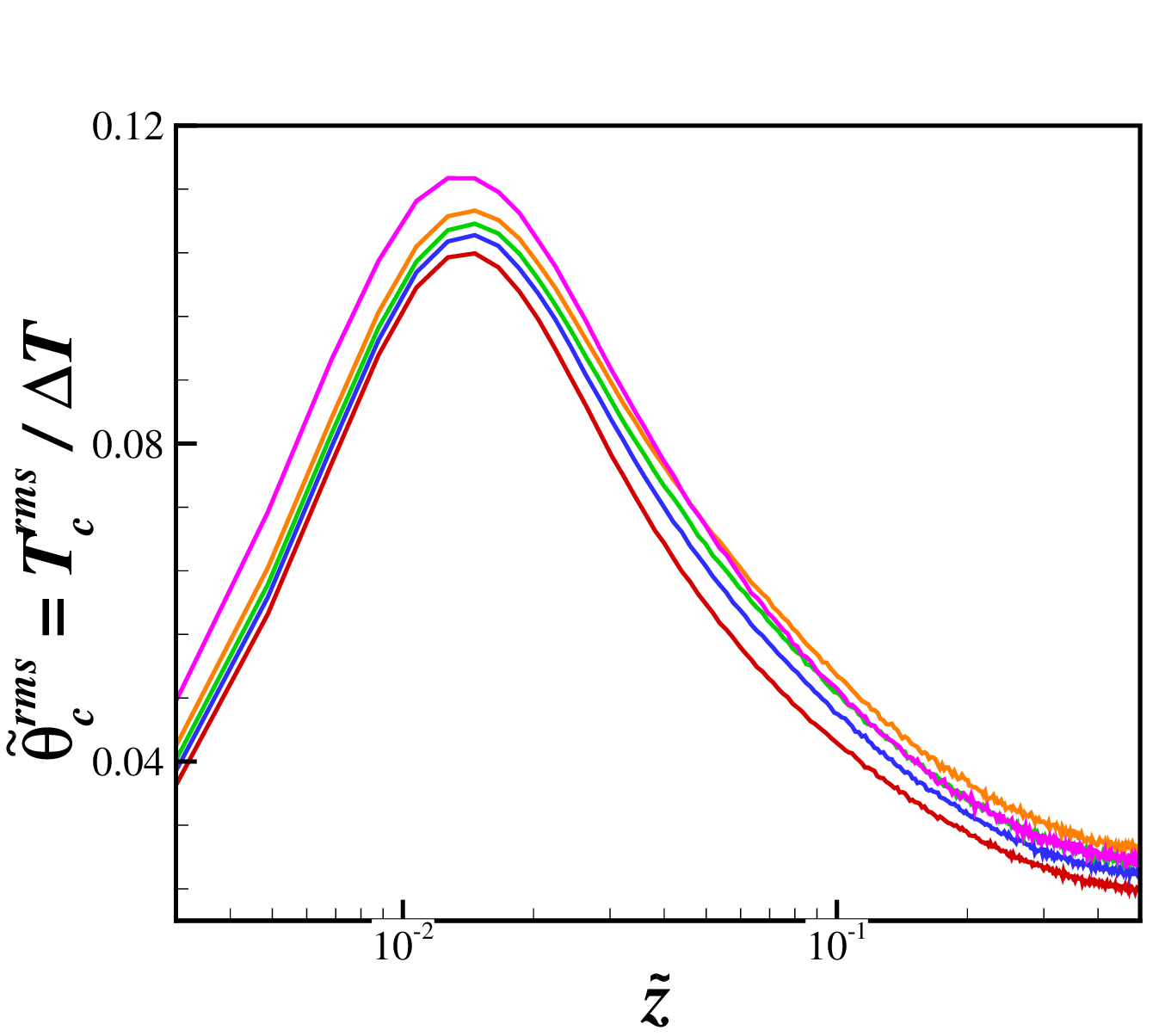}
\phantomsubcaption\label{fig:tmp_rms_c}
\end{subfigure}
\begin{subfigure}[t]{0.03\textwidth}
\centering
\fontsize{6}{9}
\textbf{(d)}
\end{subfigure}
\begin{subfigure}[t]{0.45\textwidth}
\includegraphics[width=\linewidth, valign=t]{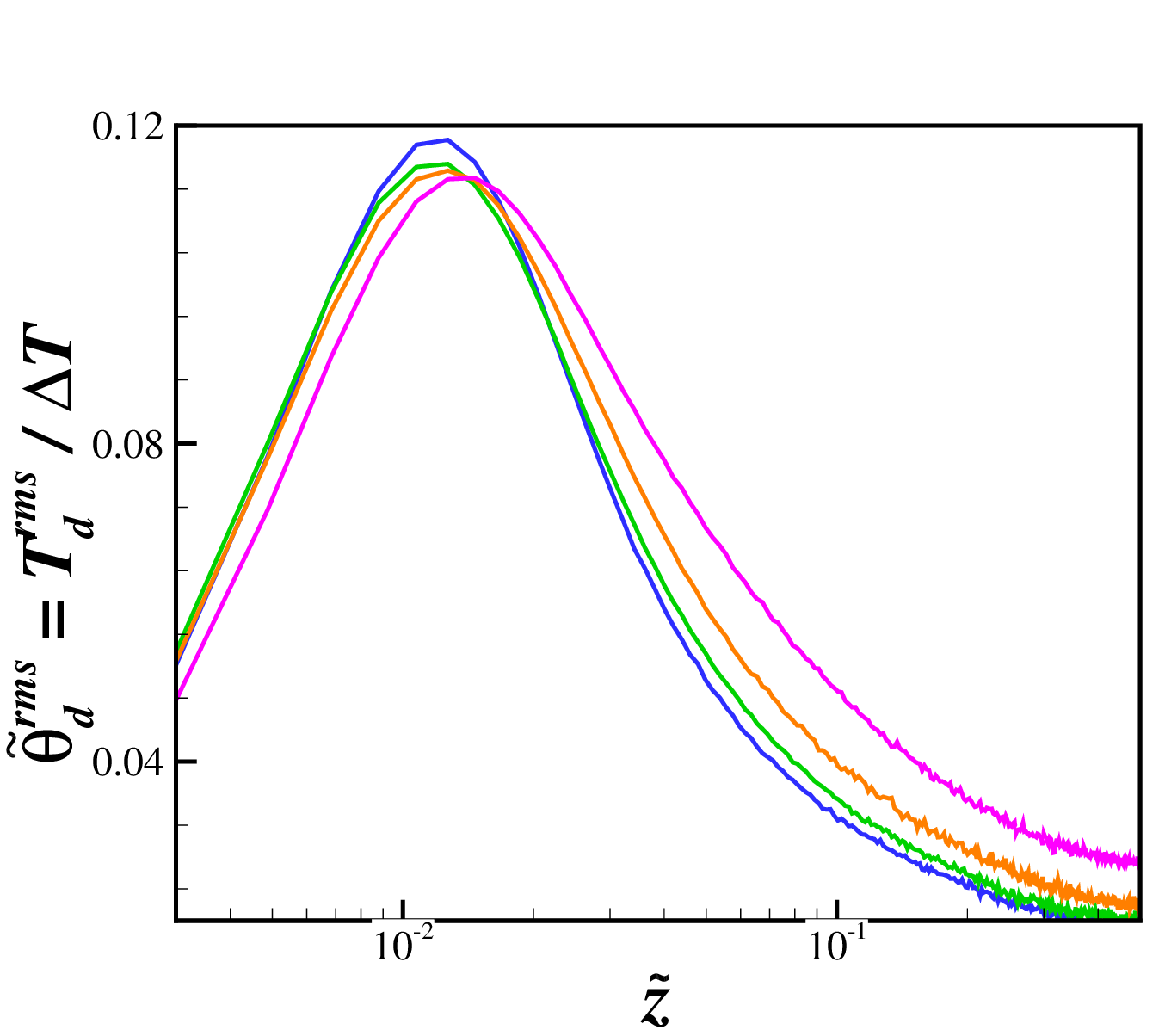}
\phantomsubcaption\label{fig:tmp_rms_d}
\end{subfigure}
\begin{subfigure}[t]{0.028\textwidth}
\centering
\fontsize{6}{9}
\textbf{(e)}
\end{subfigure}
\begin{subfigure}[t]{0.45\textwidth}
\centering
\includegraphics[width=\linewidth, valign=t]{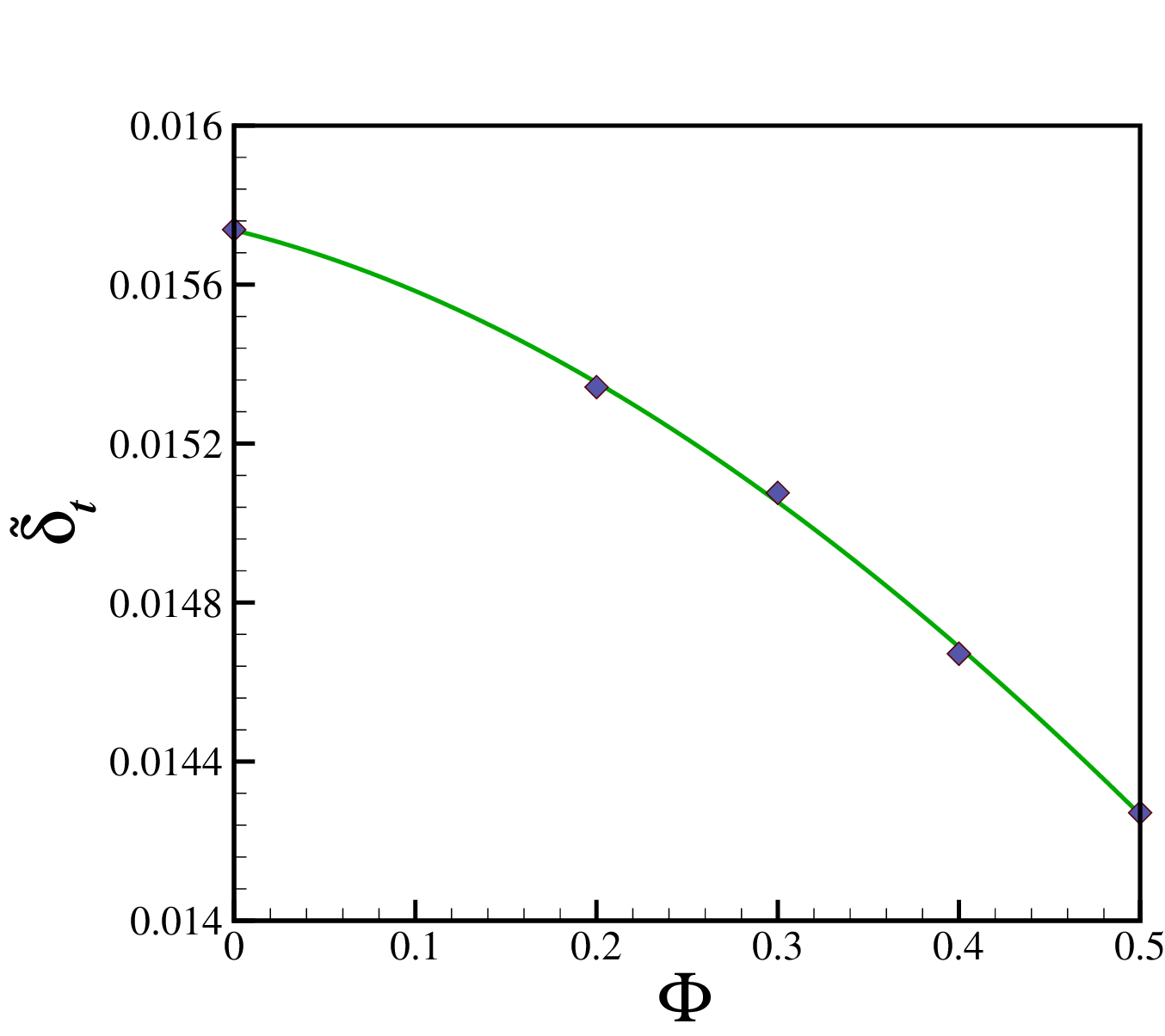}
\phantomsubcaption\label{fig:TBL_thickness}
\end{subfigure}
\caption{(a,b) mean $<\Tilde{\theta}>$ and rms ${\Tilde{\theta}}^{rms}$ temperature profiles of emulsion, (c,d) rms temperature profiles of carrier ${\Tilde{\theta}}_c^{rms}$ and dispersed ${\Tilde{\theta}}_d^{rms}$ phase, along the wall-normal direction for different droplet volume fractions; (e) thermal boundary layer thickness as a function of droplet volume fraction. In panel (a), one of the subsets shows the mean temperature per phase for a region within the TBL.}
\label{fig:tmp_mean_rms_TBL}
\end{figure}
\begin{figure}
\centering
\begin{subfigure}[t]{0.03\textwidth}
\centering
\fontsize{6}{9}
\textbf{(a)}
\end{subfigure}
\begin{subfigure}[t]{0.45\textwidth}
\includegraphics[width=\linewidth, valign=t]{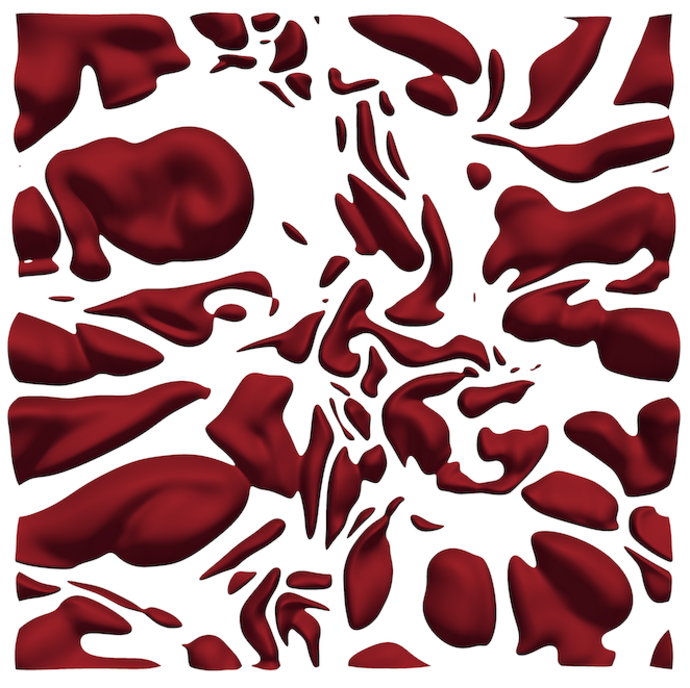}
\phantomsubcaption\label{fig:tmp_vof_isosurfaces_BL_00}
\end{subfigure}
\begin{subfigure}[t]{0.03\textwidth}
\centering
\fontsize{6}{9}
\textbf{(b)}
\end{subfigure}
\begin{subfigure}[t]{0.45\textwidth}
\includegraphics[width=\linewidth, valign=t]{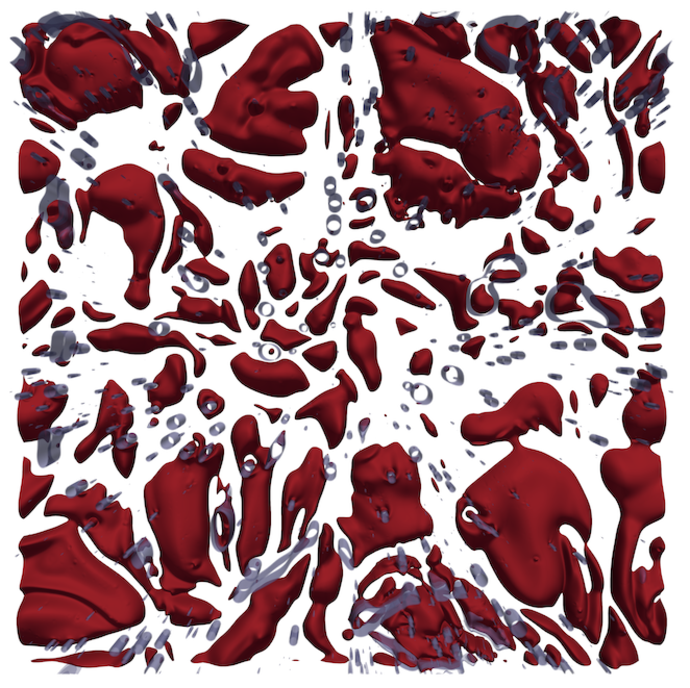}
\phantomsubcaption\label{fig:tmp_vof_isosurfaces_BL_20}
\end{subfigure}
\begin{subfigure}[t]{0.03\textwidth}
\centering
\fontsize{6}{9}
\textbf{(c)}
\end{subfigure}
\begin{subfigure}[t]{0.45\textwidth}
\includegraphics[width=\linewidth, valign=t]{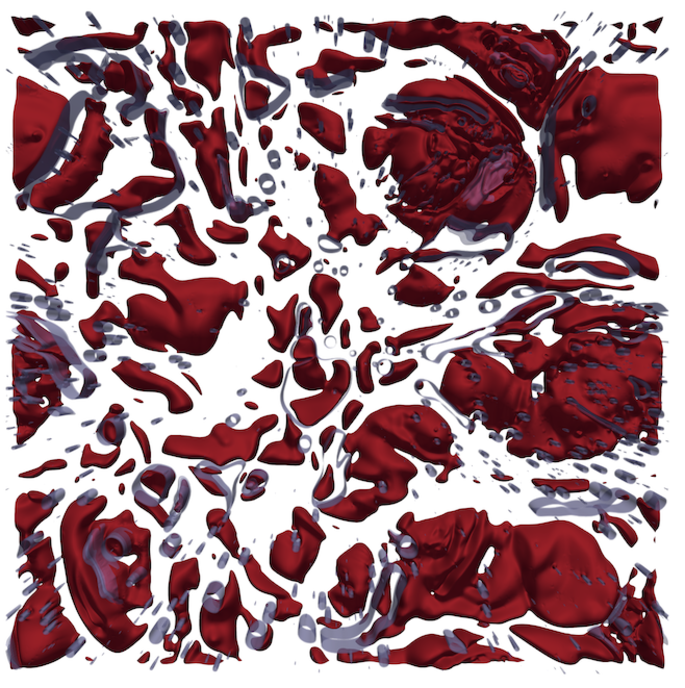}
\phantomsubcaption\label{fig:tmp_vof_isosurfaces_BL_40}
\end{subfigure}
\begin{subfigure}[t]{0.03\textwidth}
\centering
\fontsize{6}{9}
\textbf{(d)}
\end{subfigure}
\begin{subfigure}[t]{0.45\textwidth}
\includegraphics[width=\linewidth, valign=t]{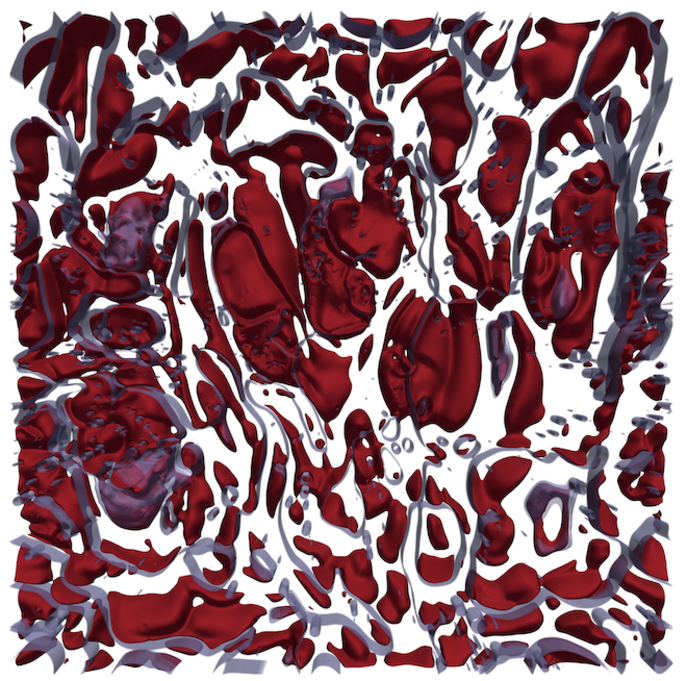}
\phantomsubcaption\label{fig:tmp_vof_isosurfaces_BL_50}
\end{subfigure}
\caption{A top view of the instantaneous temperature iso-surfaces (red color) in $\Tilde{x}-\Tilde{y}$ plane inside the hot-plate boundary layers at $\Tilde{\theta} = 0.3$ and for various dispersed-droplet volume fractions of: (a) $\Upphi= 0$, (b) $\Upphi= 0.2$, (c) $\Upphi= 0.4$ and (d) $\Upphi= 0.5$. Dispersed droplets are shown with gray color.}
\label{fig:tmp_vof_isosurfaces_BL}
\end{figure}
%
%
Figures \ref{fig:tmp_mean_tot}-\ref{fig:tmp_rms_d} report the mean and root-mean-square (rms) temperature fields along the wall-normal direction for the carrier phase, the dispersed phase, and the emulsion, where averages are taken in time and in wall-parallel planes. As one moves away from the hot (cold) wall, the average temperature gradually decreases (increases). This ultimately levels off to reach the temperature at the cavity center. The temperature fluctuation profiles exhibit a clear pattern: starting from zero at the walls, the peak near the walls defines the edge of the thermal boundary layer (TBL). Past this point, the fluctuations steadily decrease until they reach an almost constant value in the cavity's central region. 

To gain a deeper understanding of the heat transport mechanisms in turbulent Rayleigh–Bénard convection, it is crucial to access the dynamics within the thermal boundary layers  \citep{ahlers2009heat,chilla2012new,li2012boundary}. In this region, the heat transport is mainly due to conduction. Hence, we calculate the dimensionless TBL thickness, denoted as $\tilde{\delta}_{t}$, as~\citep{pope2000turbulent,ahlers2009heat,scheel2014local}
\begin{equation}
 \tilde{\delta}_{t}=\frac{{\delta}_{t}}{H} =\frac{1}{2} \left<{ \left |\frac{\partial \tilde{\theta}}{\partial \tilde{z}} \right| }_{\text{wall}}^{-1} \right>= \frac{0.5}{Nu},
  \label{eq:TBL_thickness}
\end{equation}
which indicates an inverse relationship between TBL thickness and temperature gradient at the walls. 
The TBL thickness
is reported in figure \ref{fig:TBL_thickness} for all cases under investigation.
In agreement with the data in figure \ref{fig:tmp_mean_tot}, we note that adding droplets alters the wall-normal temperature gradient. Specifically, there is a noticeable increase in the temperature gradient near the wall as $\Upphi$ increases, which corresponds to a decrease in the TBL thickness. As also documented later (see figure \ref{fig:DSD_visc_2}b), the presence of small droplets in the near wall region increases the local mixing and the global heat transfer.
The data confirm thinner thermal boundary layers at higher droplet volume fraction ($\Upphi$), from $0.0143 \leq \Tilde{z} \leq 0.0157$. At higher $\Upphi$ values, the flow experiences an increased level of heat transfer within the TBL, ultimately resulting in an improved total heat transfer rate. This conclusion can be confirmed by figure \ref{fig:tmp_vof_isosurfaces_BL}, which depicts the iso-surfaces of thermal plumes (red color) at $\Tilde{\theta} = 0.3$ and dispersed droplets (gray color) inside the boundary layer close to the hot wall, and for various dispersed-droplet volume fractions $\Upphi= [0-0.5]$. It is clear that adding the dispersed fluid to the single-phase flow enhances the mixing within the BLs and increases the amount of thermal plumes emitted from the plate and transported to the center of the cavity, which finally improves the Nusselt number and total heat transfer. Additionally, it is observed that, at higher $\Upphi$, there are more regions where thermal plumes are enclosed by the dispersed droplets, which carry the thermal plumes to the central regions by convection. 

An alternative approach to determine the TBL thickness consists in finding the maximum root-mean-square (rms) of the temperature profiles, displayed in figure \ref{fig:tmp_rms_tot}  for different droplet volume fractions. At higher $\Upphi$, the temperature fluctuations are more pronounced over the entire wall-normal direction, indicating
 increased mixing and heat transport at higher $\Upphi$ (as also observed in figure \ref{fig:tmp_vof_isosurfaces_BL}). 
 Upon closer examination of the figure, it becomes evident that the point of maximum temperature rms slightly moves closer to the wall when increasing  $\Upphi$, indicating a reduction in TBL thickness, a trend consistent with the findings presented in figure \ref{fig:TBL_thickness}. 
Moreover, figures \ref{fig:tmp_rms_c} and \ref{fig:tmp_rms_d} illustrate the temperature fluctuation profiles for each phase separately. Generally, higher fluctuations are observed at higher $\Upphi$ in the central regions of the cavity. An exception to this trend is observed within the carrier phase, where a slight reduction in fluctuation levels is discernible when $\Upphi$ exceeds $0.4$. 
Additionally, the dispersed-phase data (figure \ref{fig:tmp_rms_d}) display a decrease of the temperature fluctuation peak at higher $\Upphi$, accompanied by a shift in the peak position away from the wall.  

\begin{figure}
\center
    \includegraphics[width=0.7\linewidth]{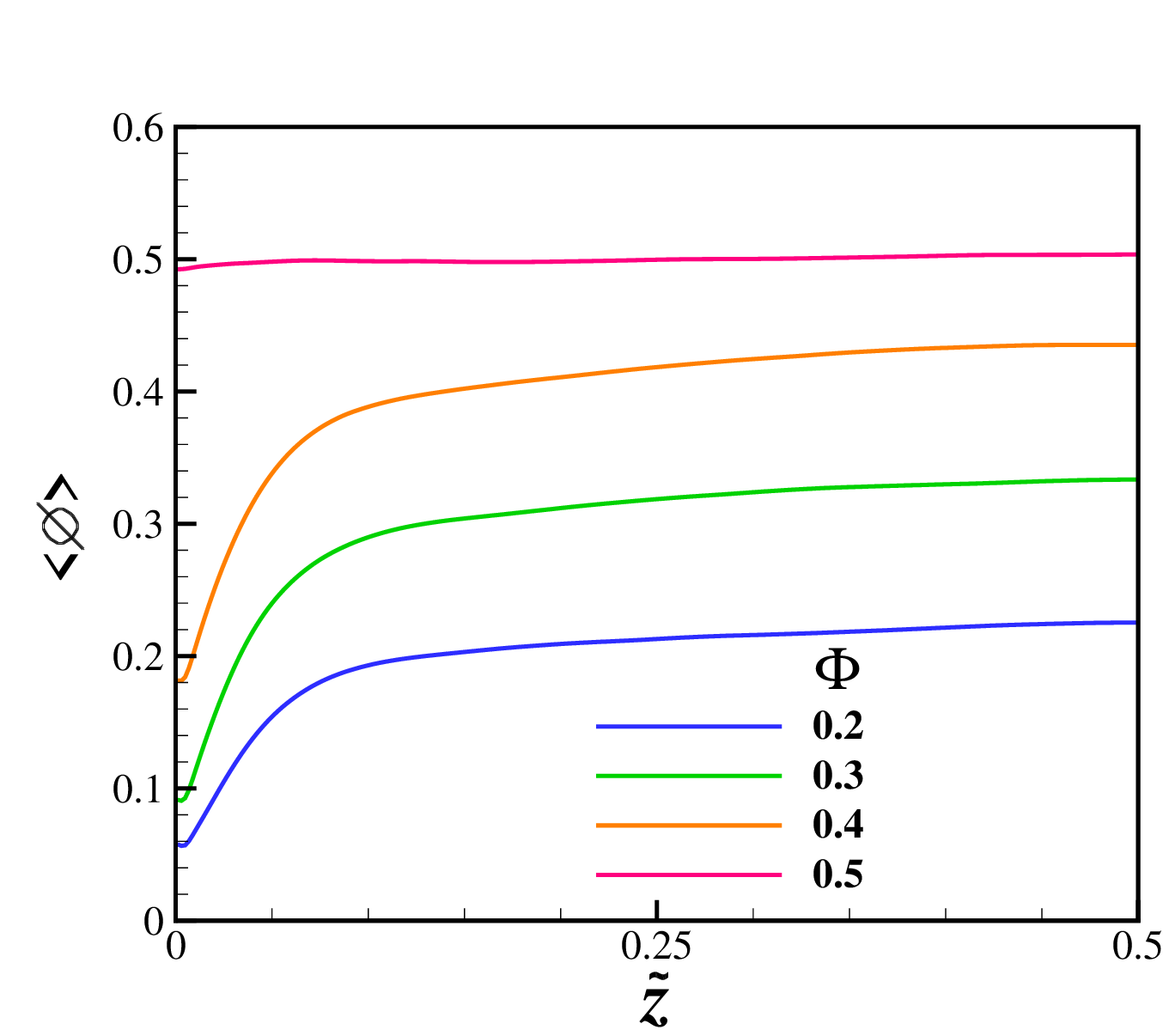} 
    \caption{Mean wall-normal distribution of local droplet volume fraction for various droplet volume fractions.} 
    \label{fig:phi_local} 
\end{figure}

To better understand these data, one crucial aspect deserving detailed analysis is the local distribution of the dispersed phase within the cavity. To compute this, we consider the local volume fraction $\phi$ is defined as the portion of a computational cell occupied by the dispersed phase. This takes the value 1 when the cell is fully occupied by a droplet and zero when the cell is fully occupied by the carrier phase. The wall-normal distributions of the spatially and temporally averaged local volume fraction, denoted as $<\phi>$, are displayed in figure \ref{fig:phi_local} for different nominal values of the dispersed-phase volume fraction, $\Upphi$. 
An approximately uniform distribution is observed within the core of the cavity for all cases. Interestingly, however, fewer droplets can be found near the wall, except for the case of a binary mixture, where droplets are evenly distributed throughout the wall-normal direction as expected by symmetry considerations ($\Upphi=0.5$ with density and viscosity ratio equal to 1 for the cases in the figure). 
Further information regarding the distributions and sizes of the dispersed droplets will be provided in section \ref{subsec:DSD}. 
In the case of TRB of a rigid-particle suspensions,
\cite{demou2022turbulent} observed a distinct near-wall peak in the dispersed-phase distribution. This peak, associated with particle layering, becomes more pronounced at higher particle volume fractions. The authors attribute this layering to the strong wall-particle lubrication interaction, which stabilizes the wall-normal position of particles after reaching the wall. Consequently, it becomes increasingly challenging for particles within the first layer to disengage from it. The opposite trend is observed in the case of deformable droplets.

\begin{figure}
\centering
\begin{subfigure}[t]{0.03\textwidth}
\centering
\fontsize{6}{9}
\textbf{(a)}
\end{subfigure}
\begin{subfigure}[t]{0.45\textwidth}
\includegraphics[width=\linewidth, valign=t]{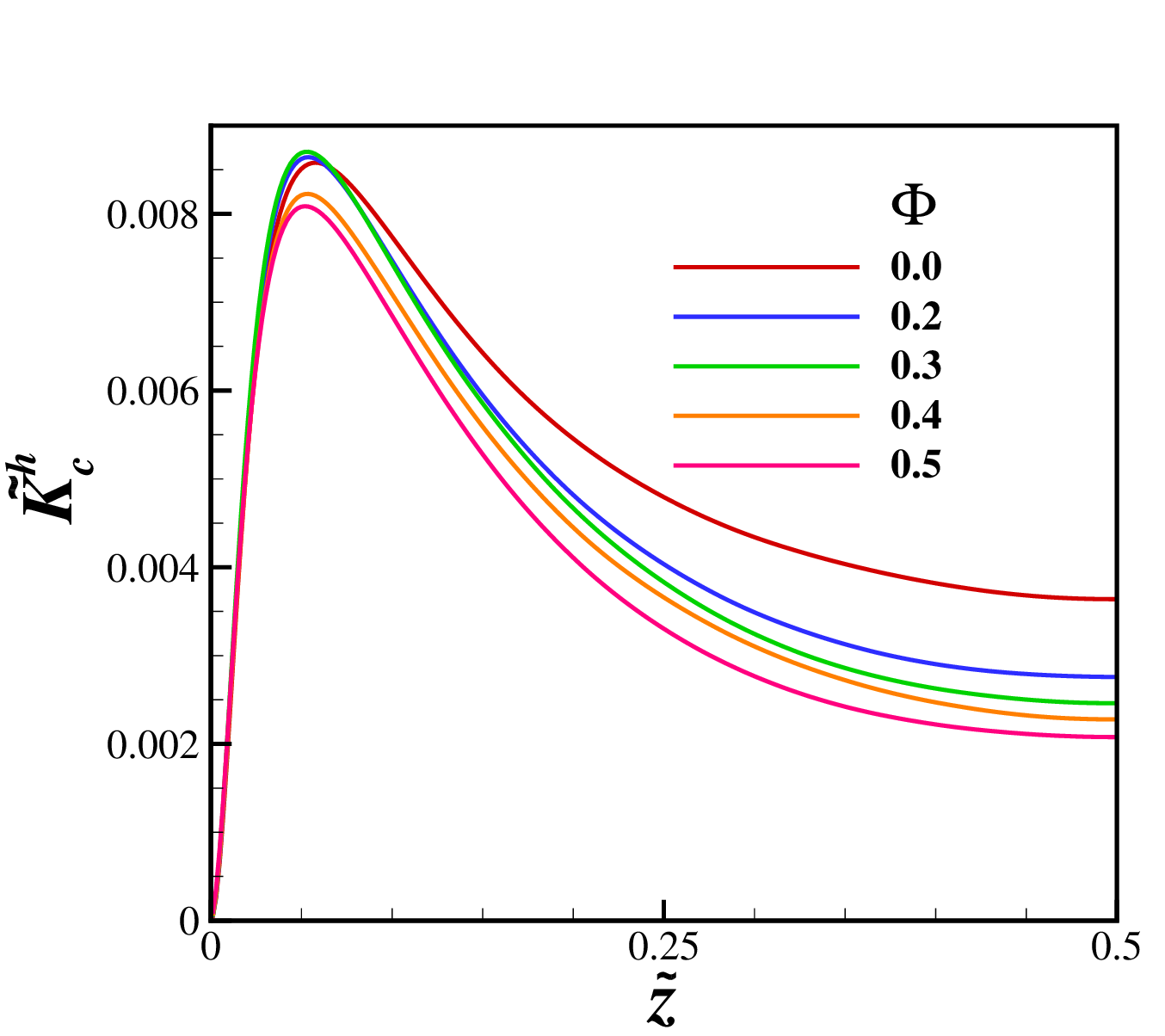}
\phantomsubcaption\label{fig:K h c sym}
\end{subfigure}\hfill
\begin{subfigure}[t]{0.03\textwidth}
\centering
\fontsize{6}{9}
\textbf{(b)}
\end{subfigure}
\begin{subfigure}[t]{0.45\textwidth}
\includegraphics[width=\linewidth, valign=t]{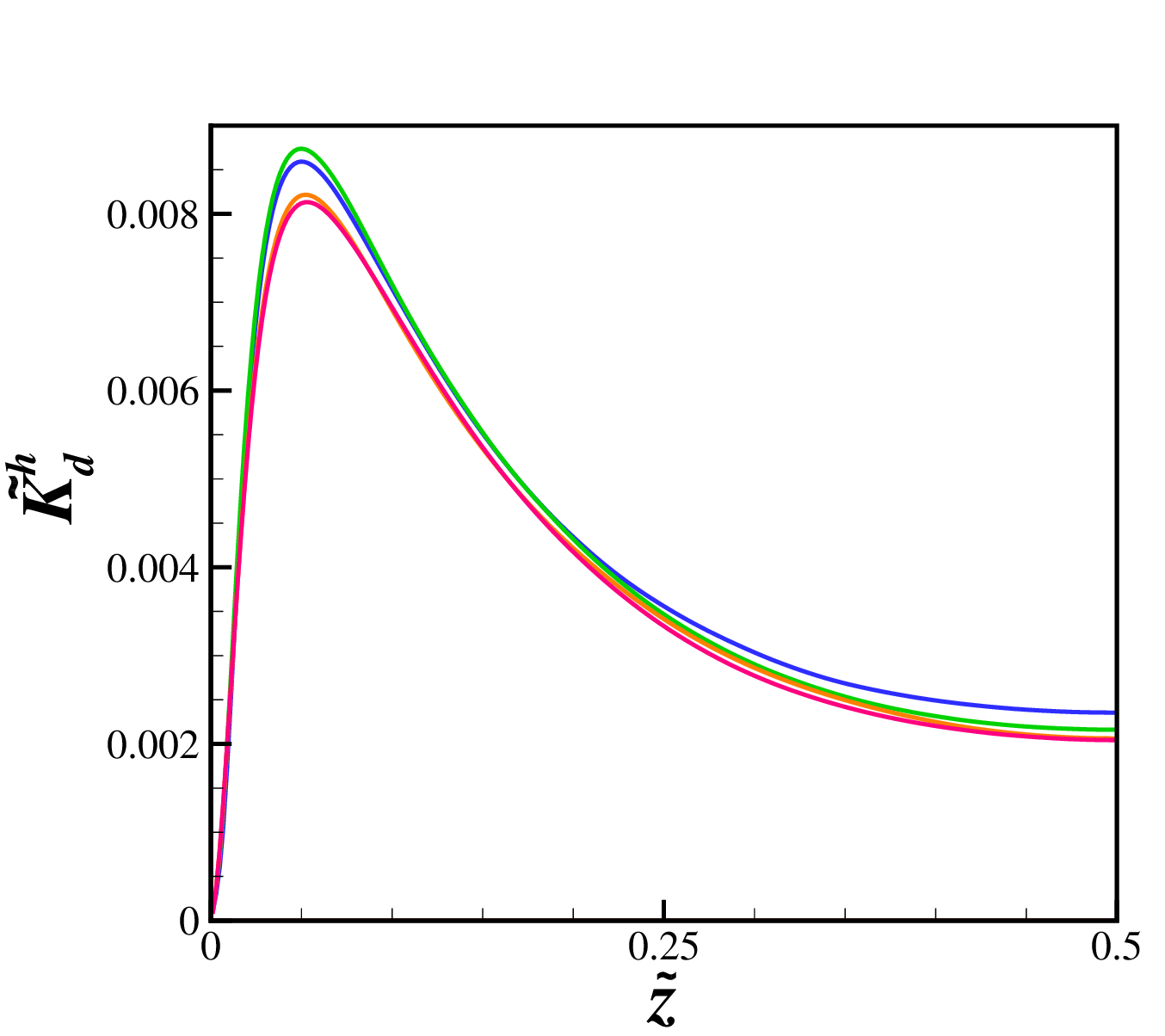}
\phantomsubcaption\label{fig:K h d sym}
\end{subfigure}
\begin{subfigure}[t]{0.028\textwidth}
\centering
\fontsize{6}{9}
\textbf{(c)}
\end{subfigure}
\begin{subfigure}[t]{0.45\textwidth}
\centering
\includegraphics[width=\linewidth, valign=t]{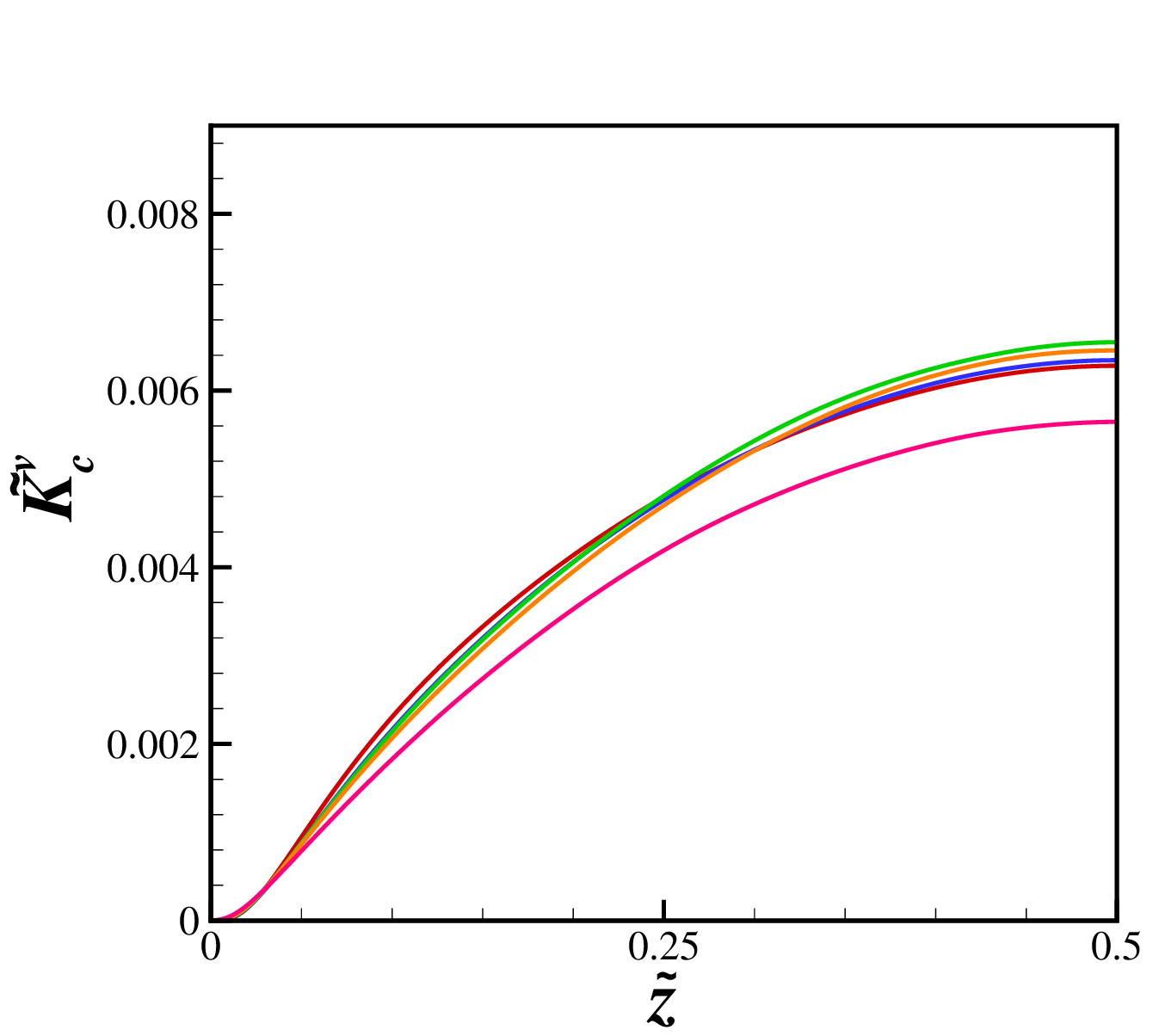}
\phantomsubcaption\label{fig:K v c sym}
\end{subfigure}
\begin{subfigure}[t]{0.048\textwidth}
\centering
\fontsize{6}{9}
\textbf{(d)}
\end{subfigure}
\begin{subfigure}[t]{0.45\textwidth}
\includegraphics[width=\linewidth, valign=t]{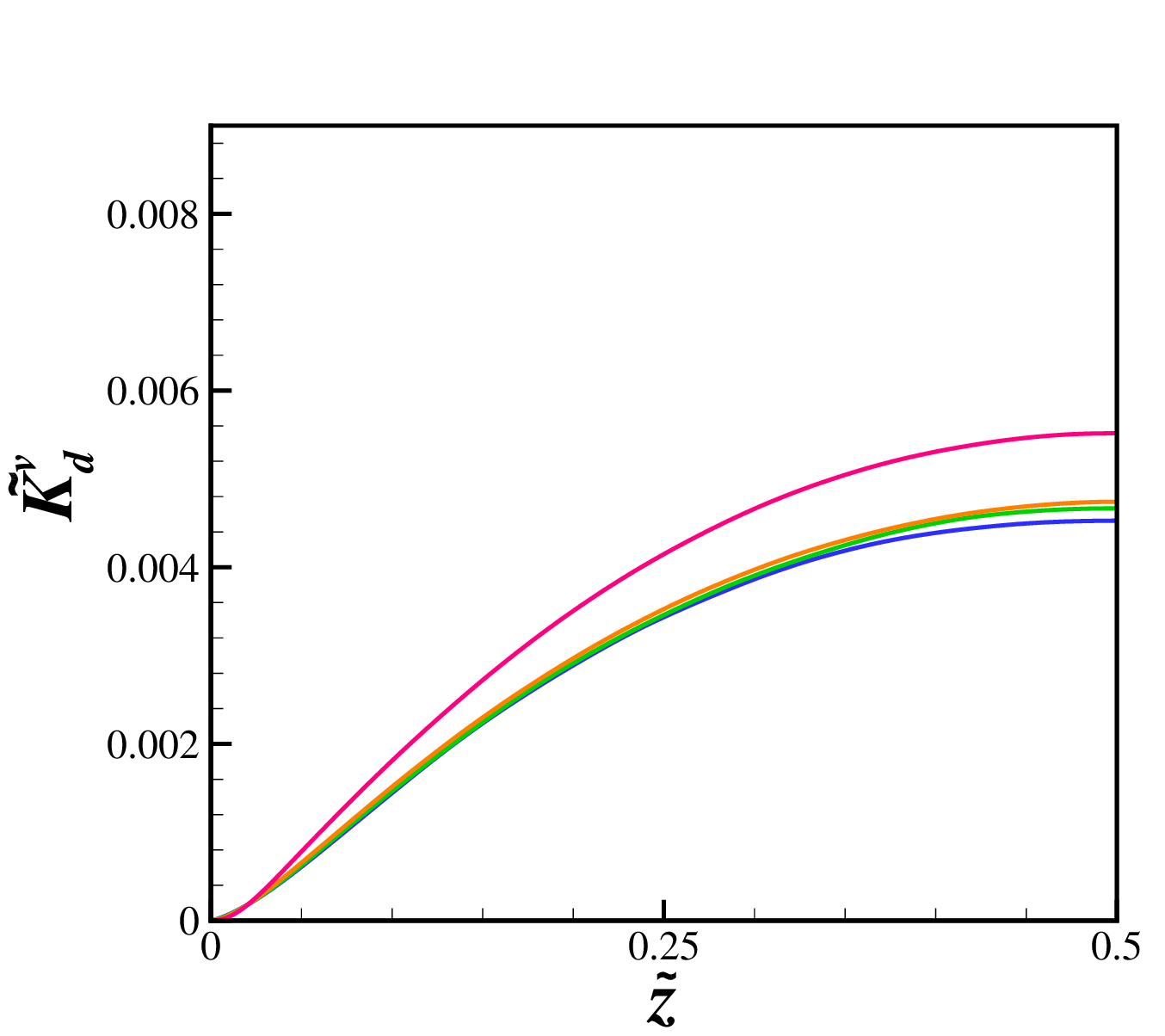}
\phantomsubcaption\label{fig:K v d sym}
\end{subfigure}
\caption{The horizontal (a,b) and vertical (c,d) components of the average kinetic energy per unit mass, derived from the velocity rms (equations \ref{eq:kinetic_energy_carrier} and \ref{eq:kinetic_energy_droplet}), as a function of the vertical direction for the different droplet volume fractions.}
\label{fig:kinetic_energy}
\end{figure}

Another global quantity of interest is the kinetic energy of the two phases. To investigate the contributions of various flow structures, it is useful to partition the average kinetic energy per unit mass into vertical and horizontal components as follows \citep{demou2022turbulent}
\begin{equation}
 \tilde{K}_c^h=\frac{1}{2} {\left[{({\tilde{u}}_c^{\text{rms}})}^2 + {({\tilde{v}}_c^{\text{rms}})}^2\right] }, \qquad \tilde{K}_c^v=\frac{1}{2} {\left[{({\tilde{w}}_c^{\text{rms}})}^2 \right] }, 
  \label{eq:kinetic_energy_carrier}
\end{equation}
\begin{equation}
 \tilde{K}_d^h=\frac{1}{2} {\left[{({\tilde{u}}_d^{\text{rms}})}^2 + {({\tilde{v}}_d^{\text{rms}})}^2 \right] }, \qquad \tilde{K}_d^v=\frac{1}{2} {\left[{({\tilde{w}}_d^{\text{rms}})}^2\right] }. 
  \label{eq:kinetic_energy_droplet}
\end{equation}
In particular, the vertical kinetic energy captures the dynamics of vertical motions within the core of the cavity, while the horizontal kinetic energy is linked to the velocity of the kinetic boundary layers near the walls. 
Figure \ref{fig:kinetic_energy} displays the variations in mean horizontal and vertical kinetic energy along the wall-normal direction for the various $\Upphi$ under investigation in both dispersed and carrier phases. 
As expected, the horizontal components (see panels \hyperref[fig:K h c sym]{\subref*{fig:K h c sym}} and \hyperref[fig:K h d sym]{\subref*{fig:K h d sym}}) exhibit a peak near the wall and gradually approach a nearly constant value in the cavity center. 
This trend is similar to that of the temperature rms. In contrast, the maximum vertical kinetic energy (depicted in panels \hyperref[fig:K v c sym]{\subref*{fig:K v c sym}} and \hyperref[fig:K v d sym]{\subref*{fig:K v d sym}}) is attained at the cavity center, with a gradual decrease towards the wall. The location of the peak of the horizontal kinetic energy corresponds to the edge of the kinetic boundary layer (KBL). Note that, at higher $\Upphi$, the KBL remains relatively unchanged, whereas the TBL decreases. This suggests that the large-scale circulation structures do not significantly  vary when 
$\Upphi$ changes. Given the value of the Prandtl number ($Pr=4$ in this study), indicating a difference in velocity and thermal boundary layers, we expect the velocity boundary layer to be thicker than the thermal boundary layer. By comparing the locations of maximum horizontal kinetic energy and temperature rms, our observations align with this expectation.

\begin{figure}
\centering
\begin{subfigure}[t]{0.03\textwidth}
\centering
\fontsize{6}{9}
\textbf{(a)}
\end{subfigure}
\begin{subfigure}[t]{0.45\textwidth}
\includegraphics[width=\linewidth, valign=t]{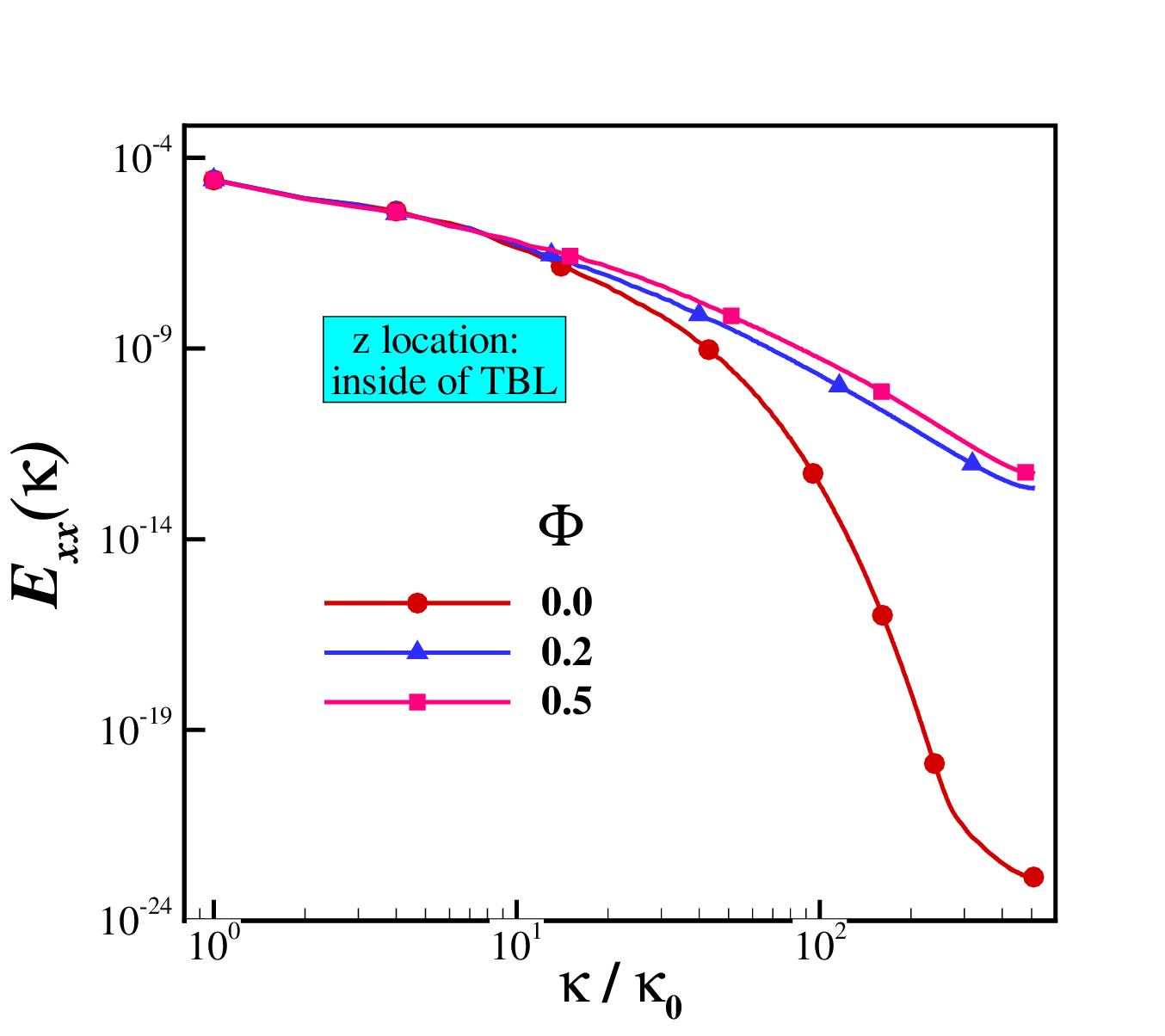}
\phantomsubcaption\label{fig:Exx_in_phi}
\end{subfigure}\hfill
\begin{subfigure}[t]{0.03\textwidth}
\centering
\fontsize{6}{9}
\textbf{(b)}
\end{subfigure}
\begin{subfigure}[t]{0.45\textwidth}
\includegraphics[width=\linewidth, valign=t]{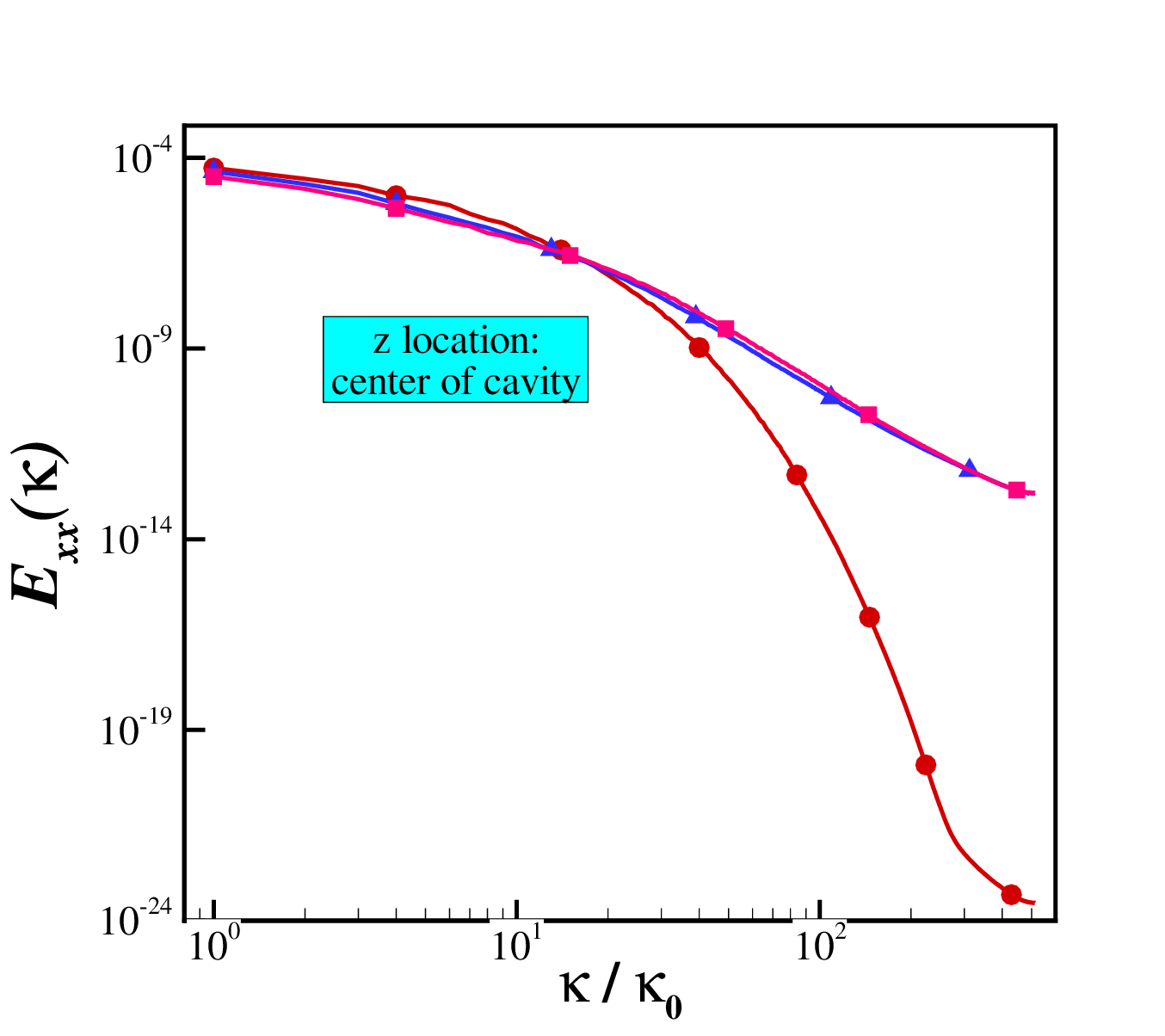}
\phantomsubcaption\label{fig:Exx_out_phi}
\end{subfigure}
\caption{The time-averaged spectrum of TKE as a function of wavenumber at different droplet volume fractions (a) inside of TBL and (b) at the center of cavity. Wavenumbers are normalized by the lowest non-zero wavenumber $\kappa_0 = \pi / H$.
}
\label{fig:spectra_phi}
\end{figure}

Considering the differences with the dispersed-phase volume fraction, figures \ref{fig:K h c sym} and \ref{fig:K h d sym} illustrate a noticeable damping of the horizontal components of the kinetic energy with the volume fraction  $\Upphi$, which is observed in both phases, however more evident for the carrier phase. Note, however, that horizontal velocity fluctuations do not directly contribute to turbulent heat transport.
Conversely, as demonstrated in figures \ref{fig:K v c sym} and \ref{fig:K v d sym}, 
the level of vertical fluctuations does not vary significantly with  $\Upphi$, except for the binary flow at $\Upphi=0.5$, when we observe a reduction for both phases. 
Referring back to figure \ref{fig:phi_local}, we recall that droplets tend to be distributed predominantly within the central region of the cavity, with fewer near the cavity walls. Furthermore, by increasing the volume fraction of the dispersed droplets, there is an increase in their absolute concentration within the cavity core. This implies that at higher $\Upphi$, a larger number of dispersed droplets actively engages in the vertical large-scale circulations of the flow. At $\Upphi=0.5$, it is not possible to define carrier and dispersed phase, and, indeed, the contribution to the vertical large-scale circulation becomes equal within statistical accuracy. 
When considering a suspension of rigid particles,  \cite{demou2022turbulent} reported a significant decrease in both the horizontal and vertical components of both phases at higher volume fractions of the secondary phase, which indicates a weakening of the large-scale circulation structures. 
We have, therefore, seen that the increase in the Nusselt number is not associated with an increase in the large-scale motions (quantified by the average turbulent kinetic energy). We therefore consider the energy spectra, in particular the one-dimensional longitudinal spectra associated with the horizontal velocity (the same trends are observed for the other components). Results pertaining to the single-phase flow, emulsions with volume fraction $\Upphi = 0.2-0.5$ are displayed in figure \ref{fig:spectra_phi} where we consider spectra $E_{hh}(\kappa_h)$ extracted inside the TBL and at the cavity midplan. As demonstrated, the presence of dispersed droplets notably enhances the energy at small scales (high wavenumbers), with this effect becoming more pronounced at higher $\Upphi$ values particularly noticeable within boundary layers (as depicted in figure \ref{fig:Exx_in_phi}). This indicates higher $Nu$ at higher $\Upphi$ as $Nu$ is evaluated at the wall. Conversely, introducing the dispersed droplets reduces the energy at large scales (low wavenumbers), a phenomenon more distinctly observed at the center of the cavity (figure \ref{fig:Exx_out_phi}). 
As reported in previous studies, the presence of the interface provides an alternative mechanism for energy transfer at small scales, typically at scales smaller than the smallest of the corresponding single phase flow \cite[see][]{Perlekar2019,crialesi2022modulation}. This enhances the small-scale mixing in the near-wall region and possibly explains the increase in global heat transfer in emulsions of two fluids with the same properties.
\subsection{Heat Transfer Budgets}\label{subsec:heat_budget}
\begin{figure}
  \centering
  \begin{subfigure}[t]{0.03\textwidth}
  \fontsize{6}{9}
    \textbf{(a)}
  \end{subfigure}
  \begin{subfigure}[t]{0.37\textwidth}
    \includegraphics[width=\linewidth, valign=t]{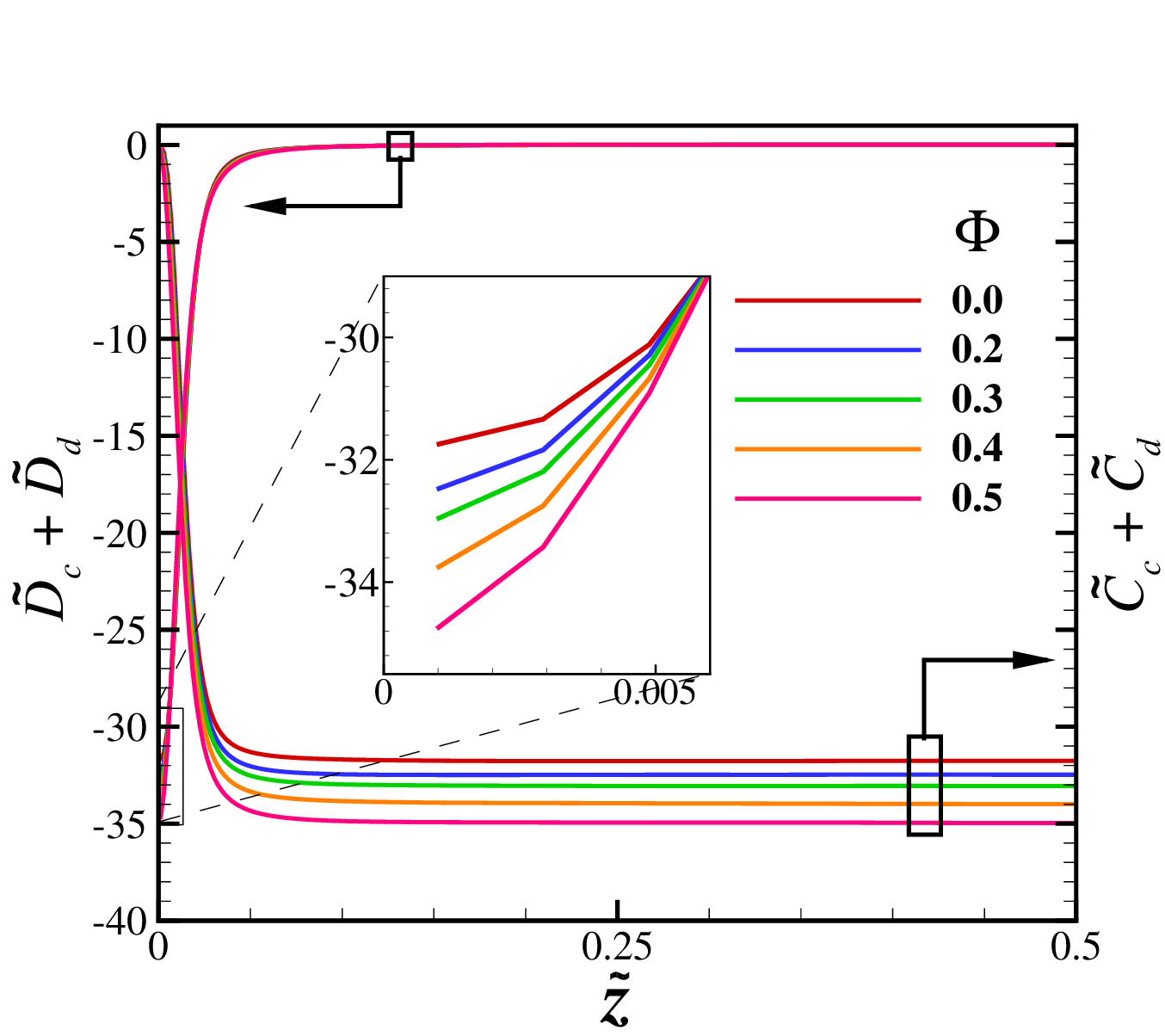}
    \phantomsubcaption\label{fig:DcPDd_CcPCd}
  \end{subfigure}
   \begin{subfigure}[t]{0.03\textwidth}
  \fontsize{6}{9}
    \textbf{(b)}
  \end{subfigure}
  \begin{subfigure}[t]{0.53\textwidth}
  \vspace{0.1cm}
    \includegraphics[width=\linewidth, valign=t]{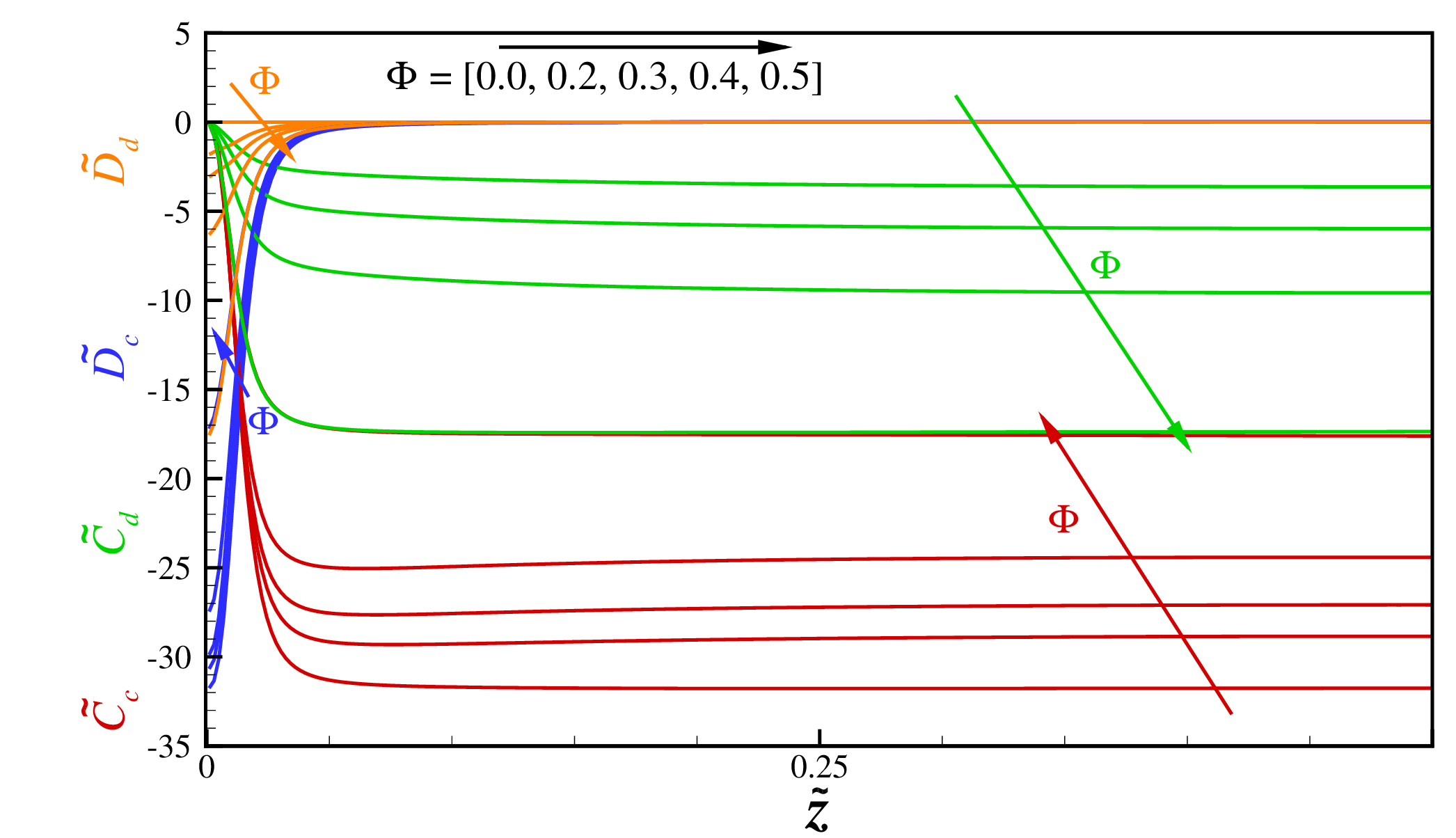}
    \phantomsubcaption\label{fig:heat_budget}
  \end{subfigure}
  \begin{subfigure}[t]{0.03\textwidth}
  \fontsize{6}{9}
    \textbf{(c)}
  \end{subfigure}
  \begin{subfigure}[t]{0.7\textwidth}
    \includegraphics[width=\linewidth, valign=t]{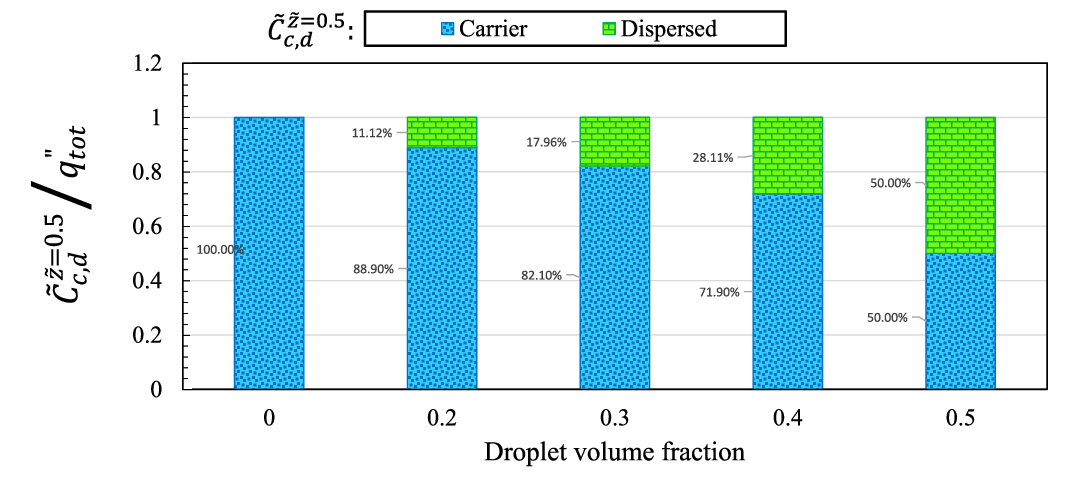}
    \phantomsubcaption\label{fig:heat_budget_barchart}
  \end{subfigure}
  \caption{(a) The total convection and total diffusion heat transfer, and (b) the convection and diffusion heat fluxes for dispersed and carrier phases, along the wall-normal direction and for various droplet volume fractions. (c) The carrier- and dispersed-fluid convection heat fluxes at the center of cavity ($\Tilde{z}=0.5$), normalized by the total heat flux of each case with various $ \Upphi$.}
  \label{fig:heat_budget_graphs}
\end{figure}

This section examines the heat transfer budget, which is derived by applying the phase-ensemble-averaging to the heat transfer equation. Using the framework developed and employed in \cite{marchioro1999mixture,zhang2010physics}, the different contributions to the total heat transfer are written in terms of convective and diffusive fluxes in each phase\citep{ahlers2009heat,ardekani2018heat,demou2022turbulent} as
%
\begin{equation}
q_{tot}'' = C_c + C_d + D_c + D_d,
  \label{eq:heat_budget}
\end{equation}
where,
\begin{subequations}
\begin{gather}
 C_c = - (1-<\phi>)<w_c'T_c'> ,
  \label{eq:Cc_heat_budget_a}\\
 C_d = -<\phi><w_d'T_d'>,
  \label{eq:Cd_heat_budget_b}\\
 D_c = (1-<\phi>)\alpha_{c}<\frac{dT_c}{dz}>,
  \label{eq:Dc_heat_budget_c}\\
D_d = <\phi>\alpha_{d}<\frac{dT_d}{dz}>\mathrm{.}
  \label{eq:Dd_heat_budget_d}
\end{gather}
\label{eq:heat_budget_terms}
\end{subequations}
Here, $C$ and $D$ refer to the convection and diffusion heat fluxes, respectively. Also, $w'=w-<w>$ and $T'=T-<T>$ are the wall-normal components of the dimensional velocity and temperature fluctuations. Given that $\alpha_d=\alpha_c=\alpha$ in cases 1-8 (refer to table \ref{table:list_of_simulations}), equations (\ref{eq:heat_budget}) and (\ref{eq:heat_budget_terms}) can be re-expressed in the following dimensionless form
\begin{equation}
\tilde{q}_{tot}'' = \tilde{C_c} + \tilde{C_d} + \tilde{D_c} + \tilde{D_d},
  \label{eq:heat_budget_dimless}
\end{equation}
where,
\begin{subequations}
\begin{gather}
 \tilde{C_c} = - \sqrt{RaPr}(1-<\phi>)<\tilde{w}_c'\tilde{\theta}_c'>,
  \label{eq:Cc_heat_budget_dimless_a}\\
 \tilde{C_d} = - \sqrt{RaPr}<\phi><\tilde{w}_d'\tilde{\theta}_d'>,
  \label{eq:Cd_heat_budget_dimless_b}\\
 \tilde{D_c} = (1-<\phi>)<\frac{d\tilde{\theta}_c}{d\tilde{z}}>,
  \label{eq:Dc_heat_budget_dimless_c}\\
\tilde{D_d} = <\phi><\frac{d\tilde{\theta}_d}{d\tilde{z}}>.
  \label{eq:Dd_heat_budget_dimless_d}
\end{gather}
\end{subequations}
The data of the heat transfer budget are reported in figure \ref{fig:heat_budget_graphs}. In figure \ref{fig:DcPDd_CcPCd}, we report the total convection and diffusion heat fluxes, confirming that diffusion dominates at the wall and approaches zero in the center of the cavity, whereas convection is the mechanism active in the bulk. The sum of the two is constant in the wall-normal coordinate and equal to the total heat flux through the cavity, used in the definition of the Nusselt number, see figure \ref{fig:nu_phi},
%
where the maximum enhancement was found to be about 10\% for the highest $\Upphi$ considered. 

Figure \ref{fig:heat_budget} shows the variation of diffusion and convection heat fluxes for both carrier and dispersed phases along the wall-normal direction. As $\Upphi$ increases from $0$ to $0.5$, the relative contributions of carrier-phase convection and diffusion heat transfer decrease, whereas the dispersed-phase contributions steadily rise, eventually reaching the same as the carrier phase for the binary mixture, i.e. $\Upphi=0.5$.
%
%
%
%

To examine the contribution to the heat transfer of carrier and dispersed phase independently of the increase in volume fraction,  we display in figure \ref{fig:heat_budget_barchart} the convection heat transfer at the center of the cavity ($\Tilde{z}=0.5$)
for dispersed and carrier fluids, normalized by the total heat transfer of each case with different $\Upphi$.
This shows that the
contribution of the dispersed phase is not directly proportional to the 
volume fraction, $\Upphi$, but rather lower. This disparity is clarified by recalling that the dispersed-phase concentration is lower near the walls. Droplets tend to preferentially remain outside the thermal boundary layer, resulting in lower temperatures than that of the hot wall (viceversa for the cold wall). Consequently, the relative contribution of the dispersed phase to the total heat transfer is smaller than that of the carrier phase, e.g. only $17\%$ for $\Upphi=30\%$. 
%
%
%
%
%
\subsection{Turbulent kinetic energy budgets}\label{subsec:TKE_budget}
To quantify the energy production and transfer mechanisms in emulsions, we investigate the turbulent kinetic energy (TKE) budget. The derivation of the TKE budget starts from the transport equation for the fluctuating velocity $u_i'$ as
\begin{equation}
\rho\left(\frac{\partial {u}_i'}{\partial t} + \frac{\partial {u}_i' {u}_j'}{\partial x_j}\right) = - \dfrac{\partial p}{\partial x_i} + \frac{\partial {\tau}_{ij}}{\partial x_j} + \sigma \kappa \delta_{\Gamma} + \rho \mathrm{g}\left[1 - \beta_{th} (T - T_0)\right]\mathrm{,}
  \label{eq:uprim_NS_equation}
\end{equation}
where $\tau_{ij} = 2\mu D_{ij}$ denotes the stress tensor and $D_{ij}=\left(\frac{\partial u_i}{\partial x_j} + \frac{\partial u_j}{\partial x_i}\right)/2$ stands for the strain rate tensor. Upon multiplying equation (\ref{eq:uprim_NS_equation}) by the velocity fluctuation, ${u}_i'$, and performing some algebraic manipulations, the turbulent kinetic energy evolution equation can be expressed as
\begin{equation}
\begin{aligned}
\rho\left(\frac{\partial {u}_i' {u}_i'/2}{\partial t} + \frac{\partial {u}_i' {u}_i' {u}_j'/2}{\partial x_j}\right) = - \frac{\partial {u}_i' p}{\partial x_i} + \frac{\partial {{u}_i' \tau}_{ij}}{\partial x_j} - \tau_{ij}D_{ij} \\
+ \sigma \kappa \delta_{\Gamma} {u}_i' + \rho \mathrm{g} [1 - \beta_{th} (T - T_0)] {u}_i'.
  \label{eq:uprim_uprim_NS_equation}
\end{aligned}  
\end{equation}
Equation (\ref{eq:uprim_uprim_NS_equation}) can be handled in two ways: it can be time-averaged over both phases, leading to the total TKE equation (\ref{eq:TKE_budget_tot}), or it can be time- and phase-averaged with respect to a specific phase $m$ (either carrier or dispersed phase). This approach results in the TKE equation specific to one phase, reported in equation (\ref{eq:TKE_budget_phase_m}). \citep{dodd2016interaction, rosti2019droplets, crialesi2022modulation}
\begin{subequations} \label{eq:TKE_budget}
\begin{gather}
 \frac{d\mathcal{K}}{dt} + \frac{d\mathcal{T}_{j}}{dx_j}  = \mathcal{P} - \varepsilon + {\Psi}^{\sigma},
  \label{eq:TKE_budget_tot}\\
  \frac{d\mathcal{K}_m}{dt} = \mathcal{P}_m - \varepsilon_m + \Xi_m^{v} + \Xi_m^{p},
  \label{eq:TKE_budget_phase_m}
\end{gather}
\end{subequations}
In equation~\eqref{eq:TKE_budget_tot}, $\mathcal{P}$ represents the rate of turbulent kinetic energy production resulting from buoyancy forces, $\varepsilon$ denotes the dissipation of kinetic energy due to the viscous effects, ${\Psi}^{\sigma}$ is the power of the surface tension (due to droplet deformation, breakup and coalescence), and $\mathcal{T}_j$ is the flux responsible for the spatial redistribution of TKE. Note that after volume averaging equation \eqref{eq:uprim_uprim_NS_equation} over both phases, the transport terms vanish. Moreover, the second term on the left-hand side of the equation (\ref{eq:uprim_uprim_NS_equation}) is omitted during phase averaging because,  as discussed in \cite{dodd2016interaction} and \cite{rosti2019droplets}, turbulent eddies cannot transport TKE across the interface of carrier and droplet fluids in immiscible fluids. Finally, the terms $\Xi_m^{v}$ and $\Xi_m^{p}$ in equation \eqref{eq:TKE_budget_phase_m} denote the viscous and pressure work rates on phase $m$, which represent the transport of TKE by viscous stresses and pressure, respectively. Note finally that under statistically steady state conditions $\frac{d\mathcal{K}}{dt} = \frac{d\mathcal{K}_m}{dt} = 0$. 
To summarize, each term in equations \eqref{eq:TKE_budget_tot} and \eqref{eq:TKE_budget_phase_m} reads
\begin{subequations} \label{eq:TKE_budget_components}
\begin{gather}
\mathcal{K} = <{u}_i' {u}_i'>/2
\label{eq:TKE_rate_tot}\\
\mathcal{P} = <\mathrm{g}[1-\beta(T-T_0)]{u}_i'>
\label{eq:production_rate_tot}\\
\varepsilon = <2 \nu D_{ij}D_{ij}>; \quad
{\Psi}^{\sigma} = <\frac{1}{\rho} \sigma \kappa \delta_{\Gamma} {u}_i'>
\label{eq:power_surface_tension}\\
\frac{d\mathcal{T}_{i}}{dx_j}  = \mathcal{P} - \varepsilon + {\Psi}^{\sigma} = \frac{\partial <{u}_i' {u}_i' {u}_j'>/2}{\partial x_j}
  \label{eq:spatial_redistribution_tot}\\
\mathcal{K}_m = {<{u}_i' {u}_i'>}_m /2
\label{eq:TKE_rate_m}\\
\mathcal{P}_m = {<\mathrm{g}[1-\beta(T-T_0)]{u}_i'>}_m
\label{eq:production_rate_m}\\
\varepsilon_m = {<2 \nu D_{ij}D_{ij}>}_m
\label{eq:dissipation_rate_m}\\
\Xi_m^{v} = {<\frac{1}{\rho} \frac{\partial{u}_i' p}{\partial{x_i}}>}_m
\label{eq:power_viscous_m}\\
\Xi_m^{p} = {<\frac{1}{\rho} \frac{\partial{u}_i' \tau_{ij}}{\partial{x_j}}>}_m
\label{eq:power_pressure_m}
\end{gather}
\end{subequations}
%

Some considerations are needed on the surface tension term (${\Psi}^{\sigma}$). This accounts for the work done by the surface tension forces. It can be either a source of TKE (positive) or a sink of TKE (negative). It relates to the rate of change of the surface energy at the droplet interface and is inversely proportional to the rate of change of the droplet surface area, i.e. ${\Psi}^{\sigma}(t) \propto - \frac{1}{We} \frac{dA(t)}{dt}$, meaning that an increase in droplet surface area leads to a decrease in TKE (${\Psi}^{\sigma}(t) < 0$), and vice versa \cite[see also discussion in][]{dodd2016interaction}. At steady state, when the surface area is on average constant, the work rate of the surface tension forces becomes zero for homogeneous isotropic and shear turbulence, as proved in \cite{dodd2016interaction,trefftz-posada_ferrante_2023}. 
Here, one can replicate the derivation outlined in \cite{trefftz-posada_ferrante_2023}
and show that ${\Psi}^{\sigma} = 0$ holds for RB convection as well, given that the mean wall-normal velocity is zero at steady state. Finally, the integral across the cavity of the nonlinear transfer terms also goes to zero, as expected by the conservative nature of the nonlinearities of the Navier-Stokes equations.

%

\begin{figure}
  \centering
  \begin{subfigure}[t]{0.03\textwidth}
  \fontsize{6}{9}
    \textbf{(a)}
  \end{subfigure}
  \begin{subfigure}[t]{0.45\textwidth}
    \includegraphics[width=\linewidth, valign=t]{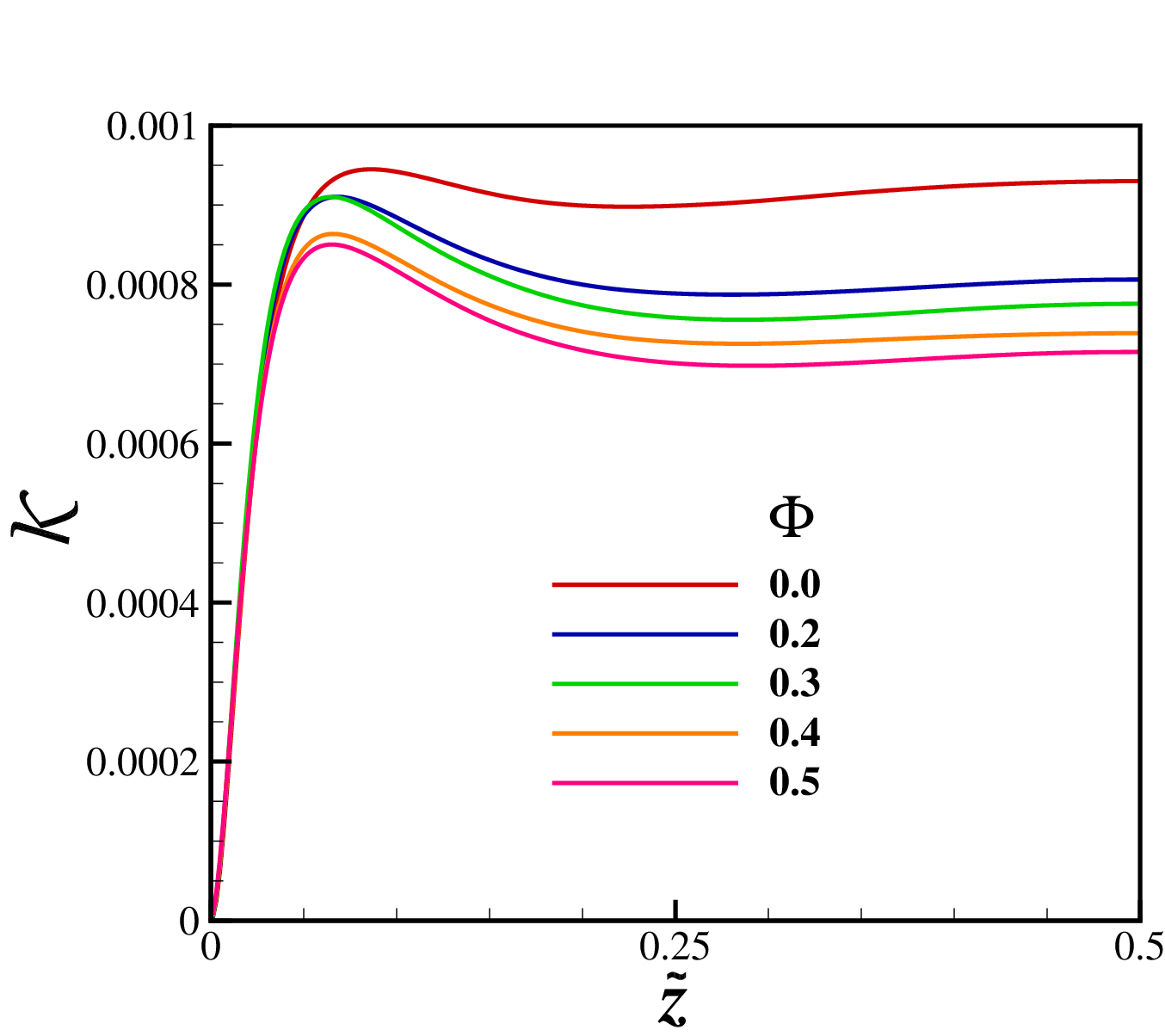}
    \phantomsubcaption\label{fig:TKE tot sym}
  \end{subfigure}
    \begin{subfigure}[t]{0.03\textwidth}
  \fontsize{6}{9}
    \textbf{(b)}
  \end{subfigure}
  \begin{subfigure}[t]{0.45\textwidth}
  \vspace{0.95cm}
  \includegraphics[width=\linewidth, valign=t]{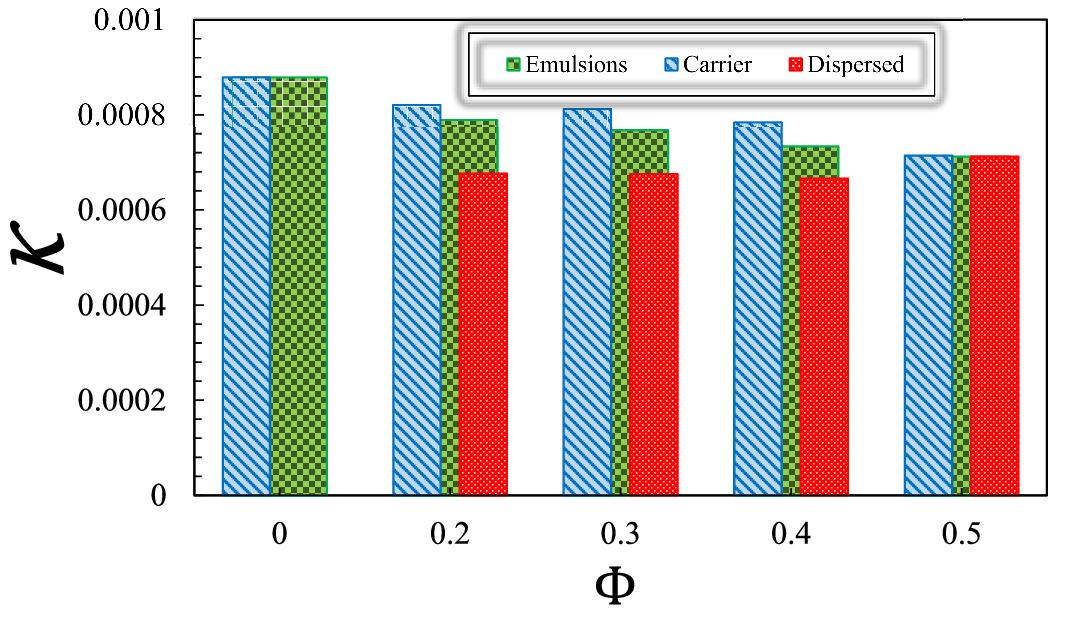}
    \phantomsubcaption\label{fig:tke vol}
  \end{subfigure}
  \caption{(a) Wall-normal profiles of average turbulent kinetic energy and (b) volume-averaged bar charts of turbulent kinetic energy at different dispersed-phase volume fractions.}
  \label{fig:TKE}
\end{figure}

We now discuss the results. Figure \ref{fig:TKE tot sym} presents the wall-normal distributions of turbulent kinetic energy density at different volume fractions. As depicted in this figure, the TKE distribution exhibits a peak close to the edge of the boundary layers, associated to the horizontal velocity components, as shown in figure \ref{fig:kinetic_energy}. Beyond this peak, we observe a gradual decrease with the approach of an approximately constant value within the central region of the cavity. Panel \ref{fig:tke vol} reports bar-charts of the volume-averaged TKE. As illustrated in both figures \ref{fig:TKE tot sym} and \ref{fig:tke vol}, introducing a dispersed phase dampens the TKE, resulting in a 20\% reduction at $\Upphi=0.5$, in agreement with the findings in \cite{dodd2016interaction} for iso-thermal emulsions in decaying turbulence. Despite the reduction, the overall heat transfer increases by about 10\%: this is attributed to the transfer of energy to small scales by the surface tension forces, which enhances near-wall mixing (cf.\ discussion of figure \ref{fig:spectra_phi}).
Indeed, as shown in \cite{crialesi2022modulation,crialesi2023interaction} for emulsions in homogeneous and isotropic turbulence, the presence of dispersed phase provides an alternative mechanism for energy transfer at small scales and reduces the size of the smallest active flow scales. In terms of global heat transfer, this more than compensates for the reduction in turbulent kinetic energy.
%


According to figure \ref{fig:tke vol}, except for the case of the binary mixture, the kinetic energy density of the dispersed phase is lower than that of the carrier phase.
This difference is due to the vertical component of the carrier phase velocity (see figure \ref{fig:kinetic_energy}) and related to the increased concentration of the carrier phase 
in the near-wall region. In other words, the carrier fluid has higher/lower temperatures than the dispersed phase on average and is, therefore, subject to stronger buoyancy.

\begin{figure}
  \centering
  \begin{subfigure}[h]{0.03\textwidth}
  \fontsize{6}{9}
    \textbf{(a)} 
  \end{subfigure}
  \begin{subfigure}[h]{0.95\textwidth}
    \includegraphics[width=\linewidth, valign=t]{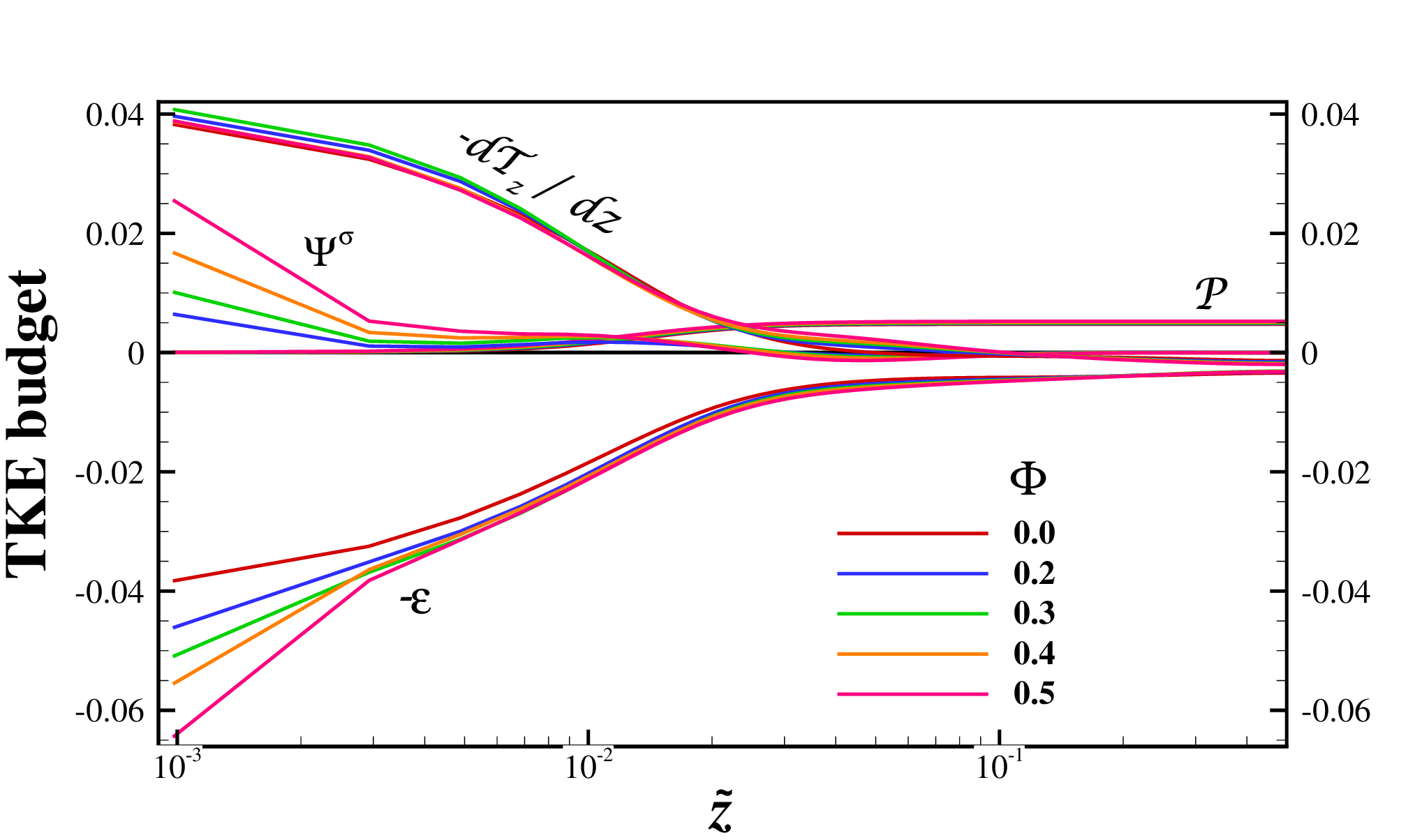}
    \phantomsubcaption\label{fig:TKE budget sym}
  \end{subfigure}\hfill
  \begin{subfigure}[h]{0.03\textwidth}
  \fontsize{6}{9}
    \textbf{(b)}
  \end{subfigure}
  \begin{subfigure}[h]{0.45\textwidth}
    \includegraphics[width=\linewidth, valign=t]{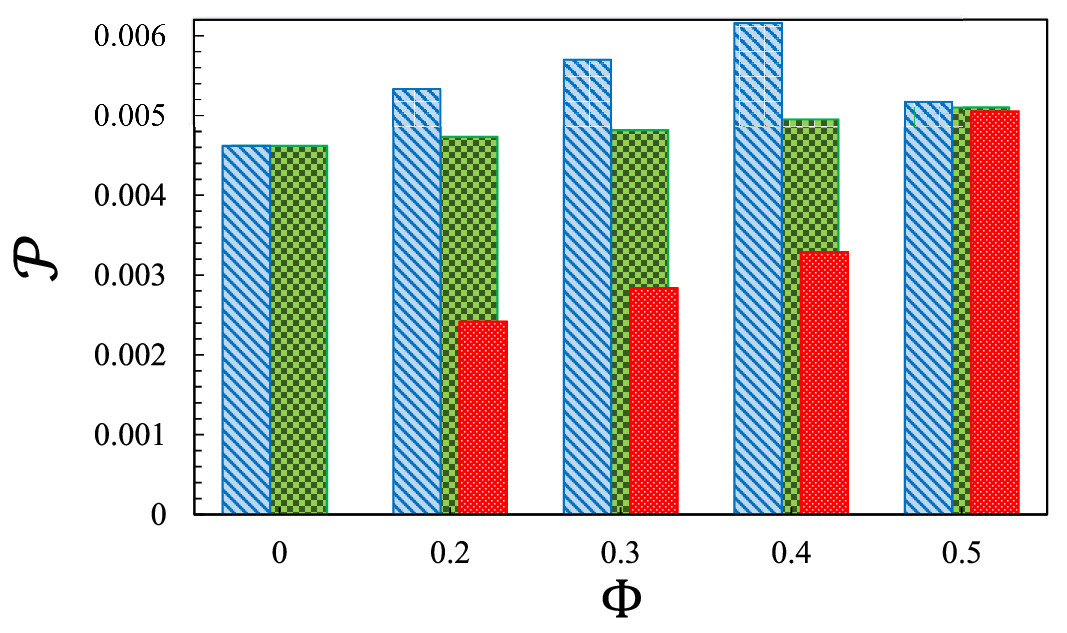}
    \phantomsubcaption\label{fig:prd vol}
  \end{subfigure}
    \begin{subfigure}[h]{0.03\textwidth}
  \fontsize{6}{9}
    \textbf{(c)}
  \end{subfigure}
  \begin{subfigure}[h]{0.45\textwidth}
    \includegraphics[width=\linewidth, valign=t]{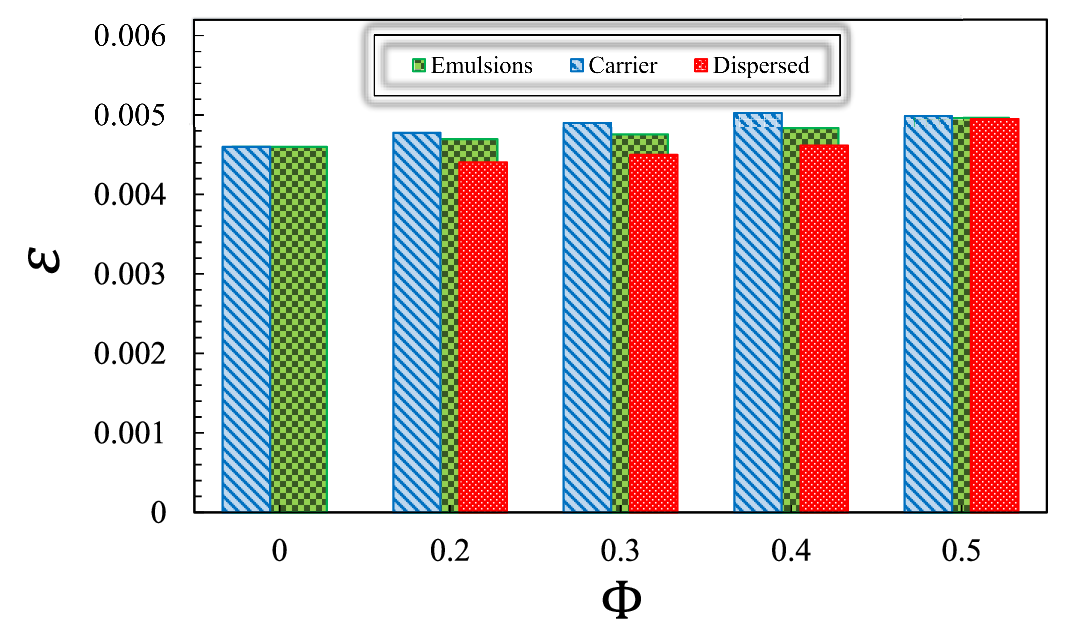}
    \phantomsubcaption\label{fig:eps vol}
  \end{subfigure}
  \caption{(a) TKE budget terms along the wall-normal direction; (b,c) bar charts representing the volume average of the different terms in TKE budget for various droplet volume fractions.}
  \label{fig:TKE_budget_results}
\end{figure}

The wall-normal profiles of the different terms of the TKE budgets are displayed in figure \ref{fig:TKE budget sym} in log-scale to highlight the near-wall dynamics. 
Before discussing the behavior of each term, we note that the differences between emulsions with different volume fractions are relatively small in the bulk of the flow and the trends of the TKE budget closely mirror the observations made for suspensions of rigid particles \citep{demou2022turbulent}. 

%
The data show that the TKE dissipation rate is highest in proximity to the wall, increasing with the volume fraction $\Upphi$, within the viscous boundary layer. It reduces towards the center of the cavity and eventually reaches an approximately constant value. In contrast, the TKE production rate, zero at the wall,
increases towards the core of the cavity where it reaches a plateau, with values increasing at higher values of $\Upphi$. We also note that the work of the interfacial stresses increases with the amount of dispersed phase. This term provides energy to the near-wall region, at the expense of the KE in the center of the cavity. This suggests that the near-wall dynamics is characterized by the relaxation and coalescence of smaller droplets in a laminar-like flow, while breakup dominates the dynamics of bigger droplets in the bulk (see also average droplet sizes across the cavity in figure \ref{fig:DSD_visc_2}). The reader is referred to \cite{crialesi2023interaction} for an analysis of the relation between scale-by-scale energy transfer and droplet dynamics.

Figure \ref{fig:TKE budget sym} also provides insights into the wall-normal distribution of $-\frac{d\mathcal{T}}{dz}$, which represents the spatial redistribution term. As can be seen in the figure, nonlinear interactions transfer TKE from the central regions of the cavity (where $-\frac{d\mathcal{T}}{dz}<0$) towards the viscous regions near the wall (where $-\frac{d\mathcal{T}}{dz}>0$), where this is dissipated by viscous forces. 
%

Next, we display in figures \ref{fig:prd vol} and \ref{fig:eps vol}, the temporally- and volume-averaged TKE production and dissipation rates. 
The bar charts reveal that the TKE production due to buoyancy forces (panel b) is larger in the carrier phase, with values increasing with the volume fraction $\Upphi$ for both phases (except the special case of binary mixture).  This is explained by the increased buoyancy of the carrier phase, on average located closer to the walls, as discussed explaining the larger vertical component of the kinetic energy. 
As mentioned above, the carrier phase is most likely to stay within the thermal boundary layers; its temperature approaches that of the nearby walls (see inset in figure 5a), and its density is, therefore, more likely to reach low/high values. In other words, the fastest rising plumes are expected to contain more of the carrier than of the dispersed phase, as suggested by the fact that the temperature-velocity fluctuations are larger in the carrier phase. 
To conclude, we also note an overall increase in the production with the volume fraction of the dispersed phase, of the order of about 10\%. 

The viscous dissipation density is found to be only slightly larger in the carrier phase, indicating the presence of larger velocity gradients. Despite production being more pronounced in the carrier phase, the dissipation appears to be more equally distributed.
Regarding the dissipation rate enhancement with $\Upphi$, as noted in \cite{dodd2016interaction}, the addition of more droplets to the single-phase flow leads to an increase in the velocity gradient ($\frac{\partial u_i}{\partial x_j}$) near the droplet surfaces, resulting in higher viscous dissipation rates. This is in local balance with the work rate of the surface tension, as shown in \cite{crialesi2022modulation}.
Finally, for the case of a binary mixture, production and dissipation density are the same in the two phases, at least within statistical error, with the dissipation rate showing a slightly better statistical convergence.

\begin{figure}
  \centering
  \begin{subfigure}[t]{0.03\textwidth}
  \fontsize{6}{9}
    \textbf{(a)} 
  \end{subfigure}
  \begin{subfigure}[t]{0.45\textwidth}
    \includegraphics[width=\linewidth, valign=t]{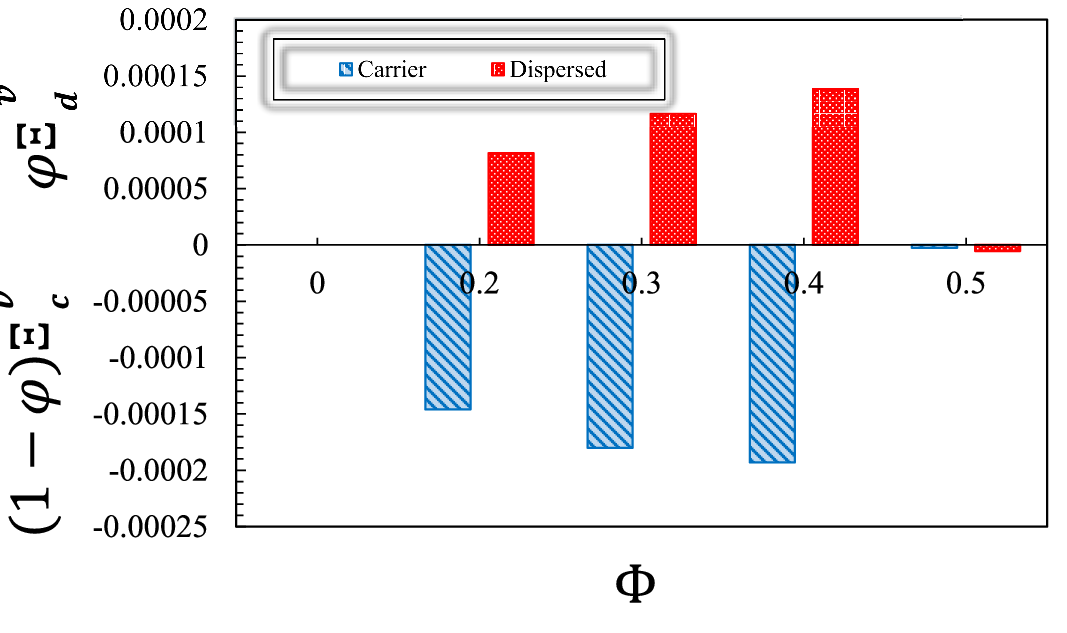}
    \phantomsubcaption\label{fig:srn vol}
  \end{subfigure}\hfill
  \begin{subfigure}[t]{0.03\textwidth}
  \fontsize{6}{9}
    \textbf{(b)}
  \end{subfigure}
  \begin{subfigure}[t]{0.45\textwidth}
    \includegraphics[width=\linewidth, valign=t]{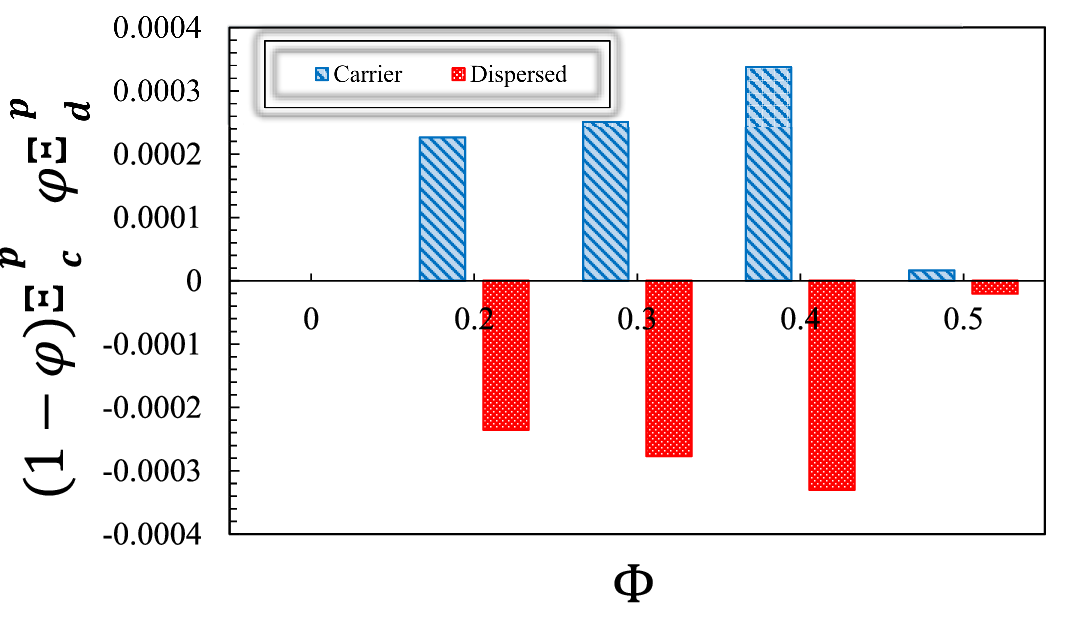}
    \phantomsubcaption\label{fig:srp vol}
  \end{subfigure}
  \caption{Bar charts denoting the volume averaged of the (a) viscous and (b) pressure power terms in the TKE budget at various dispersed-droplet volume fractions.}
  \label{fig:TKE_srn_srp_barchart}
\end{figure}

The interphase energy transfer due to viscous and pressure forces are displayed in figure \ref{fig:TKE_srn_srp_barchart}.  As shown in \cite{trefftz-posada_ferrante_2023},  at steady state, when the work of the surface tension forces is on average zero, the weighted sum of the interfacial fluxes is zero,
 \[
0= (1-\phi) (\Xi_c^{v}+\Xi_c^{p}) + \phi( \Xi_d^{v}+\Xi_d^{p}).
\]
The data in the figure show that viscous stresses transfer energy from the carrier to the dispersed phase, whereas the work by the pressure forces increases the turbulent kinetic energy of the carrier phase at the expense of the dispersed one.
For the case of a binary mixture, $\Upphi=0.5$ the different contributions vanish (within statistical error) by symmetry. Note, finally, that the viscous and pressure transport terms are significantly smaller than production and dissipation, as in the case of isothermal turbulence reported in \cite{rosti2019droplets,trefftz-posada_ferrante_2023}.
%
%
%

%
%
%
%
\subsection{Droplet size distribution }\label{subsec:DSD}
\begin{figure}
  \centering
  \begin{subfigure}[t]{0.03\textwidth}
  \fontsize{6}{9}
    \textbf{(a)} 
  \end{subfigure}
  \begin{subfigure}[t]{0.45\textwidth}
    \includegraphics[width=\linewidth, valign=t]{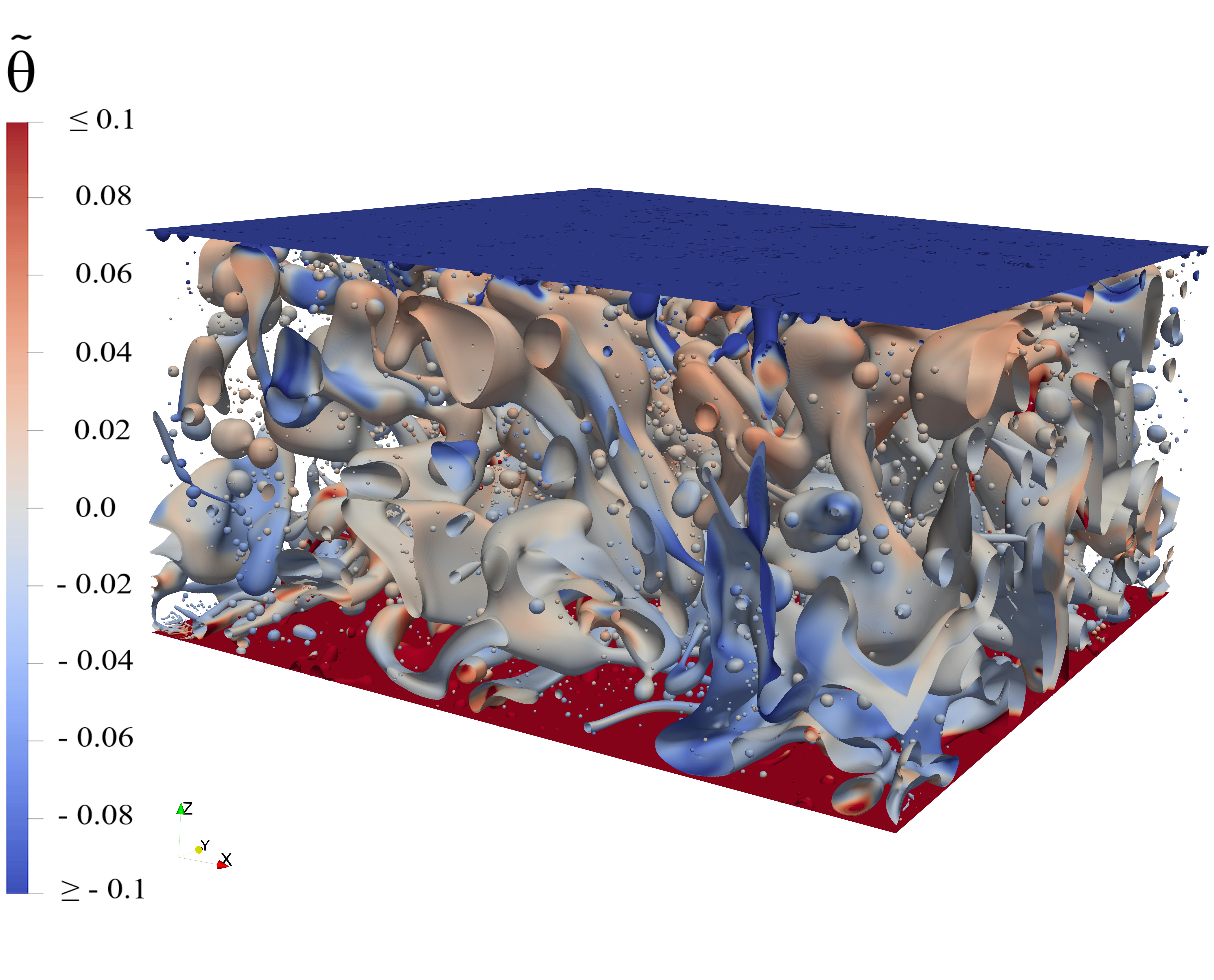}
    \phantomsubcaption\label{fig:vof_3d_20}
  \end{subfigure}\hfill
  \begin{subfigure}[t]{0.03\textwidth}
  \fontsize{6}{9}
    \textbf{(b)}
  \end{subfigure}
  \begin{subfigure}[t]{0.45\textwidth}
    \includegraphics[width=\linewidth, valign=t]{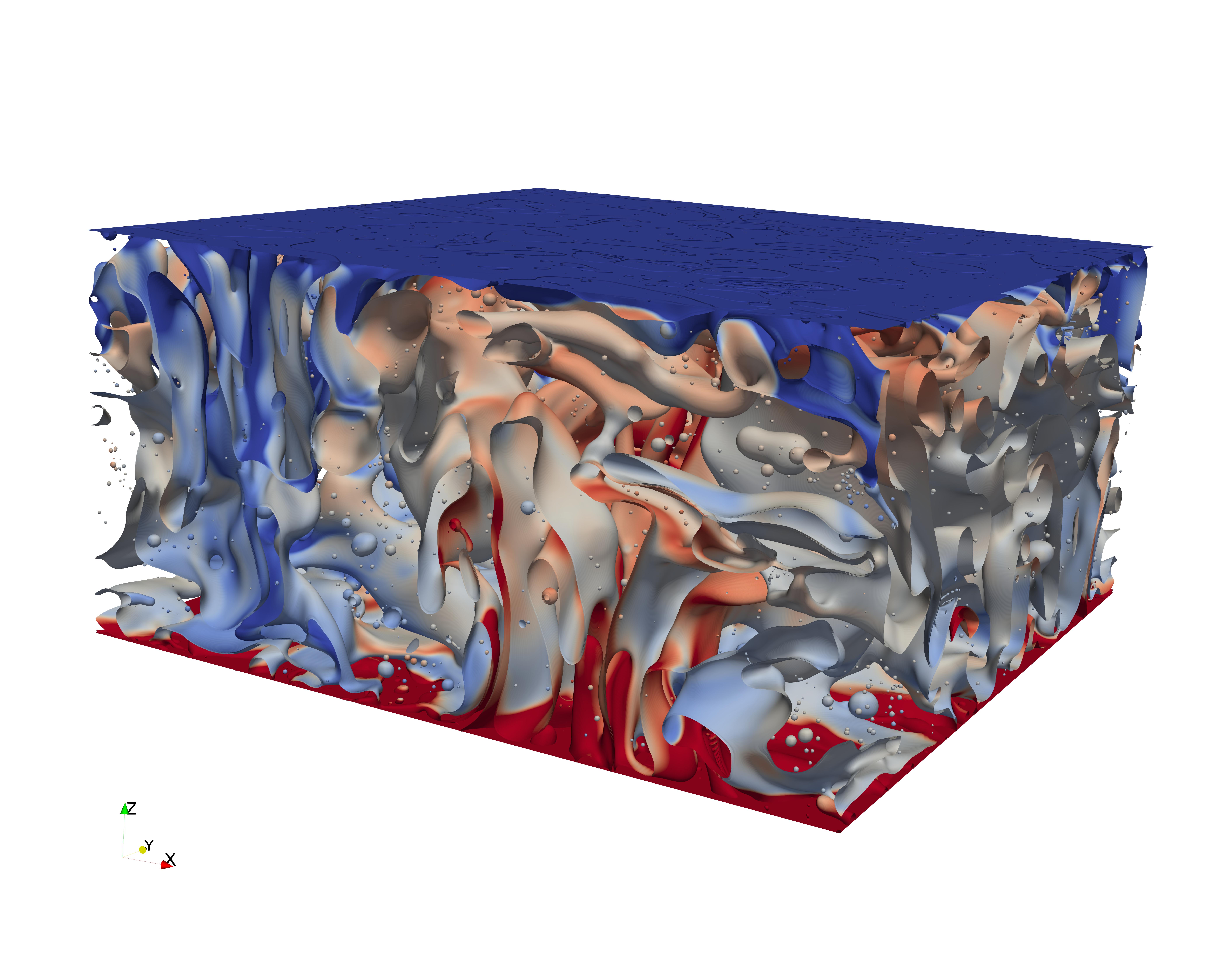}
    \phantomsubcaption\label{fig:vof_3d_50}
  \end{subfigure}
  \caption{Instantaneous distributions of dispersed droplets at (a) $\Upphi=0.2$ and (b) $\Upphi=0.5$. Dispersed droplets are colored based on their temperature. For a clear visualization of the thermal plumes between the two plates and the corresponding droplets' transport, the temperature range in the colorbar is restricted from $-0.1$ to $0.1$, i.e. any droplet with a temperature equal to or smaller than $-0.1$ and equal to or greater than $0.1$ is rendered with a uniform blue and red color, respectively.}
  \label{fig:vof_3d_iso-surface}
\end{figure}

The Droplet Size Distribution (DSD) measures the range of sizes that droplets can assume in a given multiphase system. Thermal convection may lead to a different distribution than homogeneous isotropic turbulence, as the interplay between flow patterns and dispersed droplets can alter both their sizes and overall distribution. In figure \ref{fig:vof_3d_iso-surface}, we present visualizations of the instantaneous spatial distributions of dispersed droplets during the statistical stationary state for two different dispersed-droplet volume fractions, namely $\Upphi=0.2$ and $\Upphi=0.5$. Following the release of the dispersed phase into the system, droplets undergo recurring coalescence and breakup events. Eventually, when the system reaches a stationary state, it appears as depicted in figure \ref{fig:vof_3d_iso-surface}, where at higher droplet volume fractions, dispersed droplets are more prone to coalescence, leading to the presence of larger droplets. This qualitative observation is further confirmed by examining the probability density function (p.d.f.) of droplet sizes.
%

\begin{figure}
  \centering
  \begin{subfigure}[t]{0.03\textwidth}
  \fontsize{6}{9}
    \textbf{(a)} 
  \end{subfigure}
  \begin{subfigure}[t]{0.95\textwidth}
    \includegraphics[width=\linewidth, valign=t]{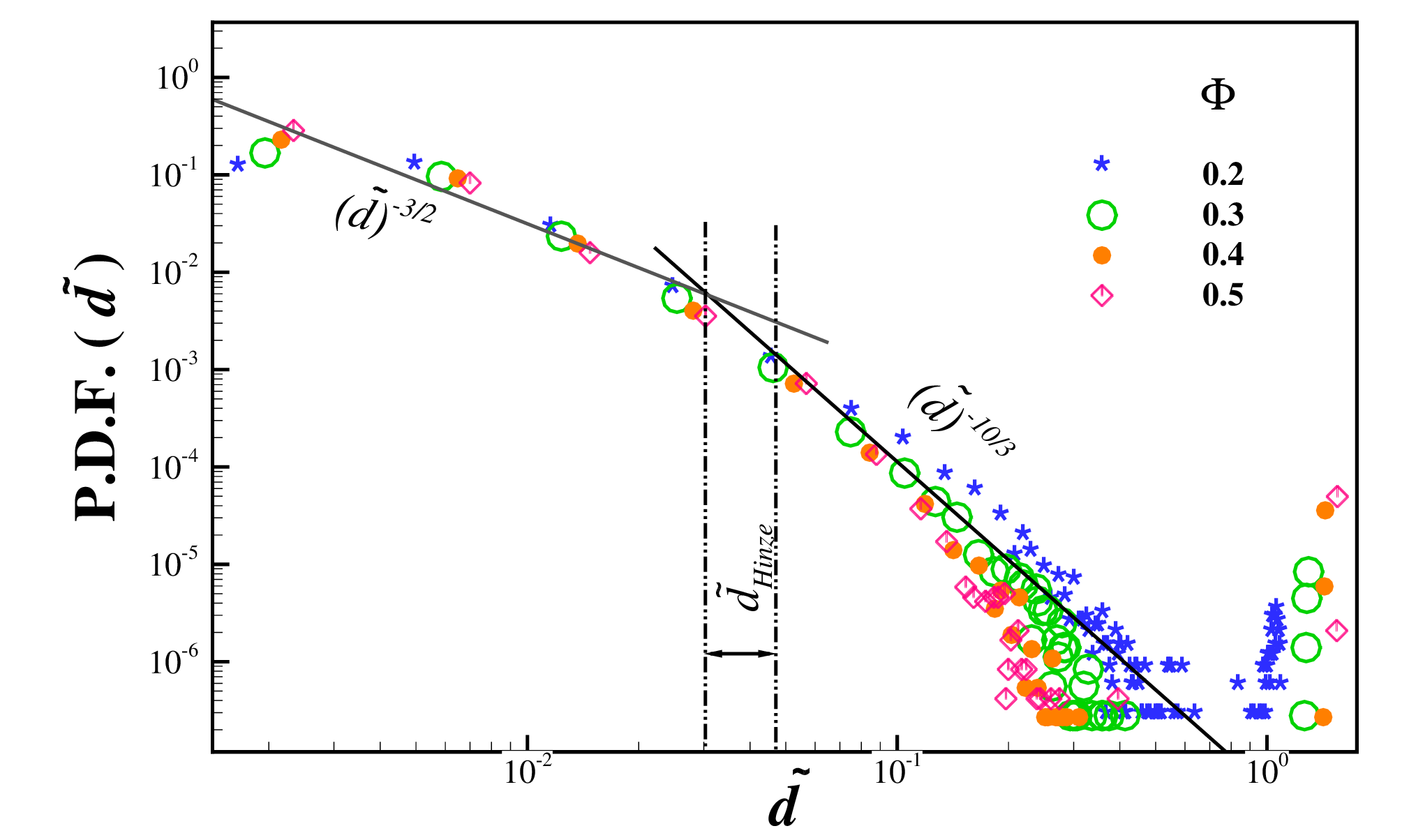}
    \phantomsubcaption\label{fig:pdf}
  \end{subfigure}\hfill
  \begin{subfigure}[t]{0.03\textwidth}
  \fontsize{6}{9}
    \textbf{(b)}
  \end{subfigure}
  \begin{subfigure}[t]{0.45\textwidth}
    \includegraphics[width=\linewidth, valign=t]{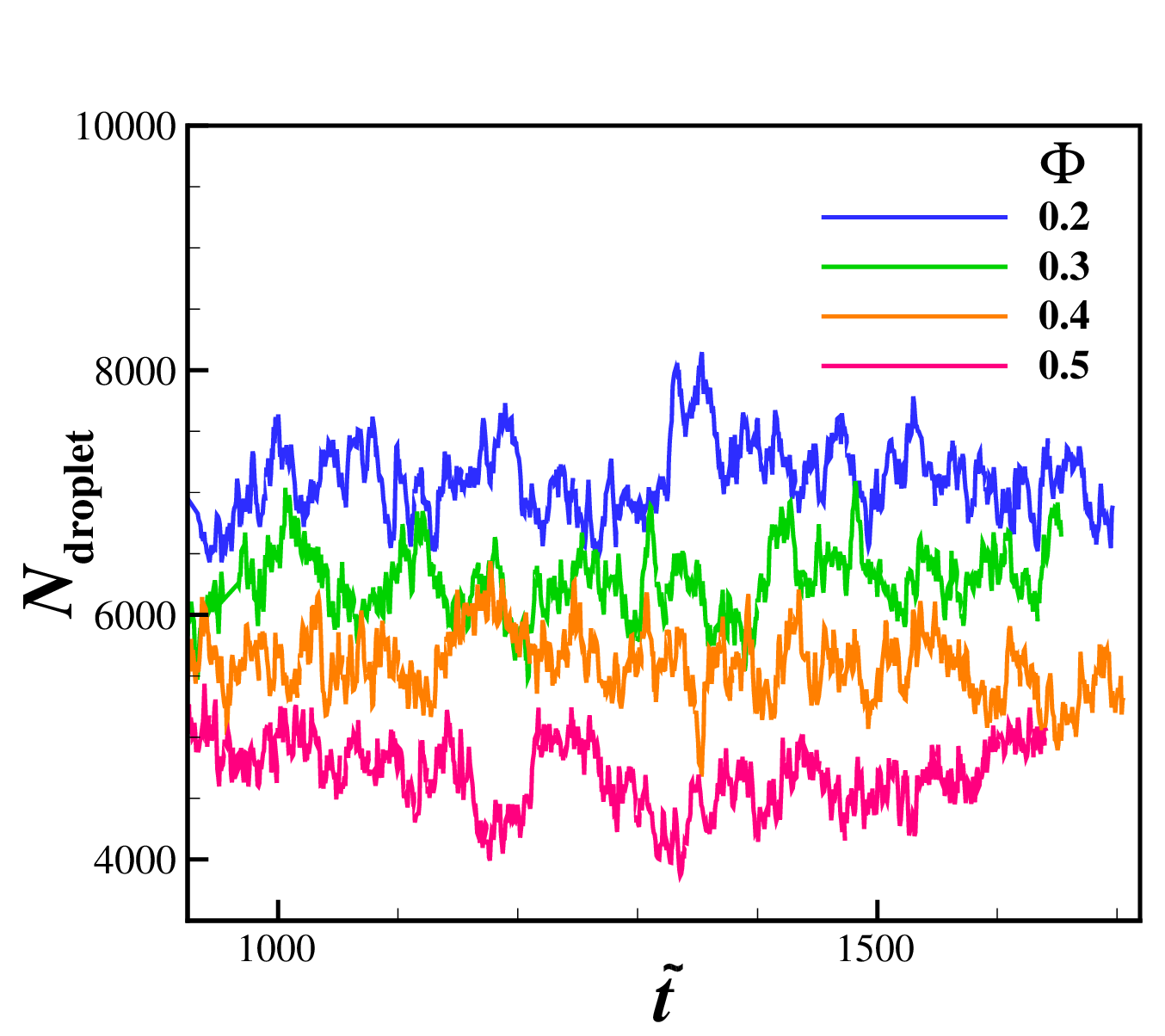}
    \phantomsubcaption\label{fig:Num of droplet over time}
  \end{subfigure}
    \begin{subfigure}[t]{0.03\textwidth}
  \fontsize{6}{9}
    \textbf{(c)}
  \end{subfigure}
  \begin{subfigure}[t]{0.45\textwidth}
    \includegraphics[width=\linewidth, valign=t]{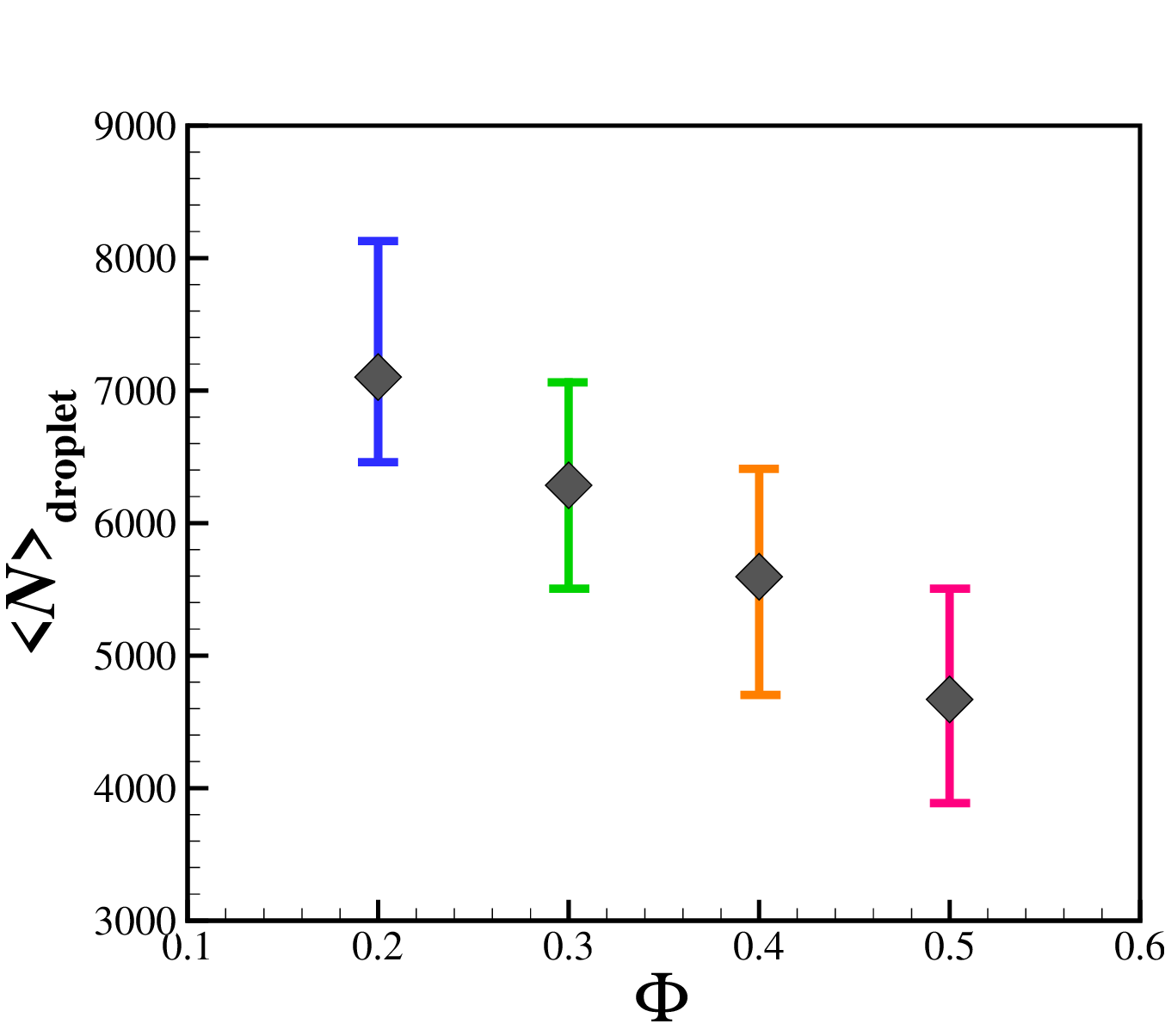}
    \phantomsubcaption\label{fig:Num of droplet range}
  \end{subfigure}
  \caption{(a) PDF of the DSD for different droplet volume fractions. The solid-black lines represent the two scaling of $d^{-3/2}$ from \cite{deane2002scale}, and $d^{-10/3}$ from \cite{garrett2000connection}; (b) temporal evolution of the number of droplets, denoted as $N_\text{droplet}$, within the domain and (c) the time-averaged number of droplets $<N>_\text{droplet}$ along with its associated fluctuation range, for distinct scenarios characterized by dispersed-phase volume fractions.}
  \label{fig:pdf_DSD}
\end{figure}

Figure \ref{fig:pdf} illustrates the probability density function of droplets of equivalent droplet diameter $d = (6V / \pi)^{1/3}$, where $V$ represents the droplet volume measured in the simulation, for different volume fractions, $\Upphi=[0.2-0.5]$. 
The equivalent droplet diameter is scaled by the cavity height. The analysis of the droplet size distribution aims to determine if it adheres to two different scaling laws: one proposed by \cite{deane2002scale} for small-size droplets ($d^{-3/2}$), and the power law  $d^{-10/3}$ introduced by \cite{garrett2000connection} for large droplets. 
Based on the Kolmogorov-Hinze criteria, there exists a critical diameter, $d_\text{Hinze}$, such that droplets with $d<d_\text{Hinze}$ mainly experience breakup, while those with diameters greater than the Hinze length scale, $d>d_\text{Hinze}$, predominantly undergo coalescence. Therefore, the Kolmogorov-Hinze scale roughly determines the transition between the two regimes mentioned above, and it is typically estimated as~\citep{hinze1955fundamentals}
\begin{equation}
 d_\text{Hinze} = \left(\frac{We_\text{cr}}{2}\right)^{3/5}\left(\dfrac{\sigma}{\rho_c}\right)^{3/5}\varepsilon^{-2/5},
  \label{eq:d_Hinze}
\end{equation}
where $\varepsilon$ is the turbulent dissipation rate that can be computed in the RB configuration as suggested in~\cite{shishkina2010boundary}
\begin{equation}
 \varepsilon = \dfrac{\nu_c^3}{L_s^4}(Nu-1)Ra{Pr}^{-2}.
  \label{eq:eps_at_d_Hinze}
\end{equation}
Equation~\eqref{eq:d_Hinze} requires a value for the critical Weber number to compute $d_\text{Hinze}$. For this study, we take $We_\text{cr} = [0.5-1.0]$, as suggested in~\cite{hinze1955fundamentals}, and based on this choice, the Hinze length scale lies in the range indicated by the two vertical lines in figure \ref{fig:pdf}.

%
%

\begin{figure}
\centering
\begin{subfigure}[t]{0.03\textwidth}
\centering
\fontsize{6}{9}
\textbf{(a)}
\end{subfigure}
\begin{subfigure}[t]{0.45\textwidth}
\includegraphics[width=\linewidth, valign=t]{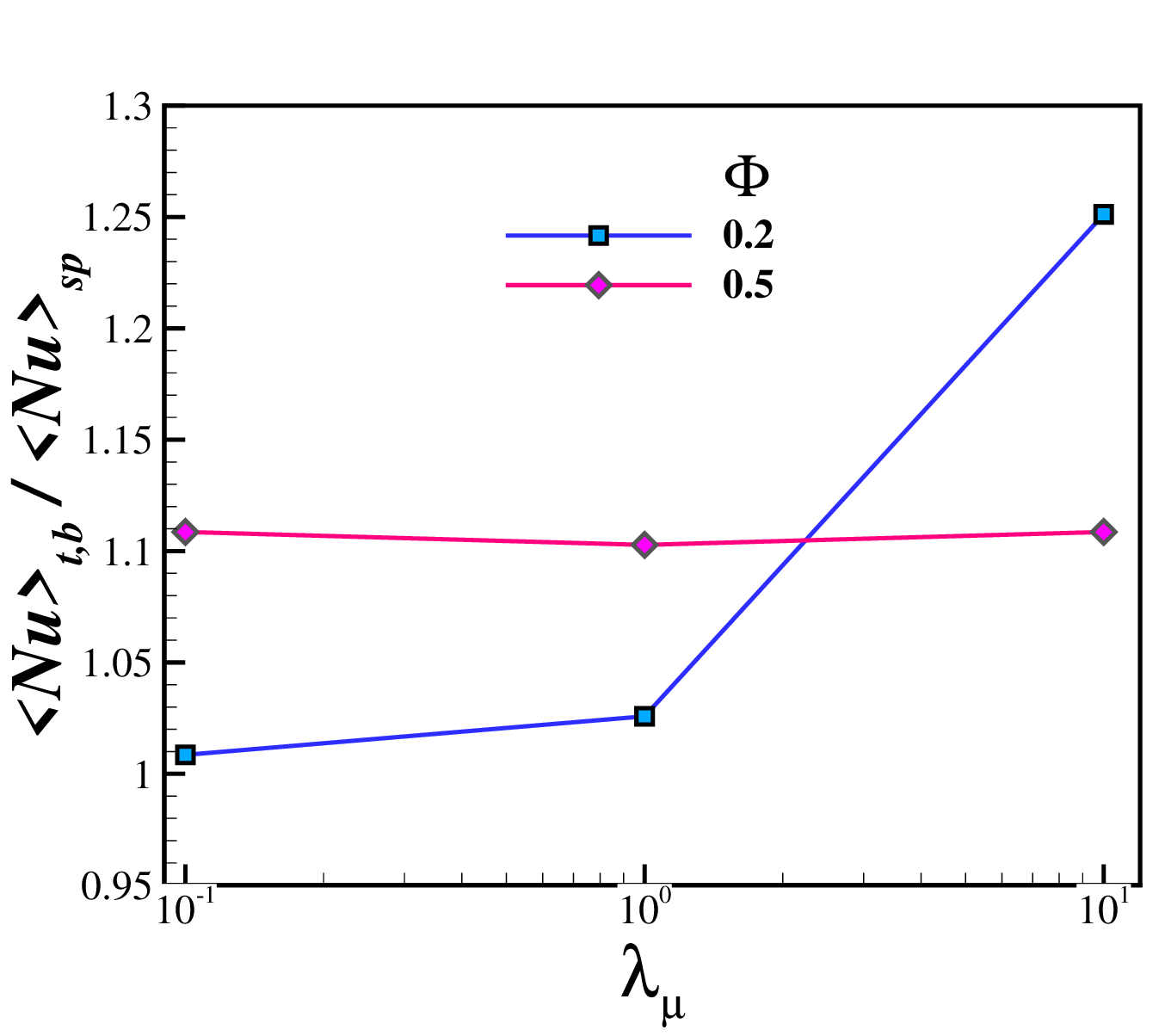}
 \phantomsubcaption\label{fig:nu visc phi}
\end{subfigure}\hfill
\begin{subfigure}[t]{0.03\textwidth}
\centering
\fontsize{6}{9}
\textbf{(b)}
\end{subfigure}
\begin{subfigure}[t]{0.45\textwidth}
\includegraphics[width=\linewidth, valign=t]{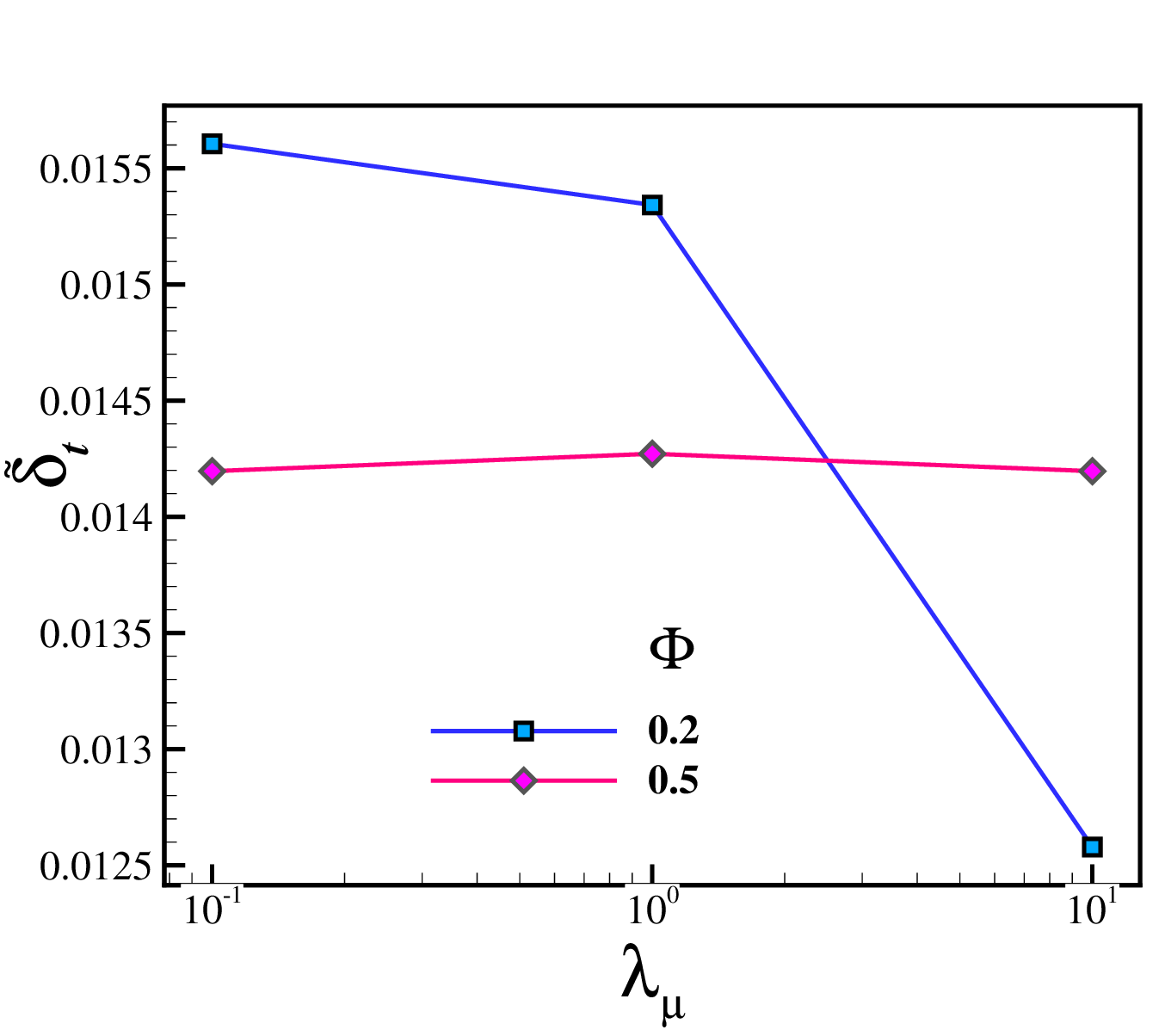}
\phantomsubcaption\label{fig:TBL thickness visc phi}
\end{subfigure}
\begin{subfigure}[t]{0.03\textwidth}
\centering
\fontsize{6}{9}
\textbf{(c)}
\end{subfigure}
\begin{subfigure}[t]{0.45\textwidth}
\centering
\includegraphics[width=\linewidth, valign=t]{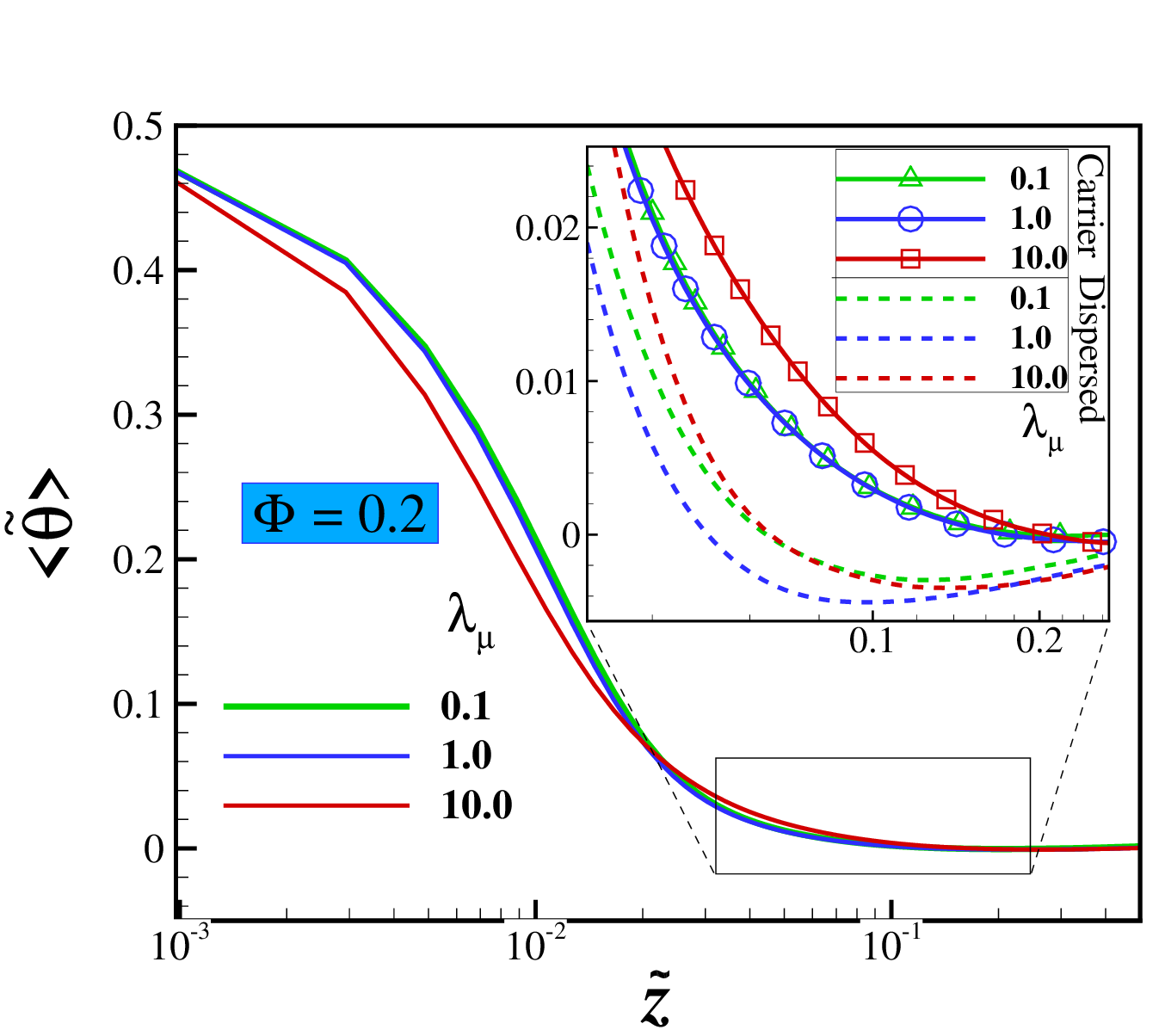}
\phantomsubcaption\label{fig:tmp mean 20 visc}
\end{subfigure}
\caption{(a) Nusselt number, (b) thermal boundary layer thickness and (c) mean temperature profiles along the wall-normal direction for the different various viscosity ratios and droplet volume fractions investigated, see legend.}
\label{fig:nu_tmp_mean_rms_visc_ratio}
\end{figure}

The DSD data confirm both power laws for small and large droplets at $\Upphi=0.2$ and $0.3$, with the smallest approximated Hinze length scale ($\tilde d_\text{Hinze} = 0.0303$) providing a good estimate of the transition between the two scaling laws. However, the $-10/3$ law becomes less apparent in more concentrated cases, i.e. $\Upphi=0.4-0.5$. It should be mentioned that, for computing the droplet equivalent diameter $\tilde{d}$, all droplets are assumed to be spherical. This assumption ceases to be valid in the most concentrated cases, where significant deviations from sphericity occur due to the formation of large filaments that follow the thermal plumes. Consequently, a deviation from the $-10/3$ law is expected and observed in the p.d.f. graphs of these cases.
Furthermore, in all scenarios, a secondary peak emerges at high values of $\tilde d$ \citep{mukherjee2019droplet}. This secondary peak indicates the presence of a few larger connected regions (large filaments) within the periodic simulation domain. Notably, this secondary peak becomes progressively more pronounced with increasing values of $\Upphi$.


Figure \ref{fig:Num of droplet over time} and \ref{fig:Num of droplet range} illustrate the temporal evolution of the number of droplets, $N_\text{droplet}$, and the averaged number of droplets $<N>_\text{droplet}$ (with their fluctuation range) for the different cases under investigation. In \ref{fig:Num of droplet over time}, the variation of $N_\text{droplet}$ is reported during the statistically stationary state. The fluctuations in the number of droplets,  $N_\text{droplet}$, indicate a competition between droplet breakup and coalescence. The large dips in these graphs denote the coalescence of a significant number of droplets, which results in the formation of large droplets. These large droplets, however, are unstable and tend to break up into smaller ones, which is reflected by the subsequent increase in $N_\text{droplet}$. Both figures~\ref{fig:Num of droplet over time} and~\ref{fig:Num of droplet range} indicate that at higher dispersed-phase volume fractions, fewer droplets are present in the flow. This is because droplets are more likely to coalesce, forming larger droplets in cases with higher $\Upphi$. This observation aligns with the p.d.f. in figure~\ref{fig:pdf}, where a more pronounced second peak establishes at larger $\Upphi$ due to enhanced coalescence. 
\begin{figure}
\centering
\begin{subfigure}[t]{0.03\textwidth}
\centering
\fontsize{6}{9}
\textbf{(a)}
\end{subfigure}
\begin{subfigure}[t]{0.45\textwidth}
\includegraphics[width=\linewidth, valign=t]{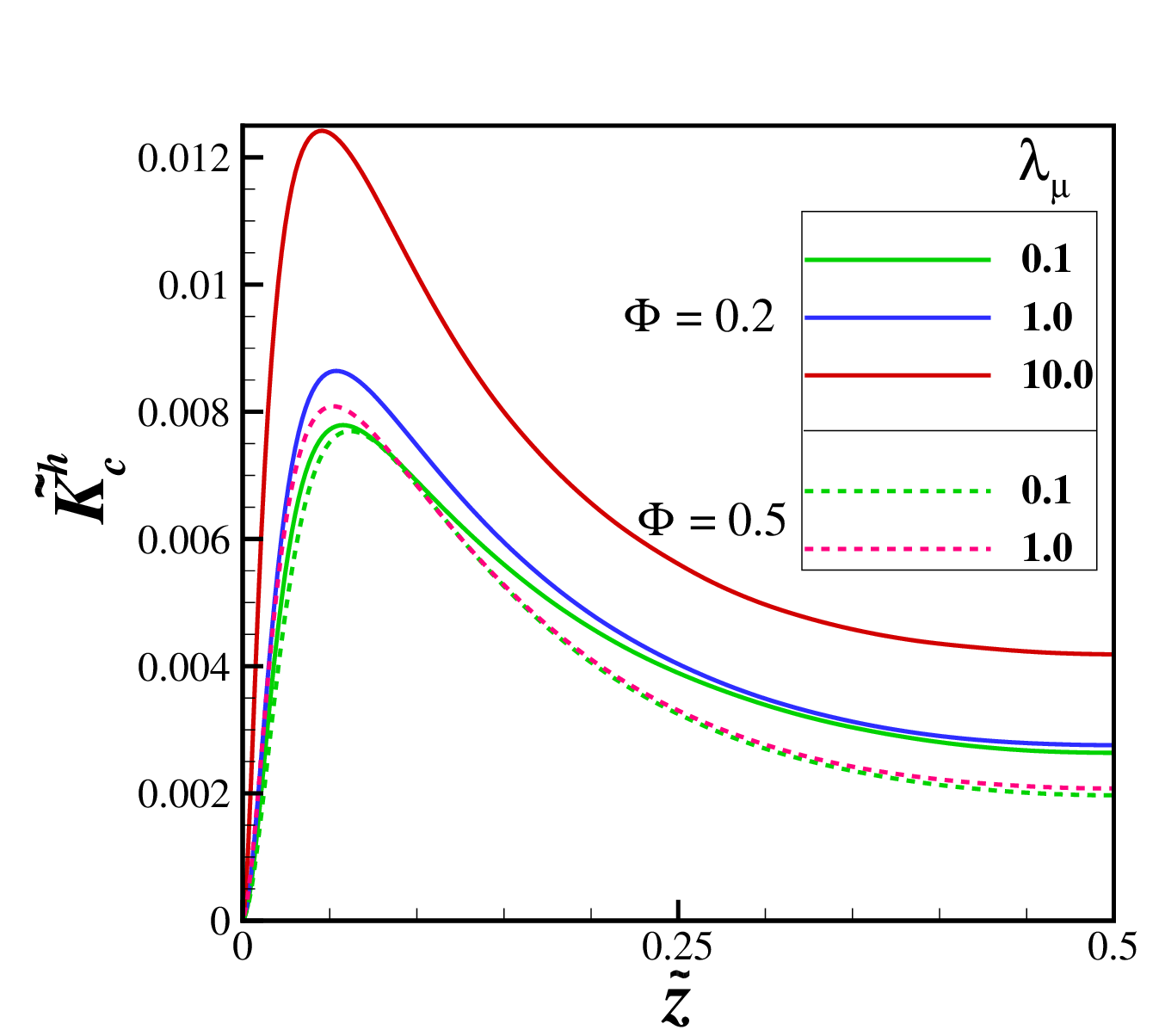}
\phantomsubcaption\label{fig:K h c sym visc}
\end{subfigure}\hfill
\begin{subfigure}[t]{0.03\textwidth}
\centering
\fontsize{6}{9}
\textbf{(b)}
\end{subfigure}
\begin{subfigure}[t]{0.45\textwidth}
\includegraphics[width=\linewidth, valign=t]{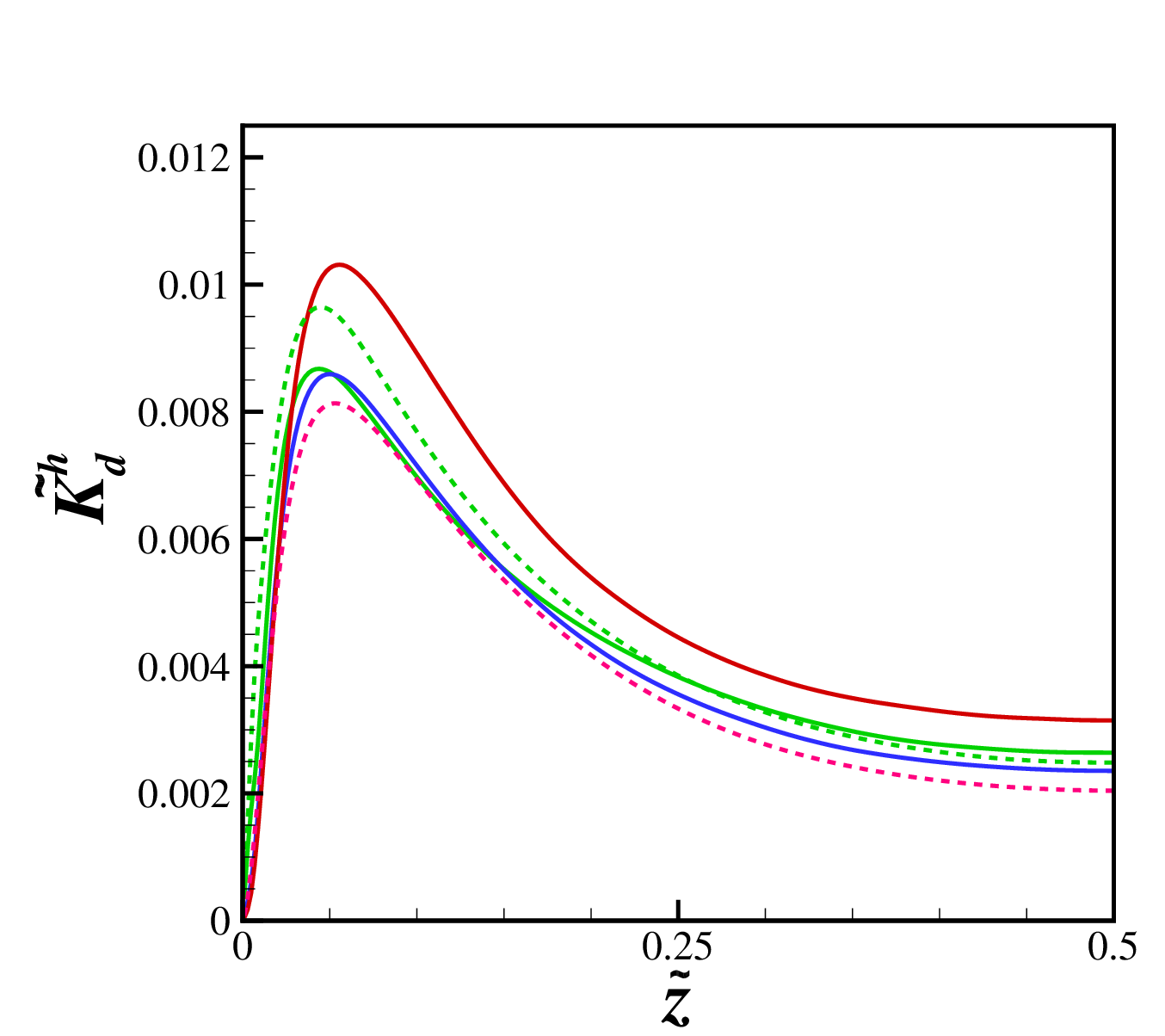}
\phantomsubcaption\label{fig:K h d sym visc}
\end{subfigure}
\begin{subfigure}[t]{0.028\textwidth}
\centering
\fontsize{6}{9}
\textbf{(c)}
\end{subfigure}
\begin{subfigure}[t]{0.45\textwidth}
\centering
\includegraphics[width=\linewidth, valign=t]{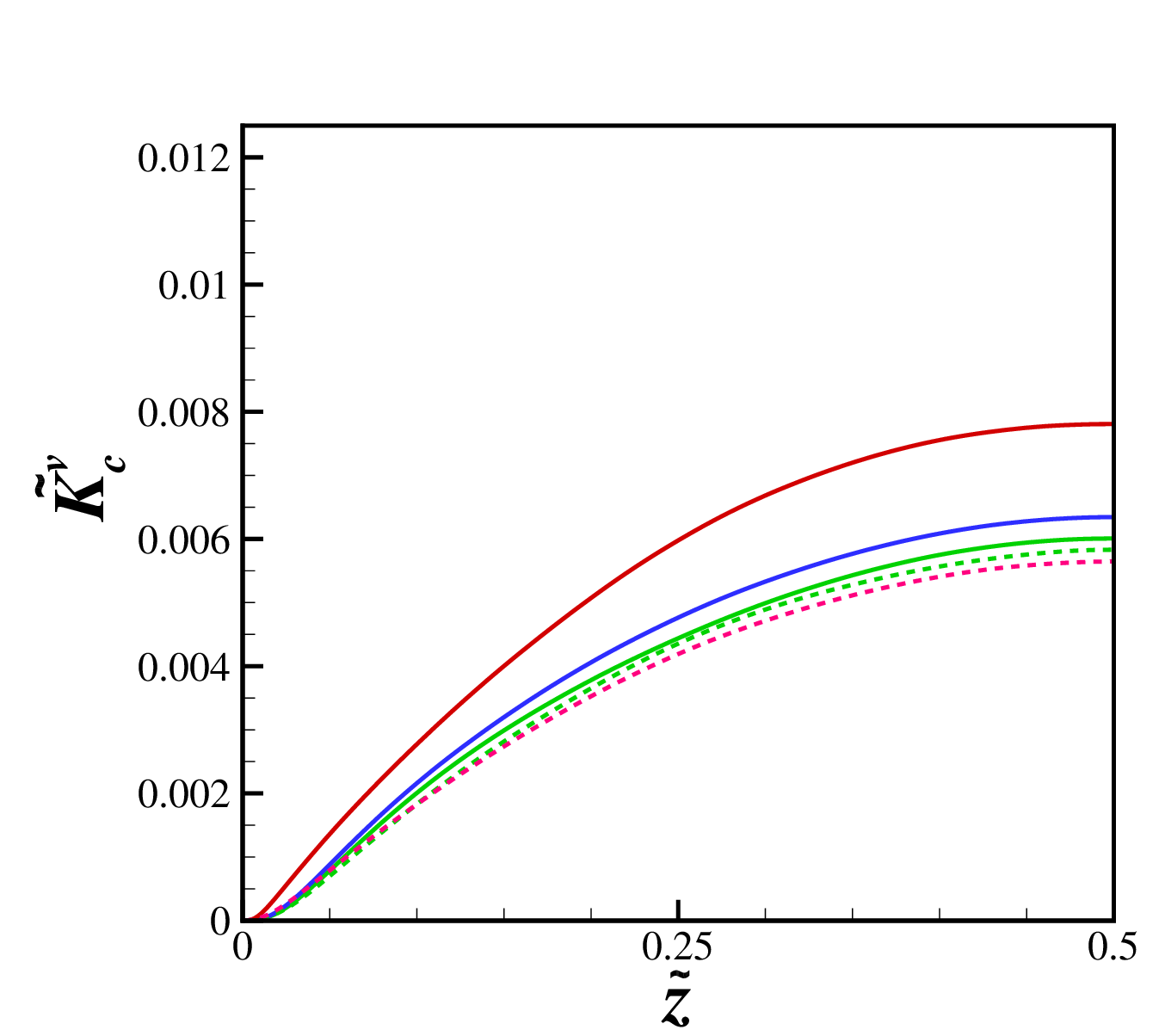}
\phantomsubcaption\label{fig:K v c sym visc}
\end{subfigure}
\begin{subfigure}[t]{0.048\textwidth}
\centering
\fontsize{6}{9}
\textbf{(d)}
\end{subfigure}
\begin{subfigure}[t]{0.45\textwidth}
\includegraphics[width=\linewidth, valign=t]{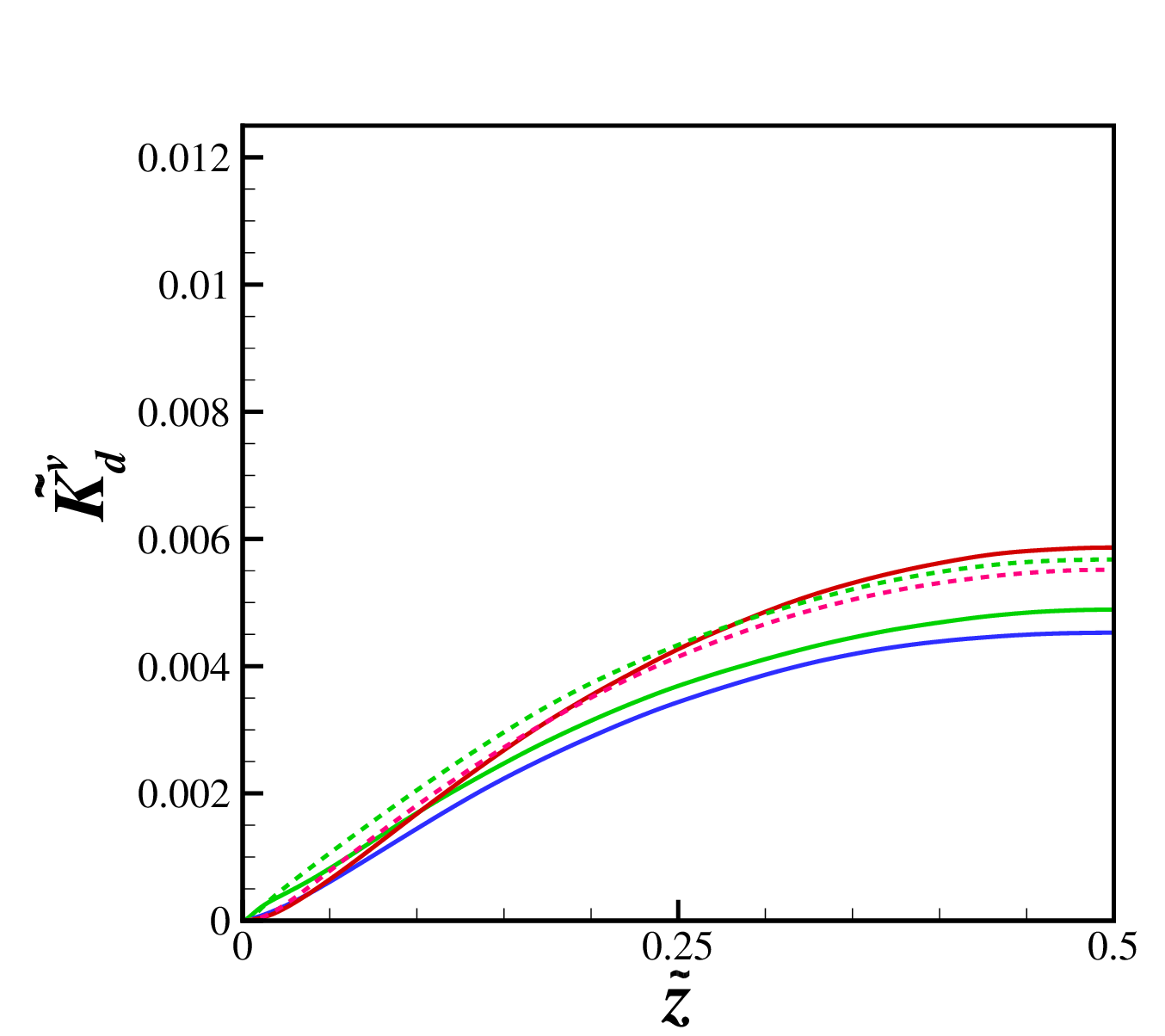}
\phantomsubcaption\label{fig:K v d sym visc}
\end{subfigure}
\caption{The horizontal (a,b) and vertical (c,d) components of the average kinetic energy per unit mass, derived from the velocity rms (equations \ref{eq:kinetic_energy_carrier} and \ref{eq:kinetic_energy_droplet}), as a function of the vertical direction for the case of $\Upphi=0.2$ and $\Upphi=0.5$ with different viscosity ratios.}
\label{fig:kinetic_energy_visc}
\end{figure}
\begin{figure}
\centering
\begin{subfigure}[t]{0.03\textwidth}
\centering
\fontsize{6}{9}
\textbf{(a)}
\end{subfigure}
\begin{subfigure}[t]{0.45\textwidth}
\includegraphics[width=\linewidth, valign=t]{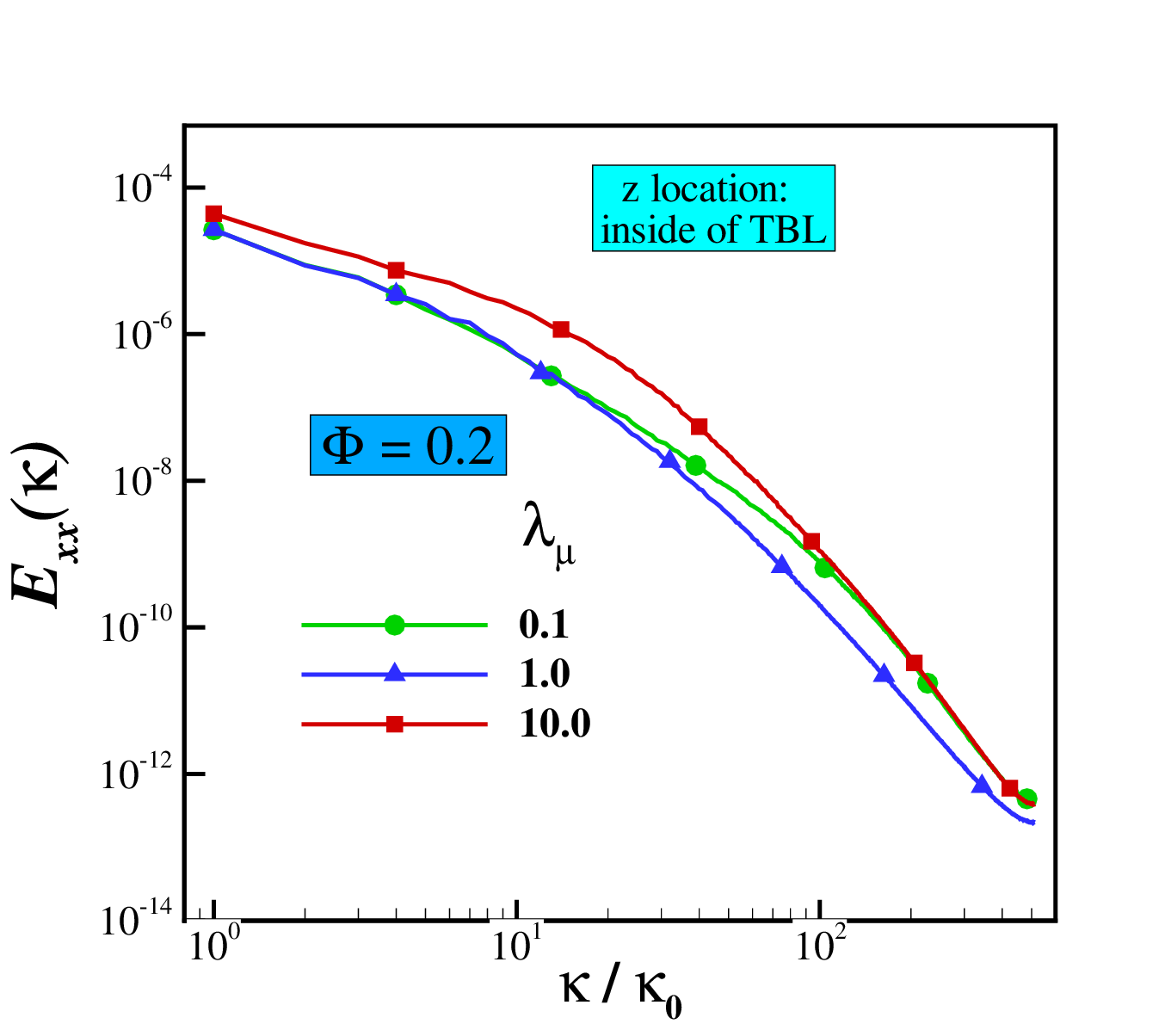}
\phantomsubcaption\label{fig:Exx_in_visc_20}
\end{subfigure}\hfill
\begin{subfigure}[t]{0.03\textwidth}
\centering
\fontsize{6}{9}
\textbf{(b)}
\end{subfigure}
\begin{subfigure}[t]{0.45\textwidth}
\includegraphics[width=\linewidth, valign=t]{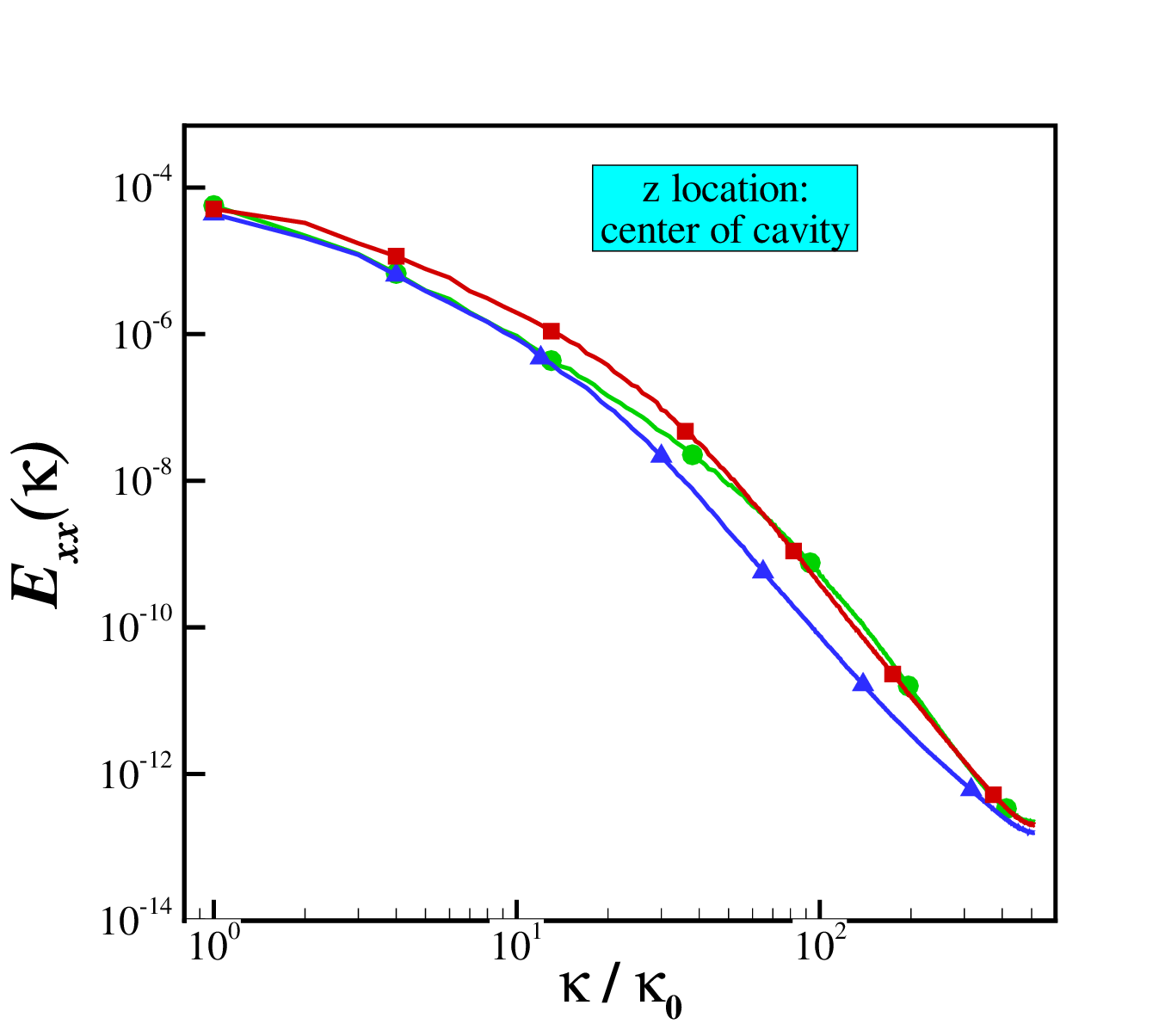}
\phantomsubcaption\label{fig:Exx_out_visc_20}
\end{subfigure}
\caption{TKE spectra $E_{xx}(\kappa)$ as a function of wavenumber at droplet volume fraction of $\Upphi=0.2$ and different viscosity ratios (a) inside of TBL and (b) at the center of cavity. Wavenumbers are normalized by the lowest non-zero wavenumber $\kappa_0 = \pi / H$.
}
\label{fig:spectra_visc_20}
\end{figure}

\subsection{Effects of different viscosity ratios on the heat transfer}\label{subsec:visc_effect}
We now consider the influence of the viscosity ratio on the flow turbulence and heat transfer, see cases 6-8 in Table \ref{table:list_of_simulations}. 
Specifically, we consider viscosity ratios $\lambda_{\mu}=(\mu_d/\mu_c)=$ 0.1 and 10, while the other dimensionless parameters remain the same as for cases 1-5. Among the various dispersed droplet volume fractions, we focus on $\Upphi=0.2$ and the case of a binary mixture $0.5$. 
As mentioned before in section \ref{subsec:governing_equations}, we define various dimensionless numbers, i.e. $Ra$, $We$, $Pr$ and $Fr$, based on an effective reference viscosity, which can be defined based on either the volumetric averaged viscosity or the viscosity of the continuous phase (traditional method). 
%
%
As a comparison between the two methods, it should be mentioned that, in defining dimensionless parameters, the traditional approach overlooks the thermophysical properties of the disperse phase, masking certain effects, particularly at high values of $\Upphi$. For instance, increasing the viscosity of the disperse phase tenfold simultaneously alters both the Rayleigh number in the disperse phase ($Ra_d$) and the viscosity ratio ($\lambda_{\mu}$), obscuring the specific impact of viscosity ratio changes. The mixture rule, however, defines dimensionless parameters based on the volumetric averaged properties, isolating the effect of the viscosity ratio. Therefore, in this section, we employ this mixture rule (cases 6-8, Table \ref{table:list_of_simulations}) to compute the effective viscosity, while the traditional method is used for cases 12 and 13 and the related results are presented in Appendix~\ref{sec:appendix_A}. The effective viscosity is calculated as the volume average of the viscosities of the two phases:
\begin{equation}\label{eqn:mu_eff}
  \mu_\text{eff}=\Upphi \mu_d + (1-\Upphi) \mu_c.
\end{equation}
In other words, the volumetric averaged viscosity, defined by the arithmetic average in equation~\eqref{eqn:mu_eff}, remains the same as the viscosity in the previous flow cases, whereas the viscosity of each phase varies. 
As an example, for viscosity ratio $\lambda_{\mu}=10$ at $\Upphi=0.2$, the viscosity of the dispersed phase is $2.8\mu_{1}$ and that of the carrier phase is $0.28\mu_{1}$, where $\mu_{1}$ is the viscosity of the cases with $\lambda_{\mu}=1$.
Furthermore, note that the cases with $\Upphi=0.5$ and $\lambda_{\mu}=0.1, 10$ are identical; we therefore report results for only one case.
With our choice, for the three ratios investigated, we expect the Nusselt number to stay the same if this is only a function of the effective emulsion viscosity: our results show that this is the case only for a viscosity ratio of $\lambda_{\mu}=0.1$. 

Regarding the choice of using the linear definition that we adopted here to define the effective viscosity, it should be mentioned that, instead of a linear definition, a more precise definition (e.g., using a relation obtained from the emulsion rheological curves) can more effectively take into account the effect of dispersed-phase in computing the effective viscosity. However, we estimated the effective viscosity for different cases using the work of \citet{de2019effect}, and it is clarified that $(\mu_\text{eff}/\mu_\text{sp})_\text{estimated}$ varies in the range of $0.55$ to $1.35$. Despite the work of \citet{de2019effect} is for shear flows, the effective viscosity obtained using that work is not substantially different from the linear definition we employed, although they are not identical.

\begin{figure}
\centering
\begin{subfigure}[t]{0.03\textwidth}
\centering
\fontsize{6}{9}
\textbf{(a)}
\end{subfigure}
\begin{subfigure}[t]{0.45\textwidth}
\includegraphics[width=\linewidth, valign=t]{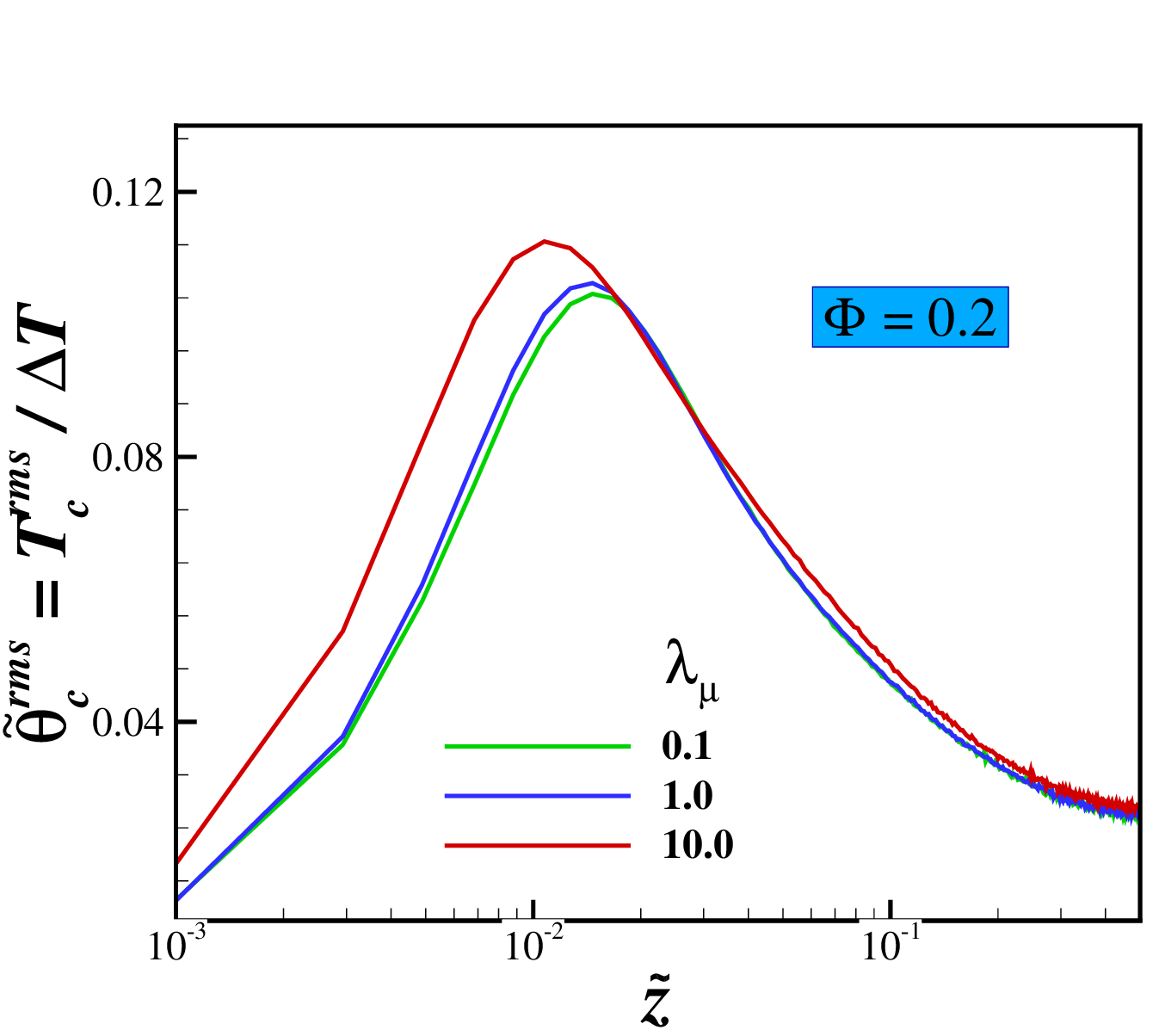}
 \phantomsubcaption\label{fig:tmp rms 20 visc carrier}
\end{subfigure}\hfill
\begin{subfigure}[t]{0.03\textwidth}
\centering
\fontsize{6}{9}
\textbf{(b)}
\end{subfigure}
\begin{subfigure}[t]{0.45\textwidth}
\includegraphics[width=\linewidth, valign=t]{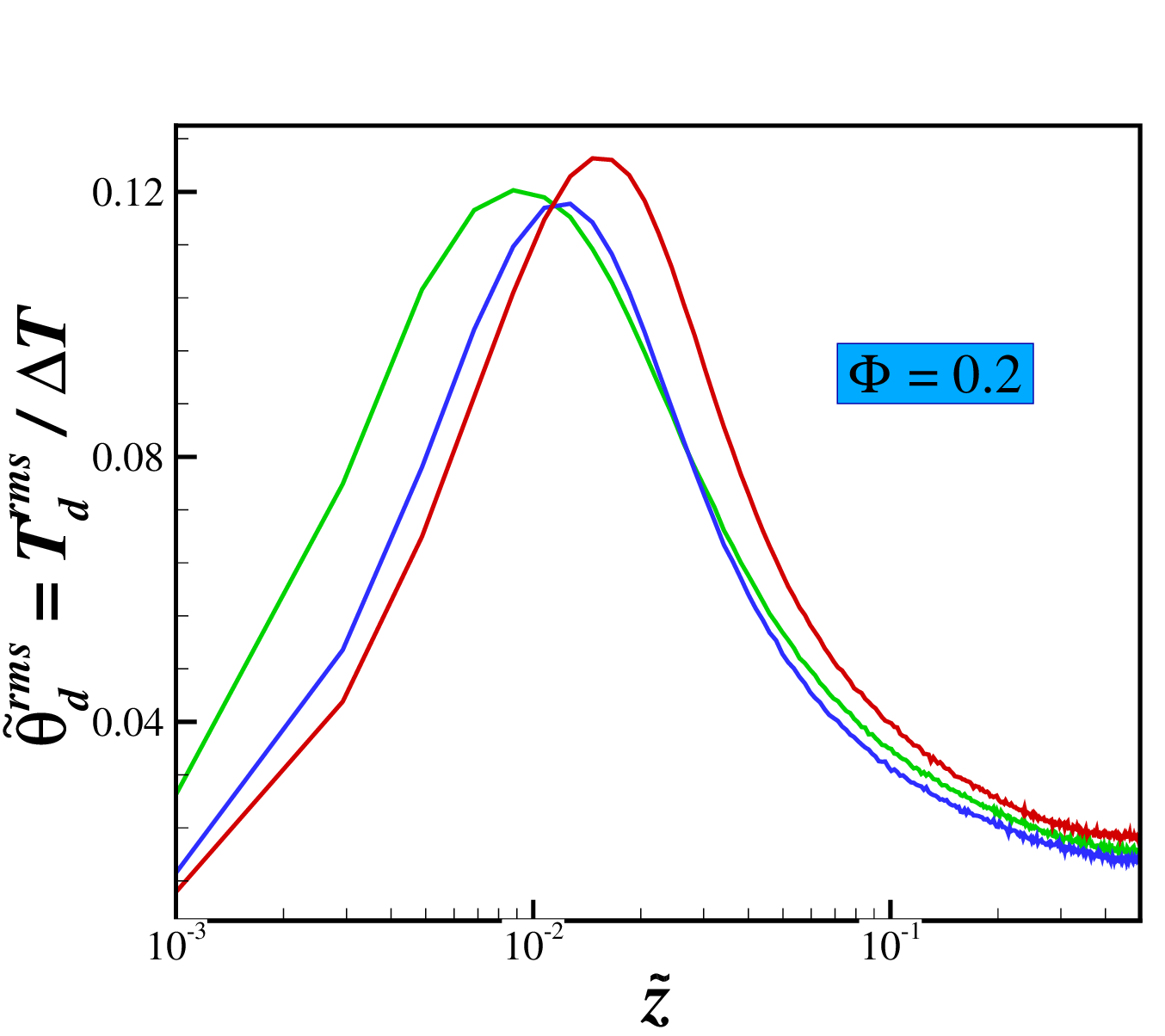}
\phantomsubcaption\label{fig:tmp rms 20 visc dispersed}
\end{subfigure}
\begin{subfigure}[t]{0.03\textwidth}
\centering
\fontsize{6}{9}
\textbf{(c)}
\end{subfigure}
\begin{subfigure}[t]{0.45\textwidth}
\centering
\includegraphics[width=\linewidth, valign=t]{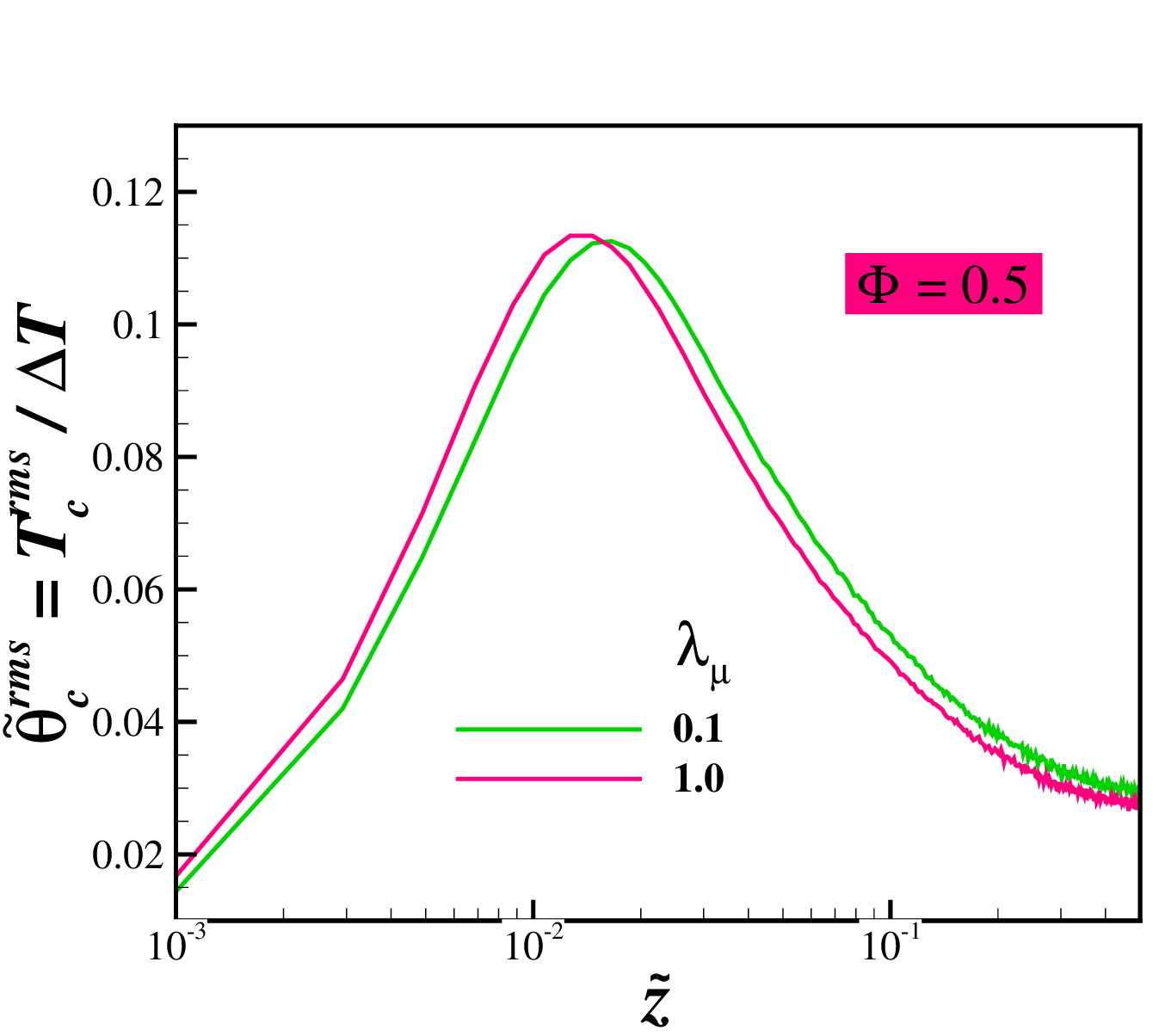}
\phantomsubcaption\label{fig:tmp rms 50 visc carrier}
\end{subfigure}
\begin{subfigure}[t]{0.03\textwidth}
\centering
\fontsize{6}{9}
\textbf{(d)}
\end{subfigure}
\begin{subfigure}[t]{0.45\textwidth}
\includegraphics[width=\linewidth, valign=t]{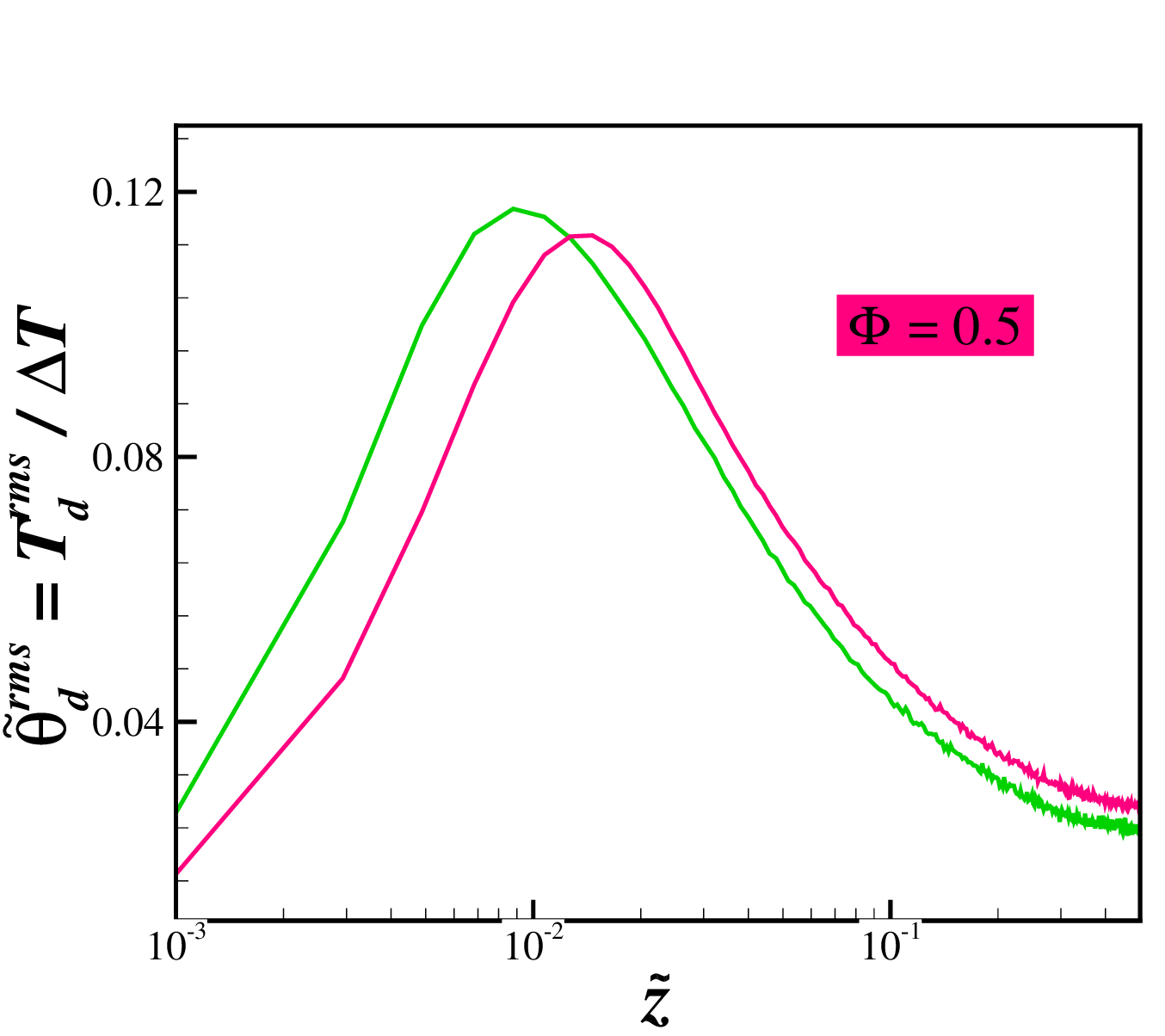}
\phantomsubcaption\label{fig:tmp rms 50 visc dispersed}
\end{subfigure}
\caption{(a-d) Carrier- and dispersed-phase rms temperature profiles along the wall-normal direction at various viscosity ratios and droplet volume fractions.}
\label{fig:tmp_rms_visc_c_d}
\end{figure}

\begin{figure}
  \centering
  \begin{subfigure}[t]{0.03\textwidth}
  \fontsize{6}{9}
    \textbf{(a)} 
  \end{subfigure}
  \begin{subfigure}[t]{0.45\textwidth}
    \includegraphics[width=\linewidth, valign=t]{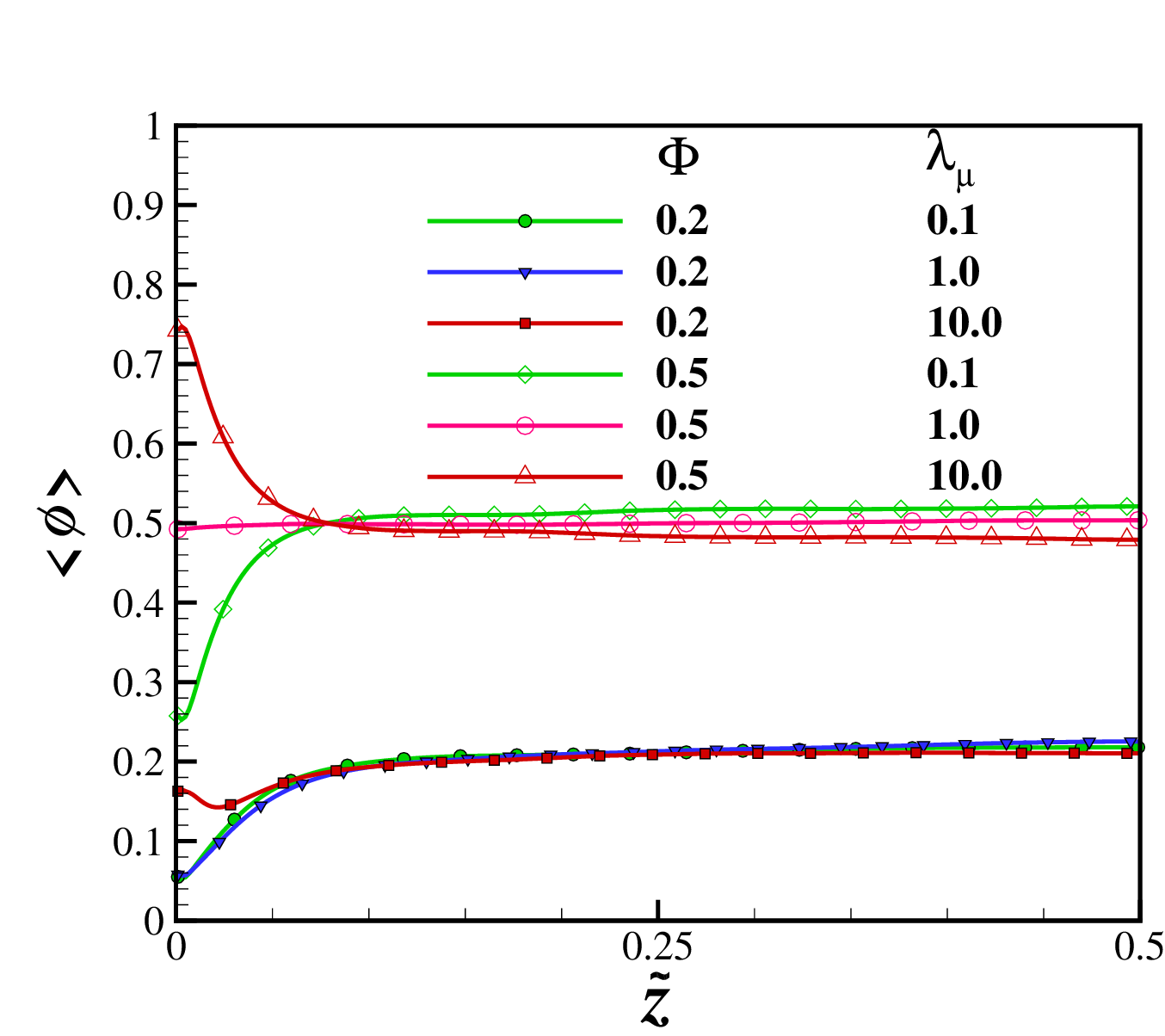}
    \phantomsubcaption\label{fig:phi local sym visc}
  \end{subfigure}\hfill
  \begin{subfigure}[t]{0.03\textwidth}
  \fontsize{6}{9}
    \textbf{(b)}
  \end{subfigure}
  \begin{subfigure}[t]{0.45\textwidth}
    \includegraphics[width=\linewidth, valign=t]{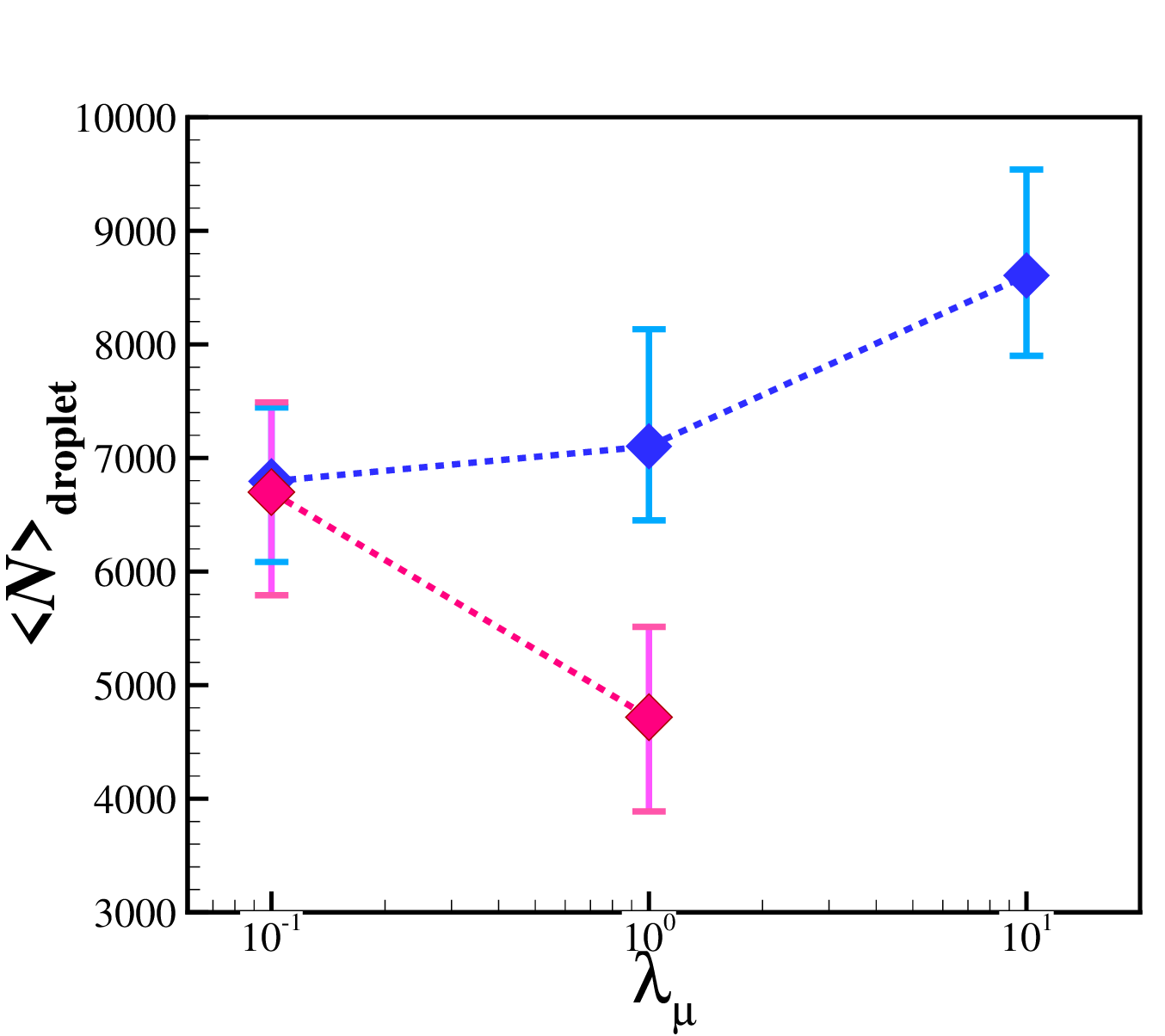}
    \phantomsubcaption\label{fig:N droplet range visc}
  \end{subfigure}
  \begin{subfigure}[t]{0.03\textwidth}
  \fontsize{6}{9}
    \textbf{(c)}
  \end{subfigure}
  \begin{subfigure}[t]{0.9\textwidth}
    \includegraphics[width=\linewidth, valign=t]{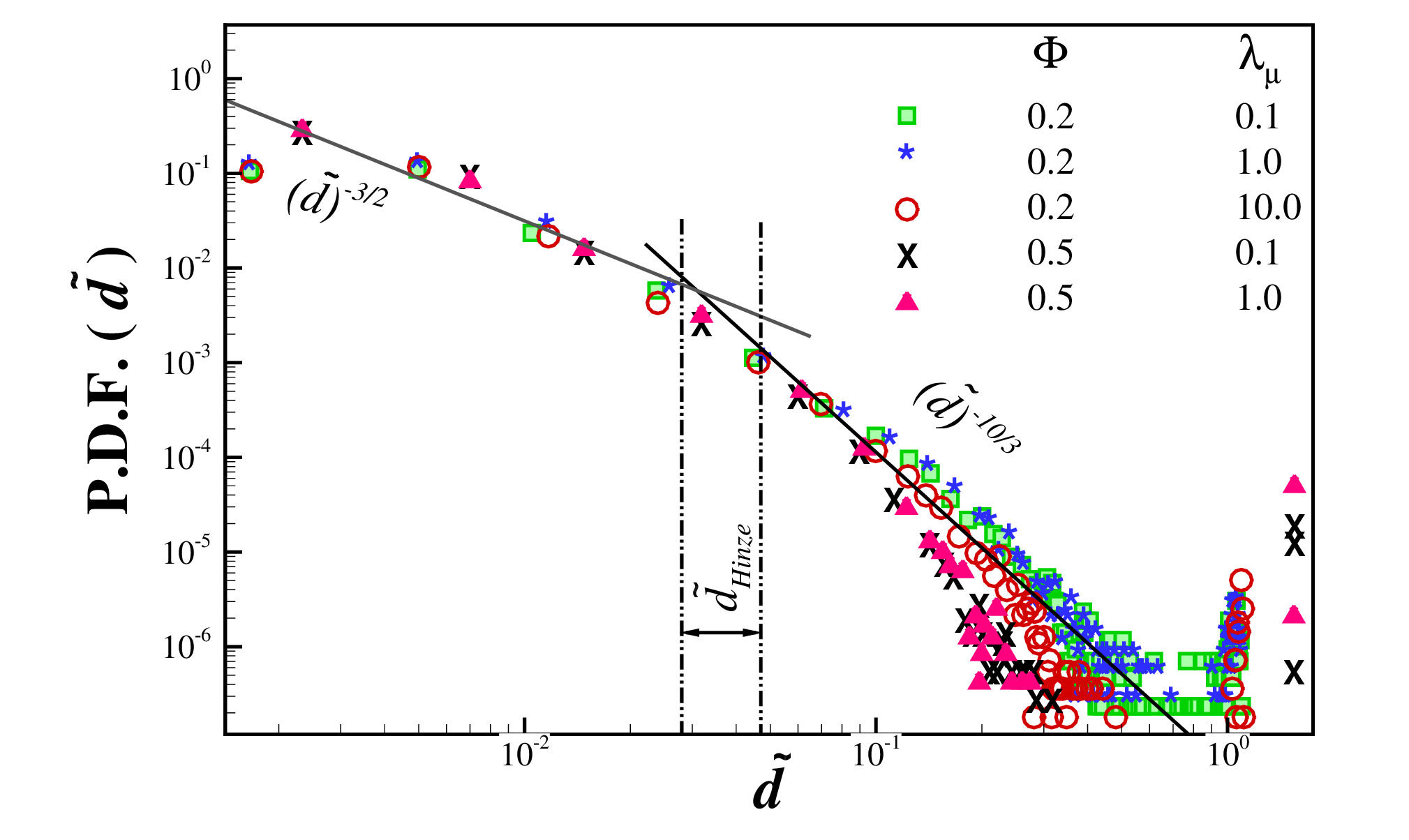}
    \phantomsubcaption\label{fig:pdf visc}
  \end{subfigure}
  \caption{(a) The local wall-normal distributions of the dispersed phase, (b) the number of droplets at the steady state condition. The blue and red dotted lines indicate $\Upphi = 0.2$ and $\Upphi = 0.5$, respectively (error bar indicating their fluctuation ranges), and (c) the p.d.f. of the DSD for the cases with different viscosity ratios and $\Upphi$.}
  \label{fig:DSD_visc}
\end{figure}

\begin{figure}
  \centering
  \begin{subfigure}[t]{0.03\textwidth}
  \fontsize{6}{9}
    \textbf{(a)} 
  \end{subfigure}
  \begin{subfigure}[t]{0.45\textwidth}
    \includegraphics[width=\linewidth, valign=t]{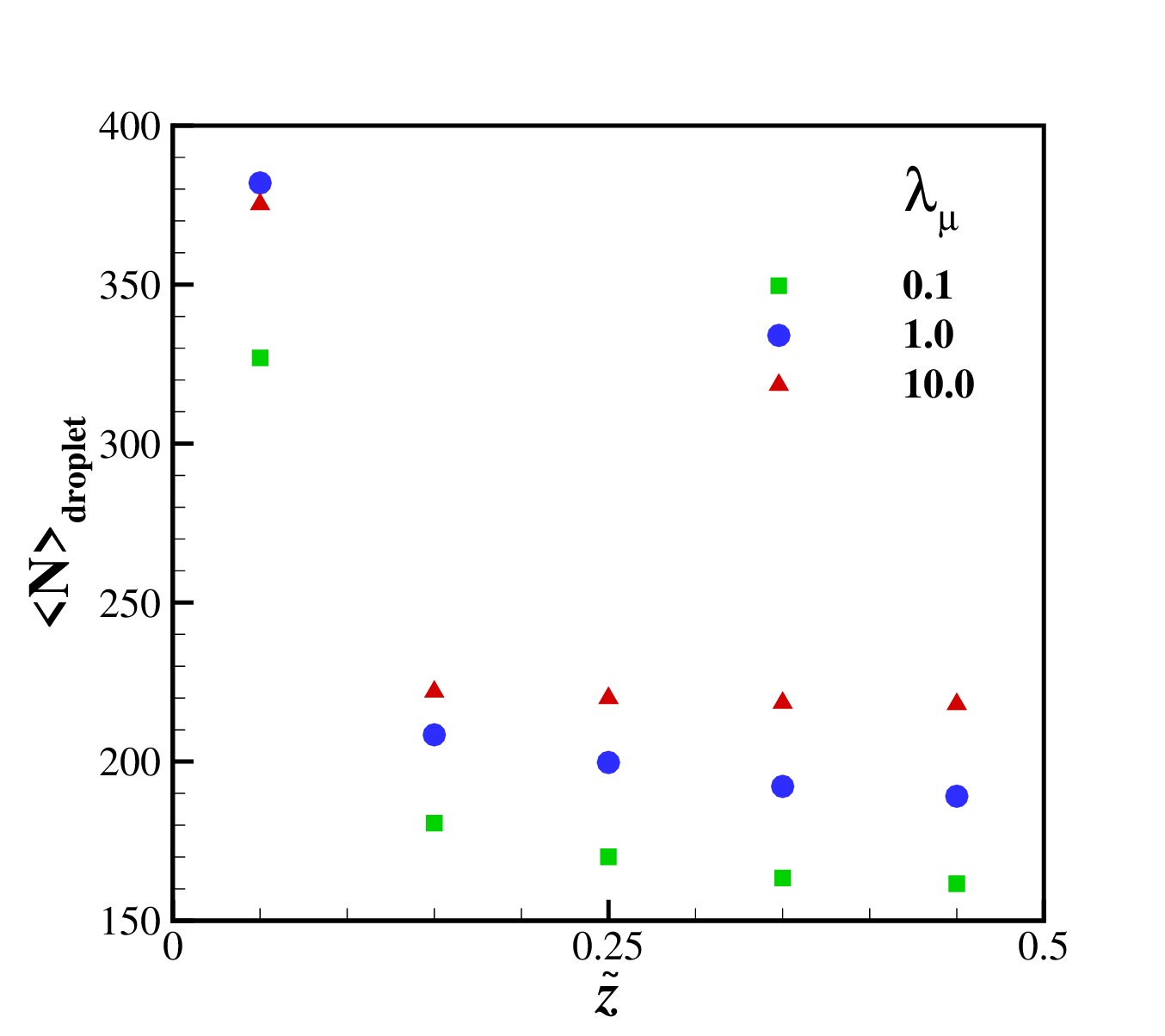}
    \phantomsubcaption\label{fig:avg_N_z_visc_20}
  \end{subfigure}\hfill
  \begin{subfigure}[t]{0.03\textwidth}
  \fontsize{6}{9}
    \textbf{(b)}
  \end{subfigure}
  \begin{subfigure}[t]{0.45\textwidth}
    \includegraphics[width=\linewidth, valign=t]{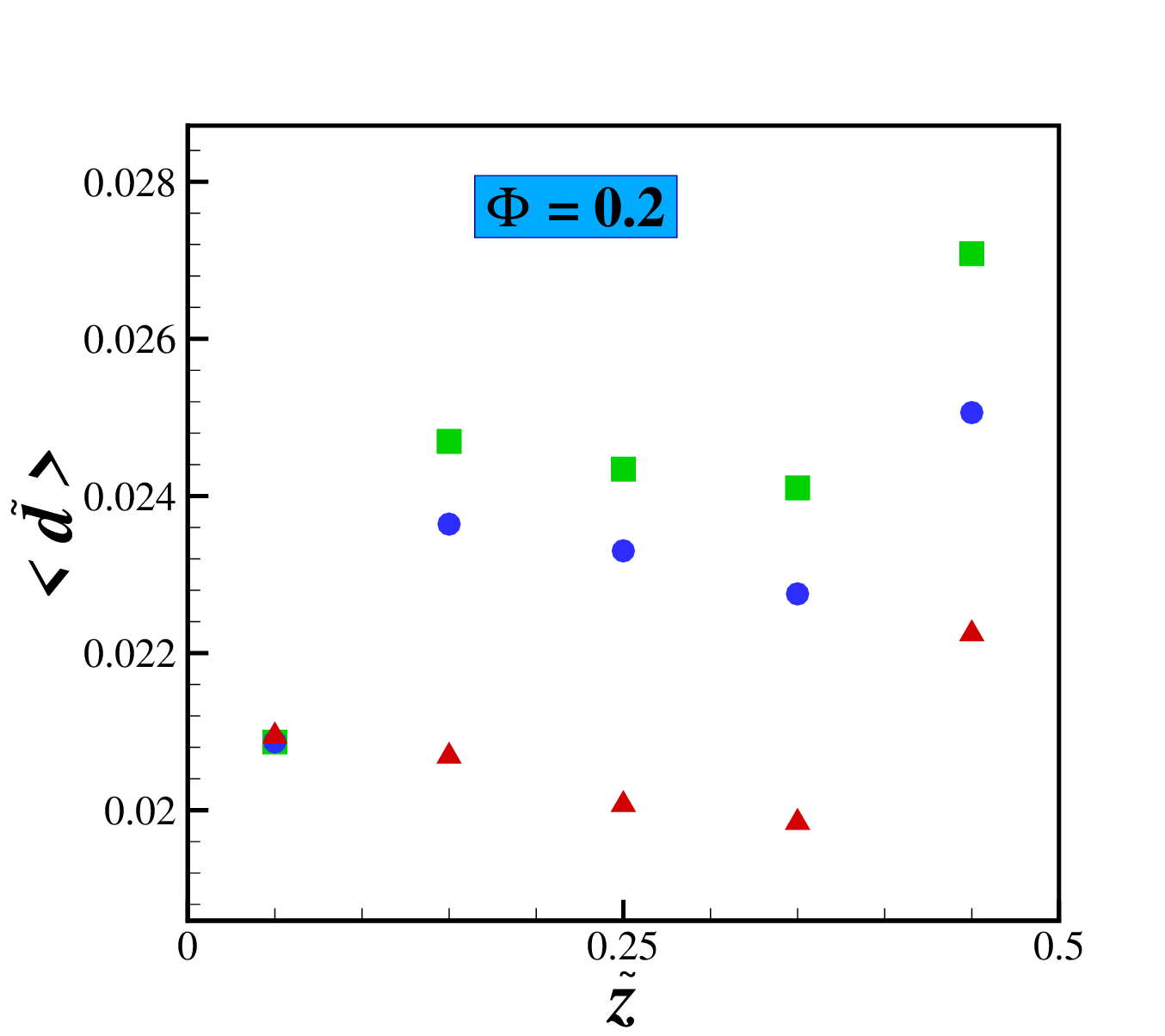}
    \phantomsubcaption\label{fig:avg_d_z_visc_20}
  \end{subfigure}
  \caption{(a) The average number of dispersed droplets and (b) the average diameter of droplets, along the wall-normal direction for the case with $\Upphi = 0.2$ and viscosity ratios of $0.1$, $1.0$ and $10$.}
  \label{fig:DSD_visc_2}
\end{figure}

 Figure \ref{fig:nu visc phi} and \ref{fig:TBL thickness visc phi} display the effect of different dynamic viscosity ratios on the Nusselt number and the thermal boundary-layer thickness. 
 For both volume fractions under examination, the Nusselt number varies weakly for $\lambda_{\mu} = 0.1$,
  while it increases for $\lambda_{\mu}=10$ and $\Upphi=0.2$. In particular, for $\Upphi=0.2$, the Nusselt number is increased by $25\%$ compared to the single-phase flow for viscosity ratio $\lambda_{\mu}= 10$, while it is enhanced only by $2.5\%$ and $1\%$ for viscosity ratios of $1$ and $0.1$. Consistently, the same trend is observed for the reduction in boundary-layer thickness, see panel (b).
 %
These observations suggest that a considerable heat transfer enhancement (around 24\%) can be achieved with a more viscous dispersed phase at the same Rayleigh number.
Figure \ref{fig:tmp mean 20 visc} reports the mean temperature for the different cases under investigation. We note that the temperature of the dispersed phase is slightly negative just outside the boundary layer closer to the hot wall, an effect more pronounced for the case with $\lambda_{\mu}=1$. This indicates that the transport of the droplet by plumes of the carrier phase occurs at a rate faster than the time required for heat to diffuse.

The increase of the global heat transfer at viscosity ratio $\lambda_{\mu}=10$ and $\Upphi=0.2$ is explained by the increased turbulence in the carrier phase. 
As shown in figure \ref{fig:kinetic_energy_visc}, the horizontal and vertical components of the turbulent kinetic energy are largest for the cases with a more viscous dispersed phase and $\Upphi=0.2$. In this case, the increase is most pronounced for the less-viscous carrier phase; nevertheless, a more turbulent carrier phase induces more intense fluctuations also inside the more-viscous dispersed phase. As discussed for the case $\lambda_{\mu}=1$, the majority of the heat transport is associated with the carrier phase as this penetrates deeper into the TBL. When this is relatively less viscous, turbulent fluctuations increase, and so does the heat transport associated with the carrier phase.
Concerning emulsions with $\lambda_{\mu}=0.1$, we note only a weak decrease in the energy of the carrier phase and an increase of the turbulence in the dispersed phase, mainly in the bulk
when comparing with the case $\lambda_{\mu}=1$. 
For the cases of binary mixtures (dotted lines in the figure), we notice an increase in the horizontal component of the less-viscous phase inside the boundary layer.

Figure \ref{fig:spectra_visc_20} displays the TKE spectrum for the flow with $\Upphi=0.2$ and at different locations: inside the TBL and in the core of the cavity. The results confirm an increase of the small-scale energy in emulsions when compared with the single-phase flow, an increase higher at viscosity ratios $\lambda_{\mu}=10$ and 1.
In the case $\lambda_{\mu}=10$, we also observe an energy increase at large scales, which we attribute to the reduced viscosity of the carrier phase. This energy enhancement at large scales is consistent with the results of figure \ref{fig:kinetic_energy_visc}. Hence, we attribute the largest increase of the Nusselt number for $\lambda_{\mu}=10$ to the combined effect of increased small-scale mixing due to the presence of an interface, as for the case $\lambda_{\mu}=1$, and of increased turbulence in the carrier phase due to its reduced viscosity. 
The wall-normal profiles of the temperature fluctuations are reported in figure \ref{fig:tmp_rms_visc_c_d} for the two-volume fractions under consideration, where values for each phase are presented. The data at $\Upphi=0.2$ show that the level of carrier-phase fluctuation increases close to the wall when this is less viscous, while those of the dispersed phase increase further from the wall when $\lambda_{\mu}=10$. In other words, the near-wall activity increases close to the wall for the case $\lambda_{\mu}=10$, as shown also by the horizontal velocity fluctuations in figure \ref{fig:K h c sym visc}. 
The results pertaining to the binary mixtures are presented in \ref{fig:tmp rms 50 visc carrier} and \ref{fig:tmp rms 50 visc dispersed} for the carrier and dispersed phase (the notation of carrier and the dispersed phase is maintained here to be able to identify the more or less viscous phases). The data indicate that the temperature fluctuation peak of the more viscous case moves towards the wall.

%


Figure \ref{fig:phi local sym visc} reports the local wall-normal distributions of the dispersed phase at different viscosity ratios for the cases with $\Upphi=0.2$ and $0.5$. It can be seen that the decrease of dispersed phase close to the wall observed in the case $\lambda_{\mu}=1$ remains when the dispersed phase is less viscous, $\lambda_{\mu}=0.1$, while the distribution is more uniform across the cavity when $\lambda_{\mu}=10$. This is attributed to the increase in turbulence and thinning of the thermal boundary layer in the case of more viscous dispersed phase. Also, at $\lambda_{\mu} = 10$ and $\Upphi=0.2$, we observe droplet layering close to the wall, as shown by a negative peak in the local distribution, roughly at the boundary layer edge. In the case of a binary mixture, we observe that the less viscous phase tends to leave the wall, whereas the more viscous is preferentially found close to it.

Next, we present the results regarding the droplet size distributions.  The steady-state number of droplets (as well as their fluctuation ranges) is depicted in figure \ref{fig:N droplet range visc}, whereas the p.d.f of the DSD in \ref{fig:pdf visc}. For the moderately concentrated case of $\Upphi=0.2$, as $\lambda_{\mu}$ increases, the turbulence intensity increases and the breakup becomes more frequent, leading to a higher number of smaller dispersed droplets and fewer larger ones (see p.d.f. graphs of figure \ref{fig:pdf visc}).
To gain insight on the DSD across the cavity, we display the average number of dispersed droplets and the average diameter of droplets along the wall-normal direction in figure \ref{fig:DSD_visc_2} for the case with $\Upphi = 0.2$: the data reveal that more droplets are found closer to the wall for the cases with $\lambda_{\mu}=1$ and 10. In the bulk, we find more and smaller droplets in the case of a more viscous dispersed phase ($\lambda_{\mu}=10$), a consequence of the more intense turbulence. Fewer and bigger droplets are observed in the bulk when $\lambda_{\mu}=0.1$. Interestingly, droplets of similar size are observed close to the wall for all viscosity ratios under investigation. The data shown in the figure confirm the presence of smaller droplets close to the wall, responsible for increased mixing, as discussed in the previous section for the flow cases at $\lambda_{\mu}=1$.
The trends in figure \ref{fig:DSD_visc_2} can be explained by the increased turbulent activity at $\lambda_{\mu}=10$ as well as depletion of the thermal boundary layer for $\lambda_{\mu}=0.1$. 
 

For the binary mixtures ($\Upphi=0.5$), the number of droplets is minimum at $\lambda_{\mu}=1$ and increases when the two fluids have different viscosity, $\lambda_{\mu}=0.1, 10$. The increase in droplet is here attributed to the preferential accumulation of the more viscous phase close to the wall and to an increase of the horizontal components of the turbulent kinetic energy.
In other words, decreasing $\lambda_{\mu}$ from $1$ to $0.1$ enhances the breakup rates, resulting in a higher number of smaller dispersed droplets with the less viscous ones migrating from the near-wall region towards the core of the cavity (see figure \ref{fig:phi local sym visc}).

\begin{figure}
  \centering
  \begin{subfigure}[t]{0.03\textwidth}
  \fontsize{6}{9}
    \textbf{(a)}
  \end{subfigure}
  \begin{subfigure}[t]{0.7\textwidth}
    \includegraphics[width=\linewidth, valign=t]{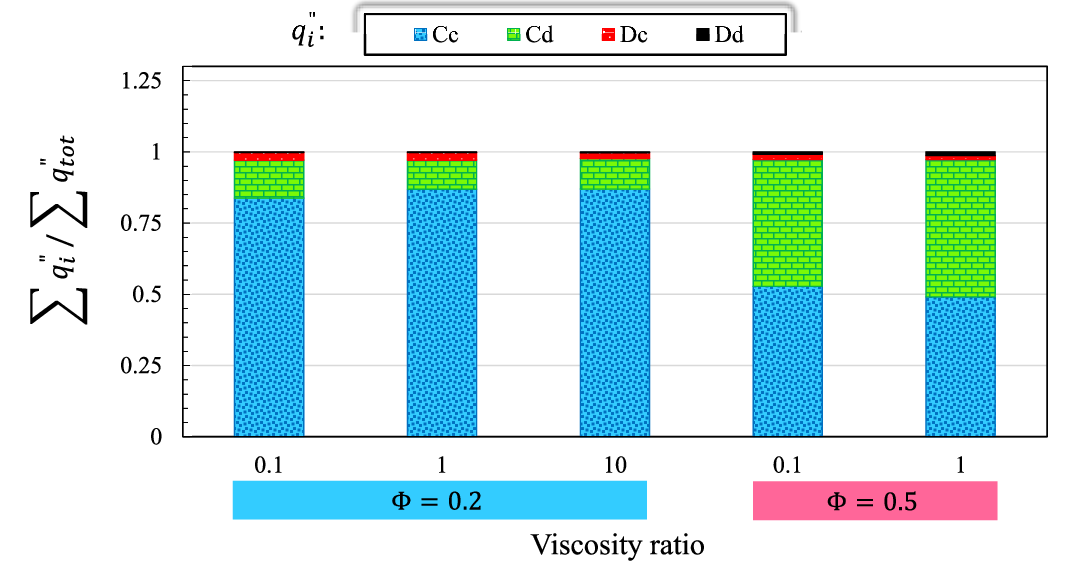}
    \phantomsubcaption\label{fig:heat budget barchart visc}
  \end{subfigure}
  \begin{subfigure}[t]{0.03\textwidth}
  \fontsize{6}{9}
    \textbf{(b)} 
  \end{subfigure}
  \begin{subfigure}[t]{0.45\textwidth}
    \includegraphics[width=\linewidth, valign=t]{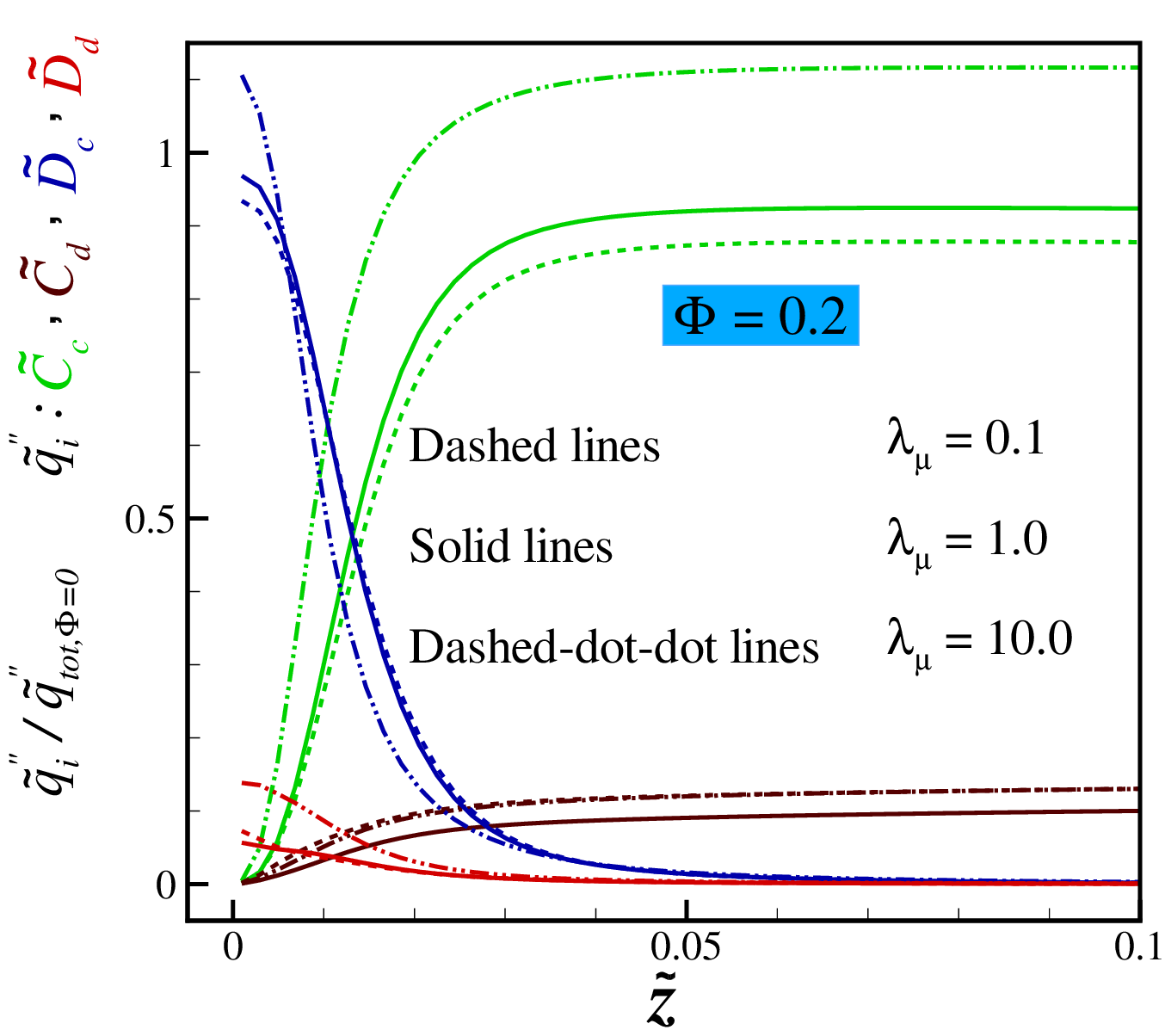}
    \phantomsubcaption\label{fig:heat budget 20 visc}
  \end{subfigure}\hfill
  \begin{subfigure}[t]{0.03\textwidth}
  \fontsize{6}{9}
    \textbf{(c)}
  \end{subfigure}
  \begin{subfigure}[t]{0.45\textwidth}
    \includegraphics[width=\linewidth, valign=t]{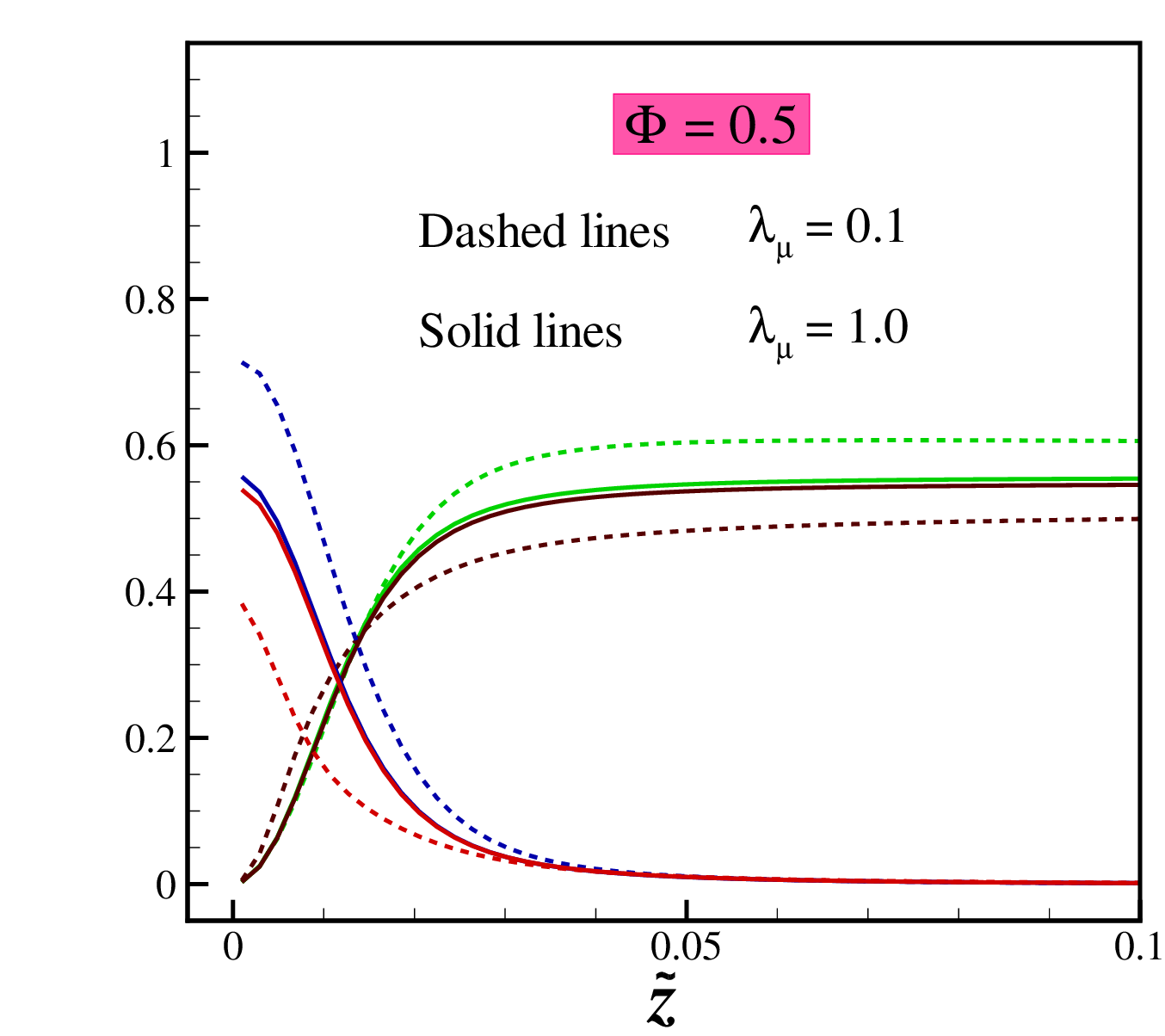}
    \phantomsubcaption\label{fig:heat budget 50 visc}
  \end{subfigure}
  \caption{(a) Wall-normal integral of heat fluxes transferred by diffusion and convection, normalized by the total heat flux of each case; (b,c) wall-normal distributions of convection and diffusion terms of both phases, normalized by the volume averaged total heat flux of the single-phase case ($\Upphi=0.0$) for volume fraction $\Upphi=0.2$ and $0.5$. }
  \label{fig:heat_budget_visc}
\end{figure}

To deepen our analysis, we report in figure \ref{fig:heat_budget_visc} the heat transfer budget for the cases with different viscosity ratios, and $\Upphi=0.2$ and $0.5$. In figure \ref{fig:heat budget barchart visc}, the barcharts indicate the wall-normal integral of the diffusion and convection heat fluxes, normalized by the total heat flux for each case. 
Figures \ref{fig:heat budget 20 visc} and \ref{fig:heat budget 50 visc} further detail the wall-normal distributions of these heat flux terms, specifically focusing on the near-wall region, $\tilde z = [0-0.1]$. This range is selected as it encompasses the region where most variations occur. 

Starting with the emulsion with volume fraction $\Upphi=0.2$,  the different plots confirm that the increase in heat fluxes at $\lambda_{\mu} = 10$ is due to the increase of the transport in the carrier phase, which is now less viscous than for $\lambda_{\mu} = 1$. The turbulent transport from the dispersed phase is similar among the different cases, being slightly lower in the case of viscosity ratio 1
(see in particular panel  \ref{fig:heat budget 20 visc}). For the case of a less viscous dispersed phase, we note a decrease in transport from the carrier phase, which is only partially compensated by the increase in fluxes associated to transport by the droplets.

To conclude, we have shown that, at fixed Rayleigh number and effective viscosity, the Nusselt number increases when the viscosity ratio increases from $1$ to $10$ for volume fraction $\Upphi=0.2$, i.e.\ when the carrier phase becomes less viscous at $\lambda_{\mu} = 10$. In this case, in addition to the increase of small-scale mixing discussed for the case $\lambda_{\mu} = 1$, the turbulent fluctuations increase, the thermal boundary layer is thinner, and the more viscous droplets tend to remain closer to the wall. 
%
%
Our results indicate that the viscosity ratio needs to be considered explicitly when determining empirical correlations for the Nusselt number when larger than 1 ($1 < \lambda_{\mu} \le 10$), while the idea of an effective viscosity can be used when the dispersed phase is less viscous ($0.1 \le \lambda_{\mu} < 1$).
Note finally that the addition of a more viscous phase to the same carrier fluid would lead to an increase of the effective viscosity. In this case, the increase in Nusselt shown here will be modulated by the decrease in the effective Rayleigh number, so that both parameters, $\lambda_{\mu}$ and $Ra$, need to be considered. This has been shown in the Appendix~\ref{sec:appendix_A} where the traditional method has been employed, and a reduction in Nusselt number is observed at higher $\lambda_{\mu}$ due to the damped turbulence level caused by the reduction in the effective Rayleigh number.

\subsection{Effects of various thermal diffusivity ratios on heat transfer}\label{subsec:k_effect}
\begin{figure}
\center
    \includegraphics[width=0.6\linewidth]{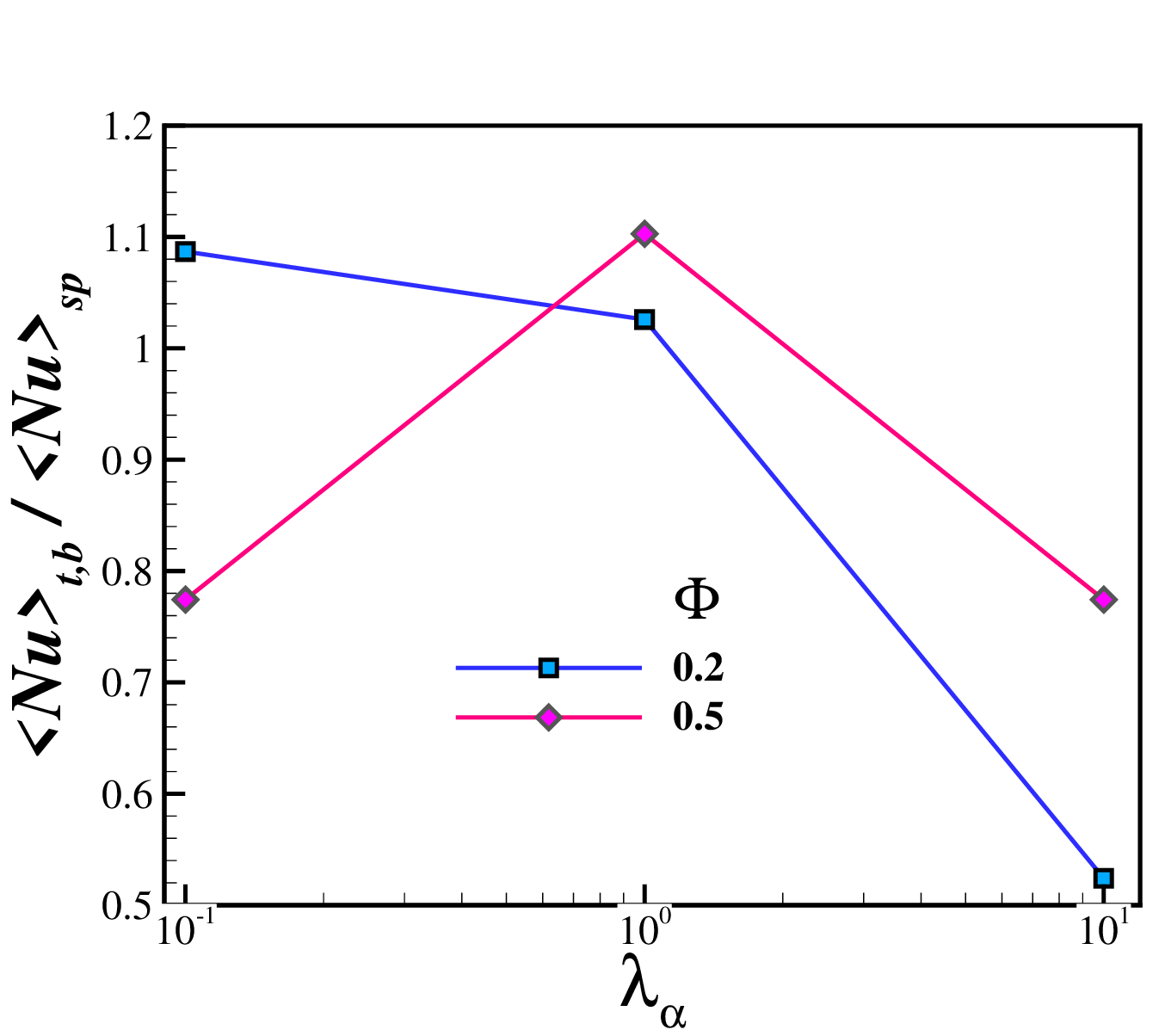} 
    \caption{Nusselt number as a function of the thermal diffusivity ratio for the emulsion with volume fraction of the dispersed phase $\Upphi=0.2$ and for binary mixture.}
  \label{fig:nu_k_phi}
\end{figure}
In this section, we examine the effects of varying thermal diffusivity ratios ($\lambda_{\alpha} = \alpha_d/\alpha_c$) on the heat transfer mechanism inside the cavity. We conduct simulations for three configurations (cases 9-11 in Table \ref{table:list_of_simulations}) within the range of $0.1 \leq \lambda_{\alpha} \leq 10$ by changing the thermal conductivity of both phases ($k_d$ and $k_c$). This is done so that the average thermal diffusivity of the emulsion remains constant and equals the thermal diffusivity of cases 1-5.

Figure \ref{fig:nu_k_phi} presents the normalized Nusselt number as a function of thermal diffusivity ratio for the three cases with the dispersed-phase volume fractions of $\Upphi=0.2$ and $0.5$. The behavior differs between these two volume fractions. In particular, at $\Upphi=0.2$, the Nusselt number increases at smaller $\lambda_{\alpha}$, whereas, for the case of a binary mixture, it is maximum for $\lambda_{\alpha}=1$ and decreases when the thermal diffusivities of the two phases are not equal ($\lambda_{\alpha}=0.1, 10$).
%

%
\begin{figure}
  \centering
  \begin{subfigure}[t]{0.03\textwidth}
  \fontsize{6}{9}
    \textbf{(a)}
  \end{subfigure}
  \begin{subfigure}[t]{0.45\textwidth}
    \includegraphics[width=\linewidth, valign=t]{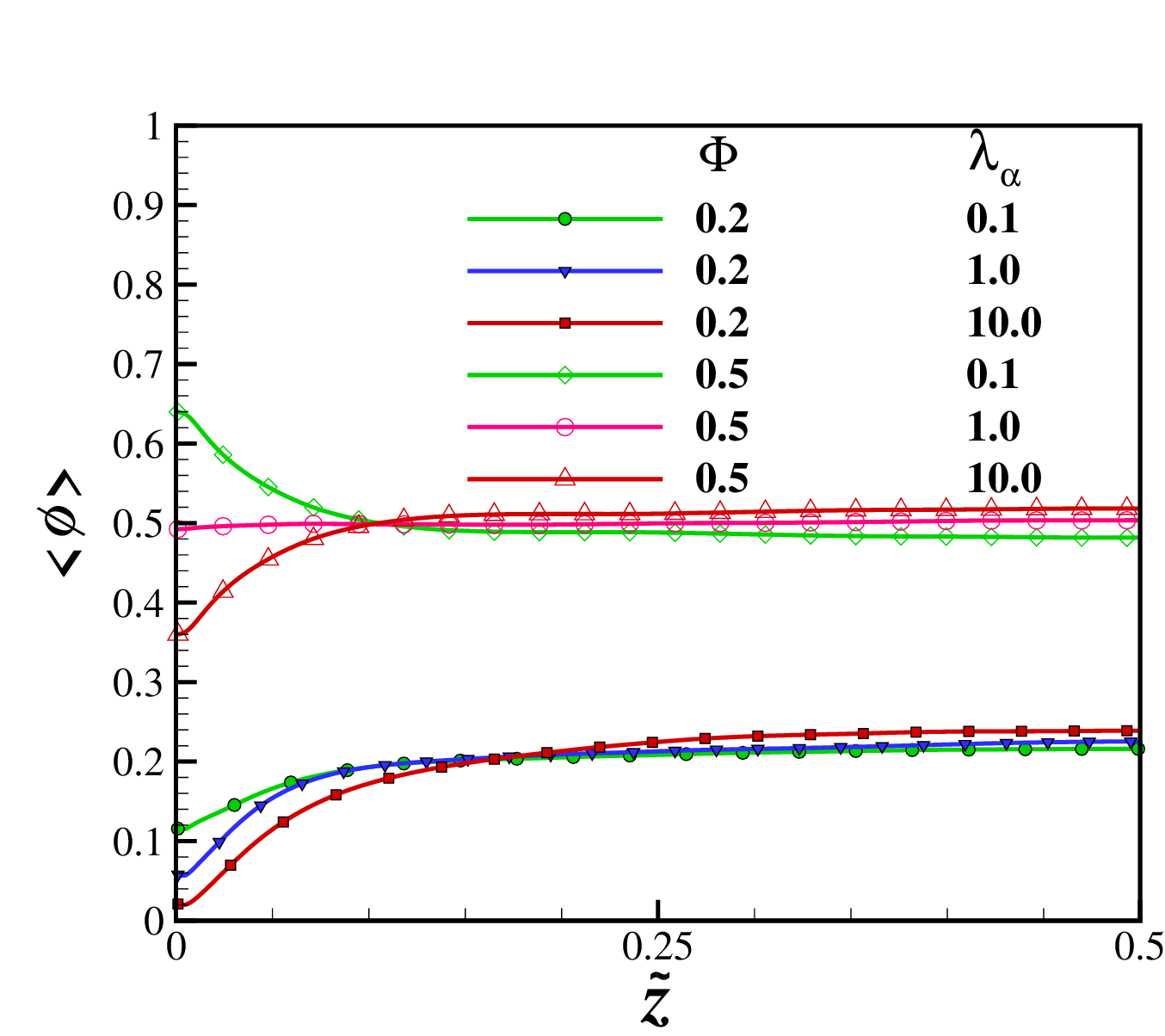}
    \phantomsubcaption\label{fig:phi local sym k}
  \end{subfigure}
  \begin{subfigure}[t]{0.03\textwidth}
  \fontsize{6}{9}
    \textbf{(b)} 
  \end{subfigure}
  \begin{subfigure}[t]{0.45\textwidth}
    \includegraphics[width=\linewidth, valign=t]{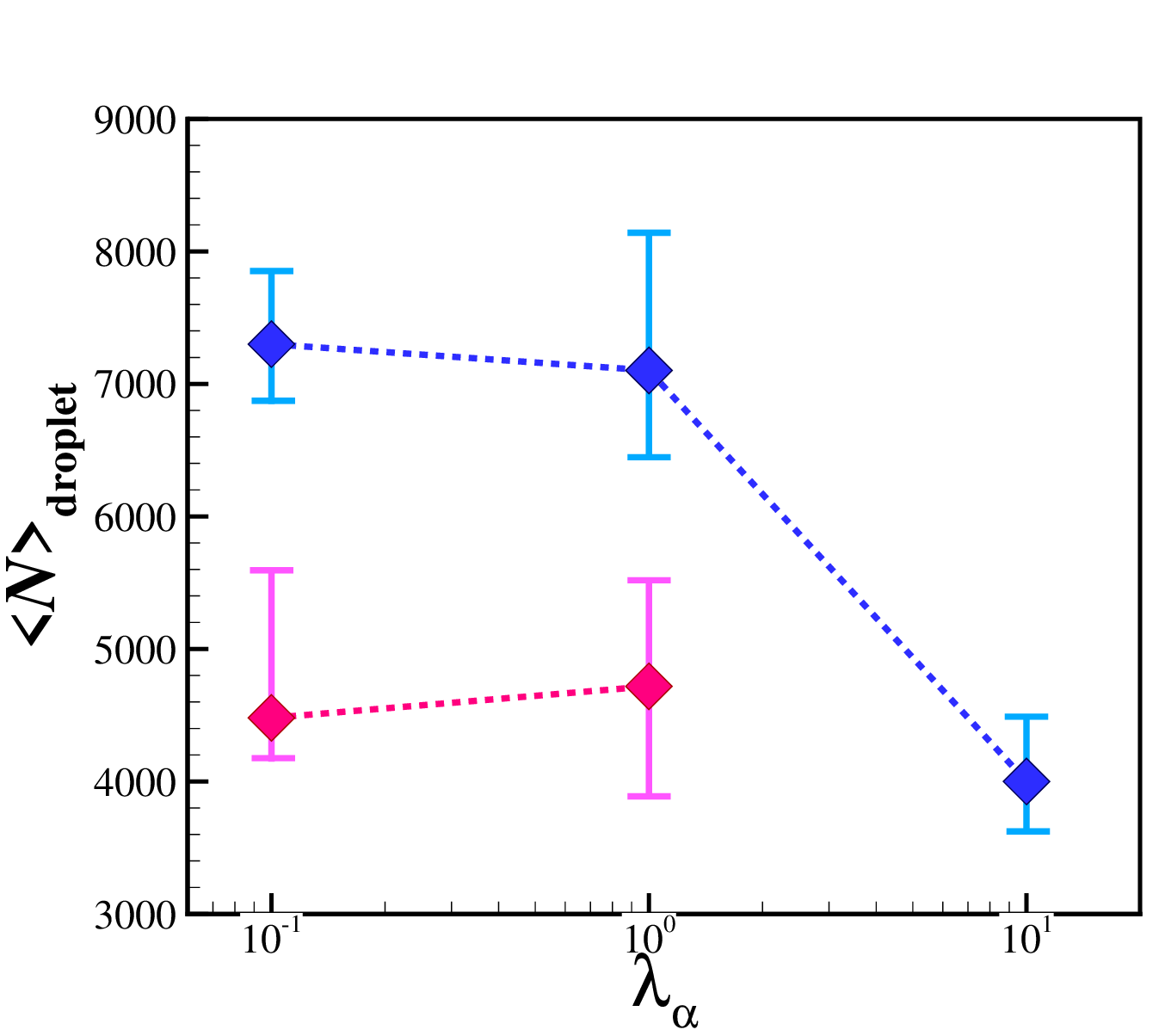}
    \phantomsubcaption\label{fig:N droplet k}
  \end{subfigure}
  \begin{subfigure}[t]{0.03\textwidth}
  \fontsize{6}{9}
    \textbf{(c)} 
  \end{subfigure}
  \begin{subfigure}[t]{0.45\textwidth}
    \includegraphics[width=\linewidth, valign=t]{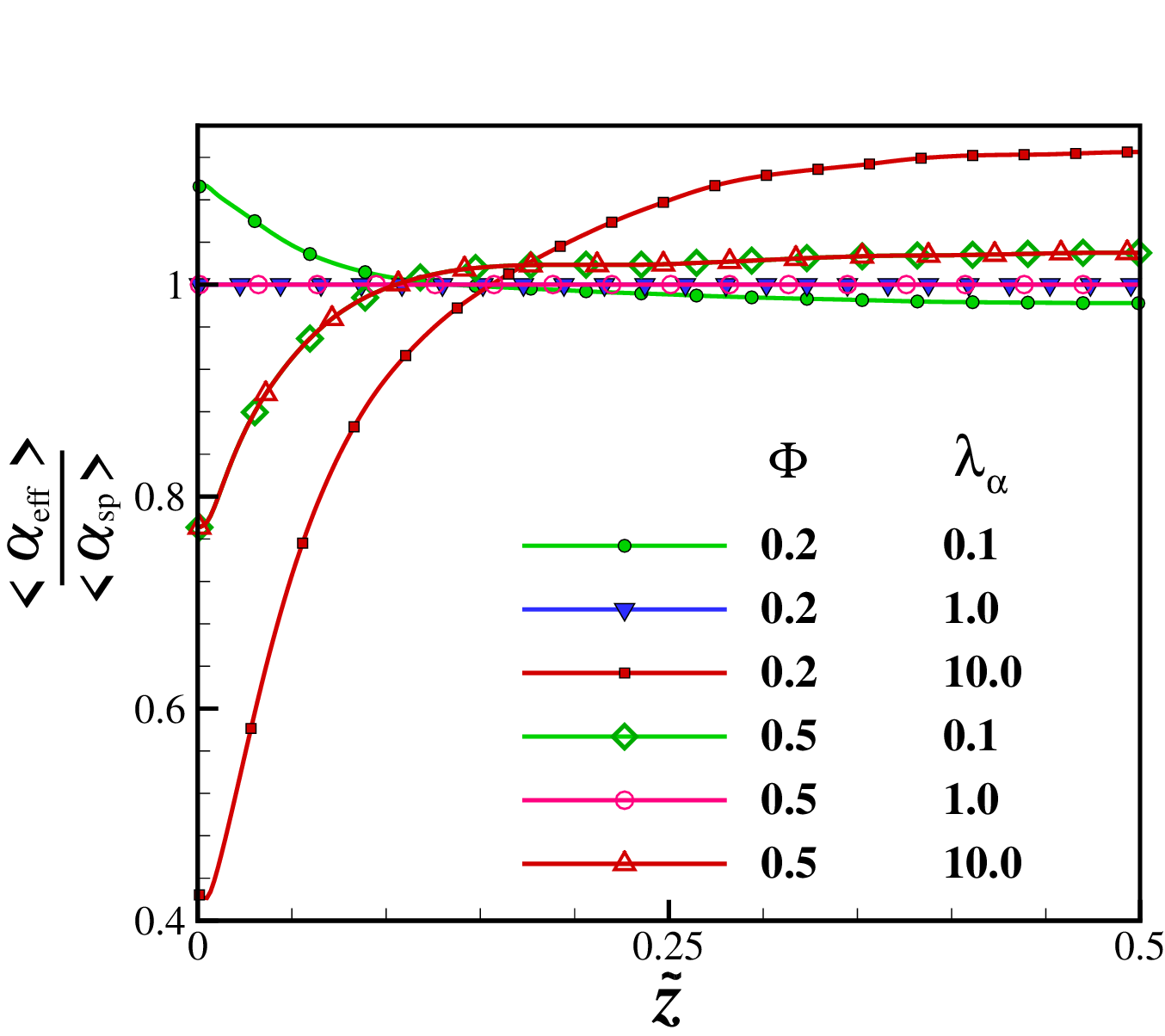}
    \phantomsubcaption\label{fig:alpha local}
  \end{subfigure}
  \caption{(a) The local wall-normal distributions of dispersed droplets, (b) the averaged number of dispersed droplets
at the steady state condition (together with their transient fluctuation ranges), for the cases with different thermal diffusivity ratios and various $\Upphi$. The blue and red dotted lines indicate $\Upphi = 0.2$ and $\Upphi = 0.5$, respectively, and (c) the local wall-normal distribution of thermal diffusivity.}
  \label{fig:phi_local_N_droplet_k}
\end{figure}

To understand the effect of the thermal diffusivity ratio, we first display the wall-normal distributions of the dispersed phase and the total number of droplets, see figure \ref{fig:phi_local_N_droplet_k}.
As shown in figure \ref{fig:phi local sym k}, we observe an increase in the local volume fraction $\phi$ close to the wall for $\lambda_{\alpha}=0.1$ and $\Upphi=0.2$, and a corresponding decrease in the core of the cavity. Conversely, when $\Upphi=0.2$ and $\lambda_{\alpha}=10$, the dispersed phase is, on average, very seldom in contact with the walls.
To explain this, note that, when $\lambda_{\alpha}=10$, 
the thermal diffusivity of the carrier fluid decreases by nearly 65\%, while that of the dispersed phase increases by a factor $3.5$ (this is to keep the nominal average thermal diffusivity the same as for the emulsion with $\lambda_{\alpha}=1$). Thus, the dispersed phase absorbs/releases heat faster and leaves sooner the near-wall region due to buoyancy. Given the reduced  
local concentration near the wall (figure \ref{fig:phi local sym k}), 
the local average diffusivity is closer to that of the carrier phase in the near wall region, so effectively less than for the case with $\lambda_{\alpha}=1$. This is confirmed by the data in panel c of the same figure where we report the local average thermal diffusivity versus the wall normal distance. 
 Here, we indeed note a significant decrease close to the wall for the case $\lambda_{\alpha}=10$, when the dispersed phase very seldom reaches the wall, and an increase of the effective diffusivity for the case $\lambda_{\alpha}=0.1$. This is however not as pronounced as the decrease observed for the case $\lambda_{\alpha}=10$, which explains why the Nusselt number is only slightly increased when reducing the thermal diffusivity of the dispersed phase ($\lambda_{\alpha}=0.1$). We speculate that this asymmetry is possibly due to the ratio between the timescale of thermal diffusion and transport: although the heat transfer is faster, 
the onset of motion is limited by the viscosity, which is constant among the different cases.

\begin{figure}
\centering
\begin{subfigure}[t]{0.03\textwidth}
\centering
\fontsize{6}{9}
\textbf{(a)}
\end{subfigure}
\begin{subfigure}[t]{0.45\textwidth}
\includegraphics[width=\linewidth, valign=t]{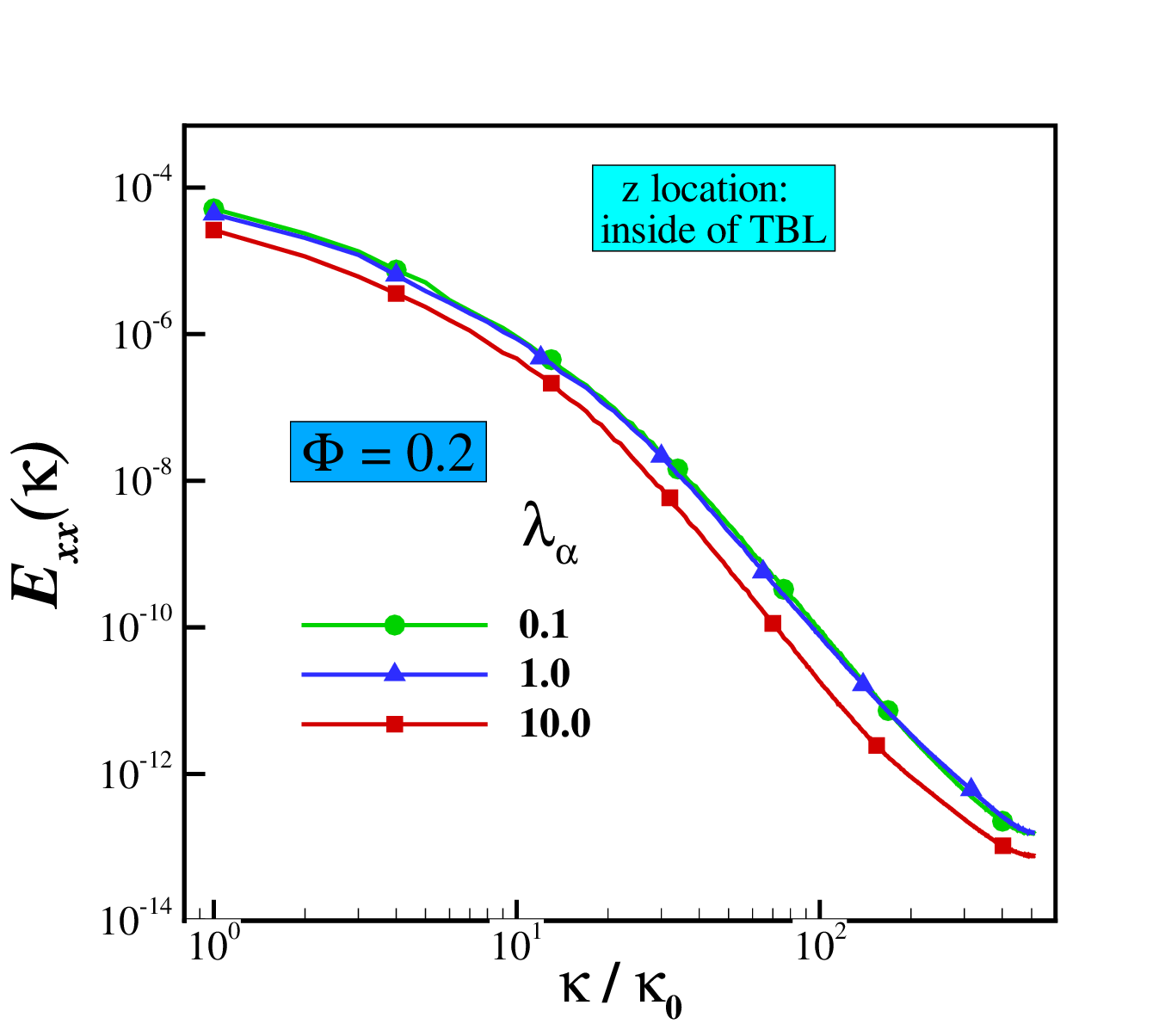}
\phantomsubcaption\label{fig:Exx_in_k_20}
\end{subfigure}\hfill
\begin{subfigure}[t]{0.03\textwidth}
\centering
\fontsize{6}{9}
\textbf{(b)}
\end{subfigure}
\begin{subfigure}[t]{0.45\textwidth}
\includegraphics[width=\linewidth, valign=t]{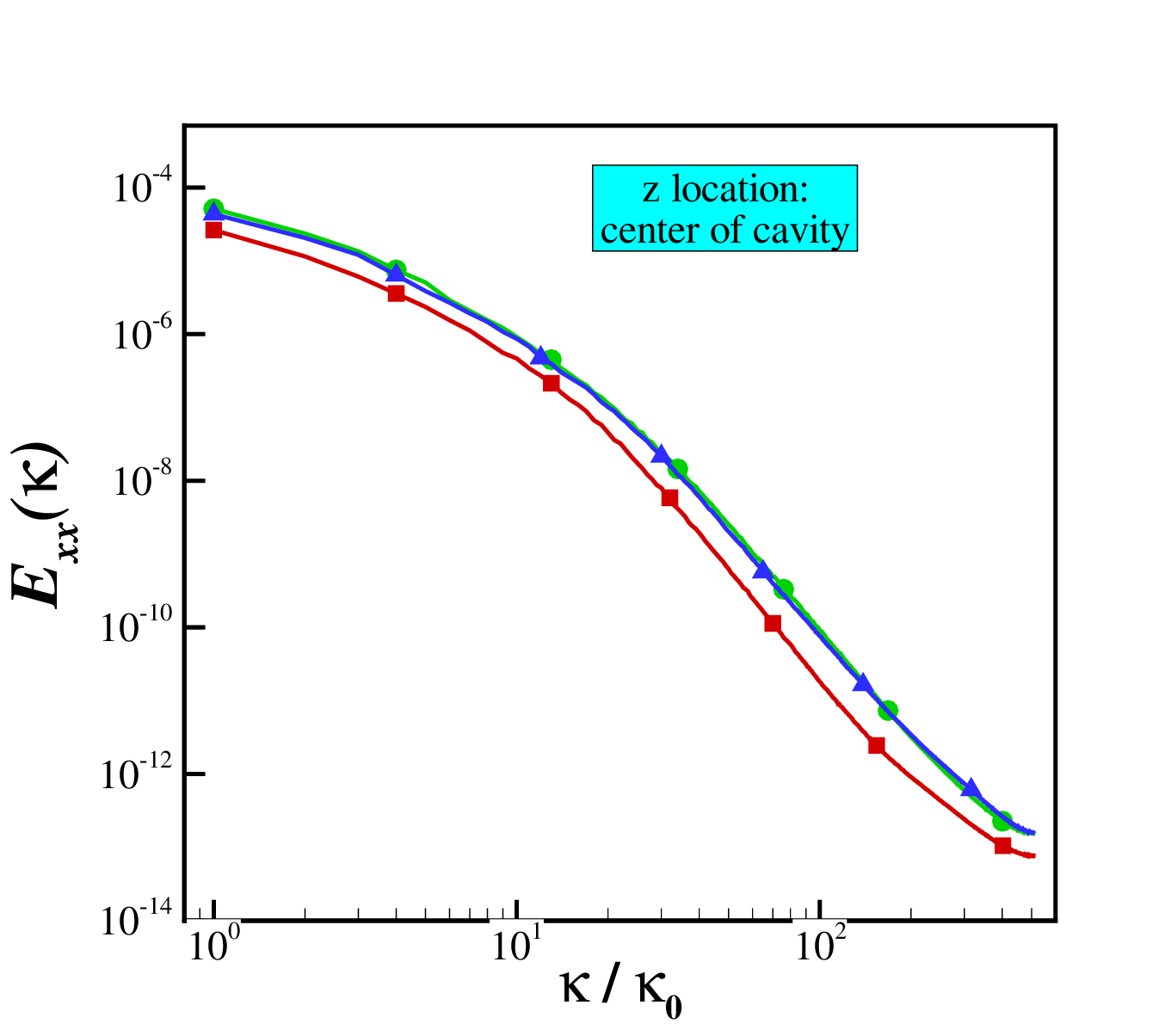}
\phantomsubcaption\label{fig:Exx_out_k_20}
\end{subfigure}
\caption{TKE spectra as a function of wavenumber at a droplet volume fraction of $\Upphi=0.2$ and different thermal diffusivity ratios (a) inside of TBL and (b) at the center of cavity. Wavenumbers are normalized by the lowest non-zero wavenumber $\kappa_0 = \pi / H$.
}
\label{fig:spectra_k_20}
\end{figure}

As mentioned above, 
owing to the increased diffusivity, the dispersed phase more quickly adjusts to the local temperature and plumes quickly form. This reduces the dispersed-phase residence time in the near-wall region. Since this is also the region with high shear, we observe a reduction in the number of droplets for the case $\lambda_{\alpha}=10$ (see panel b of the same figure).
The reduced presence of dispersed phase in the near-wall region causes reduced small-scale mixing due to the fewer smaller droplets forming inside the boundary layers, as shown by the energy spectra in figure \ref{fig:spectra_k_20}, where we observe less energy when $\lambda_{\alpha}=10$. Again, almost no difference is observed between emulsions with $\lambda_{\alpha}=1$ and 0.1. 

In the case of a binary mixture, see figure \ref{fig:phi local sym k}, again the more conductive phase is less likely to be found close to the wall and the average near-wall thermal diffusivity is effectively reduced, which explains the reduction in global heat transfer. As for the case with $\Upphi=0.2$, the reduction of the Nusselt number is caused by a reduction of the local average conductivity, see \ref{fig:alpha local}; for a binary mixture, however,  
 the  number of droplets is almost independent of $\lambda_{\alpha}$, see figure \ref{fig:N droplet k}.

\begin{figure}
\centering
\begin{subfigure}[t]{0.03\textwidth}
\centering
\fontsize{6}{9}
\textbf{(a)}
\end{subfigure}\hfill
\begin{subfigure}[t]{0.45\textwidth}
\centering
\includegraphics[width=\linewidth, valign=t]{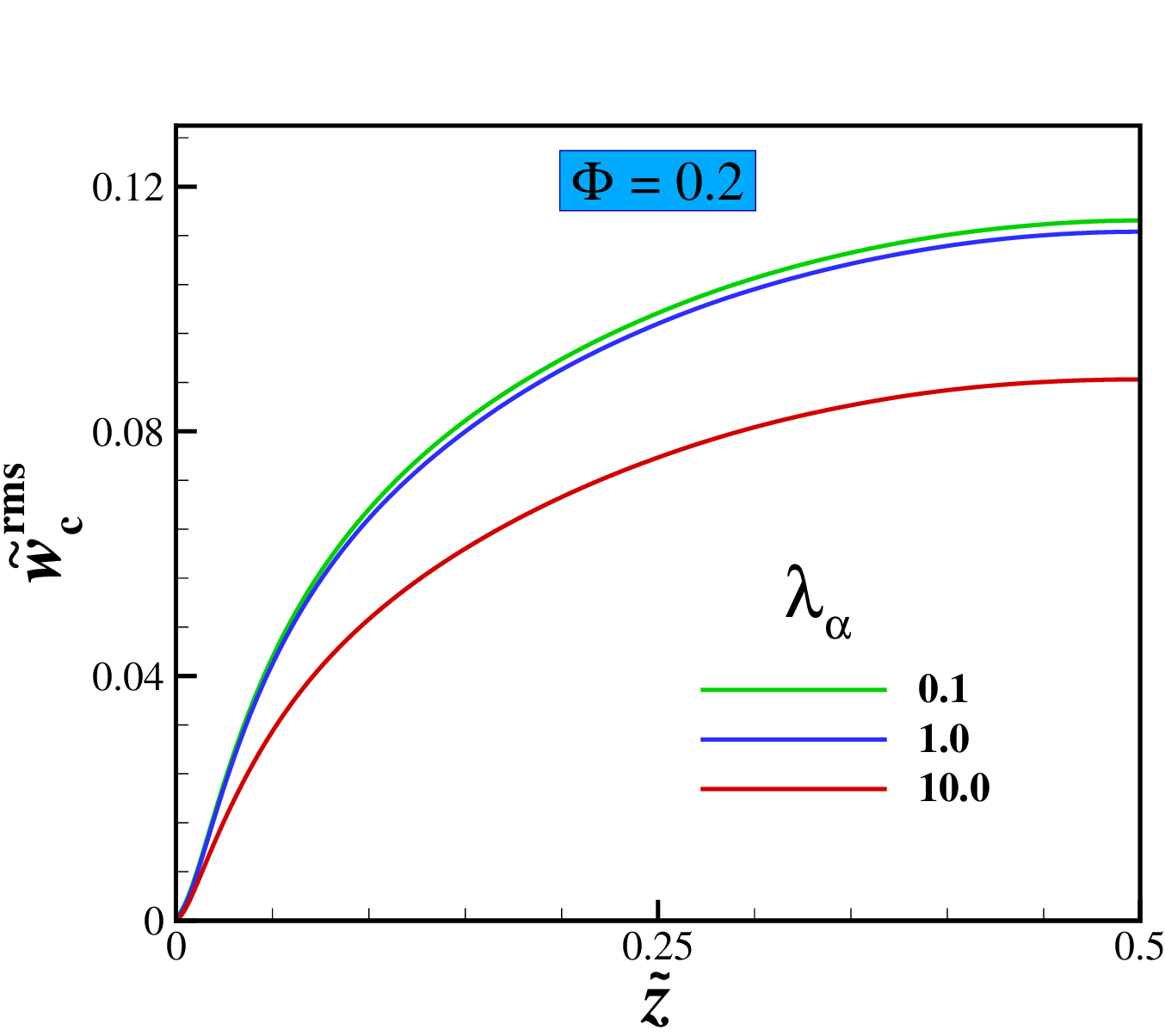}
\phantomsubcaption\label{fig:w rms 20 k carrier}
\end{subfigure}
\begin{subfigure}[t]{0.03\textwidth}
\centering
\fontsize{6}{9}
\textbf{(b)}
\end{subfigure}
\begin{subfigure}[t]{0.45\textwidth}
\includegraphics[width=\linewidth, valign=t]{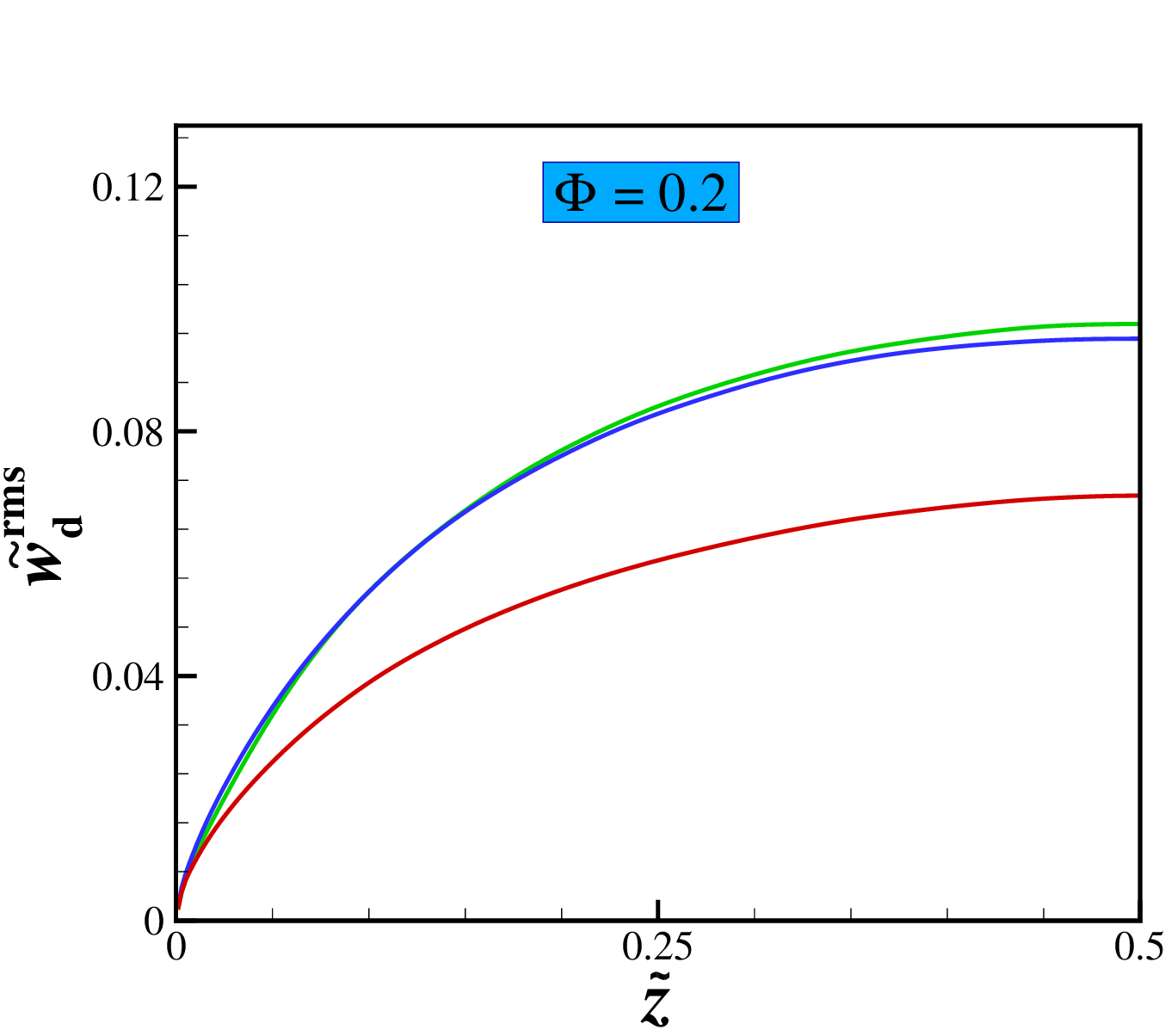}
\phantomsubcaption\label{fig:w rms 20 k dispersed}
\end{subfigure}
\begin{subfigure}[t]{0.03\textwidth}
\centering
\fontsize{6}{9}
\textbf{(c)}
\end{subfigure}
\begin{subfigure}[t]{0.45\textwidth}
\centering
\includegraphics[width=\linewidth, valign=t]{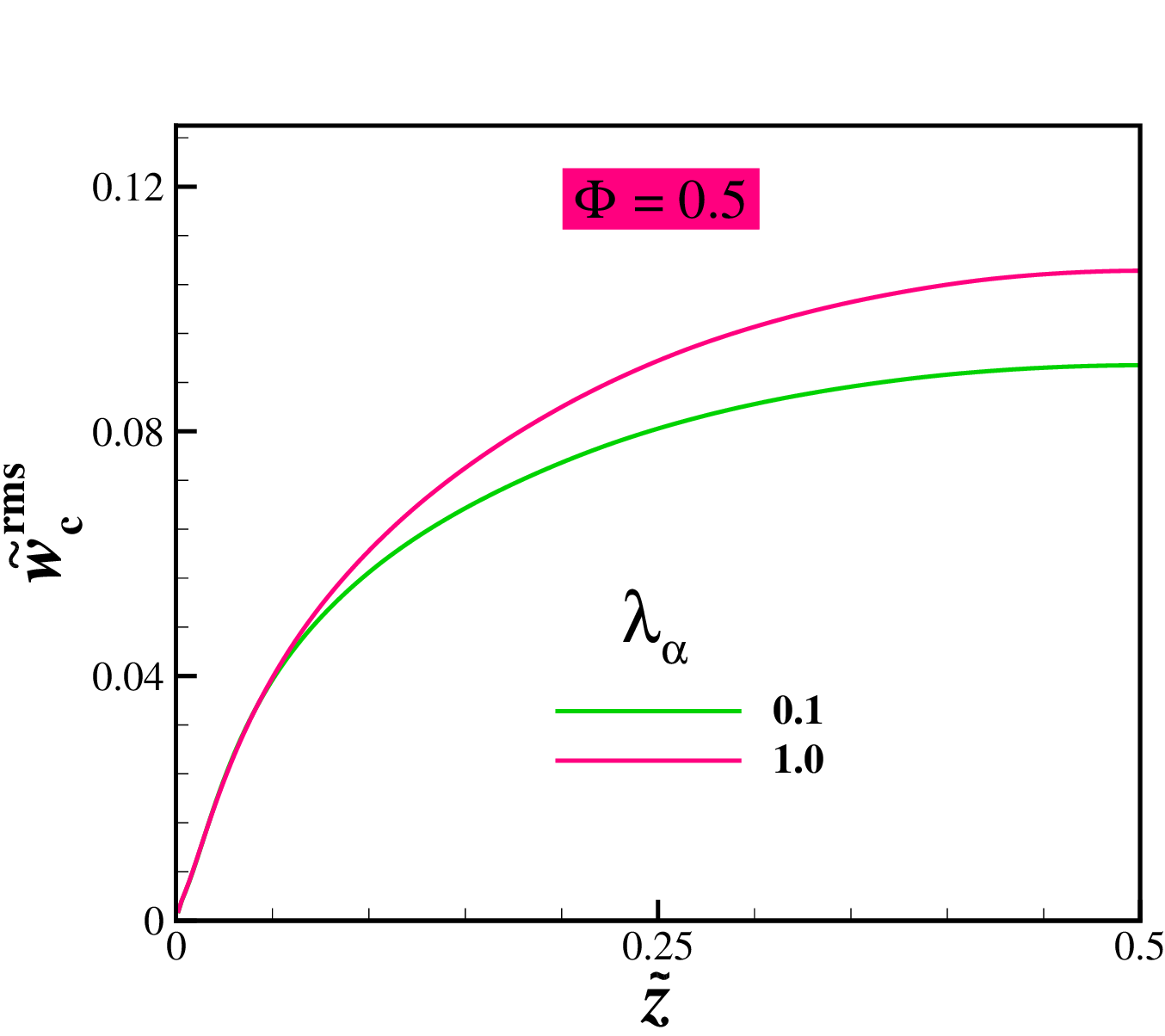}
\phantomsubcaption\label{fig:w rms 50 k carrier}
\end{subfigure}
\begin{subfigure}[t]{0.03\textwidth}
\centering
\fontsize{6}{9}
\textbf{(d)}
\end{subfigure}
\begin{subfigure}[t]{0.45\textwidth}
\includegraphics[width=\linewidth, valign=t]{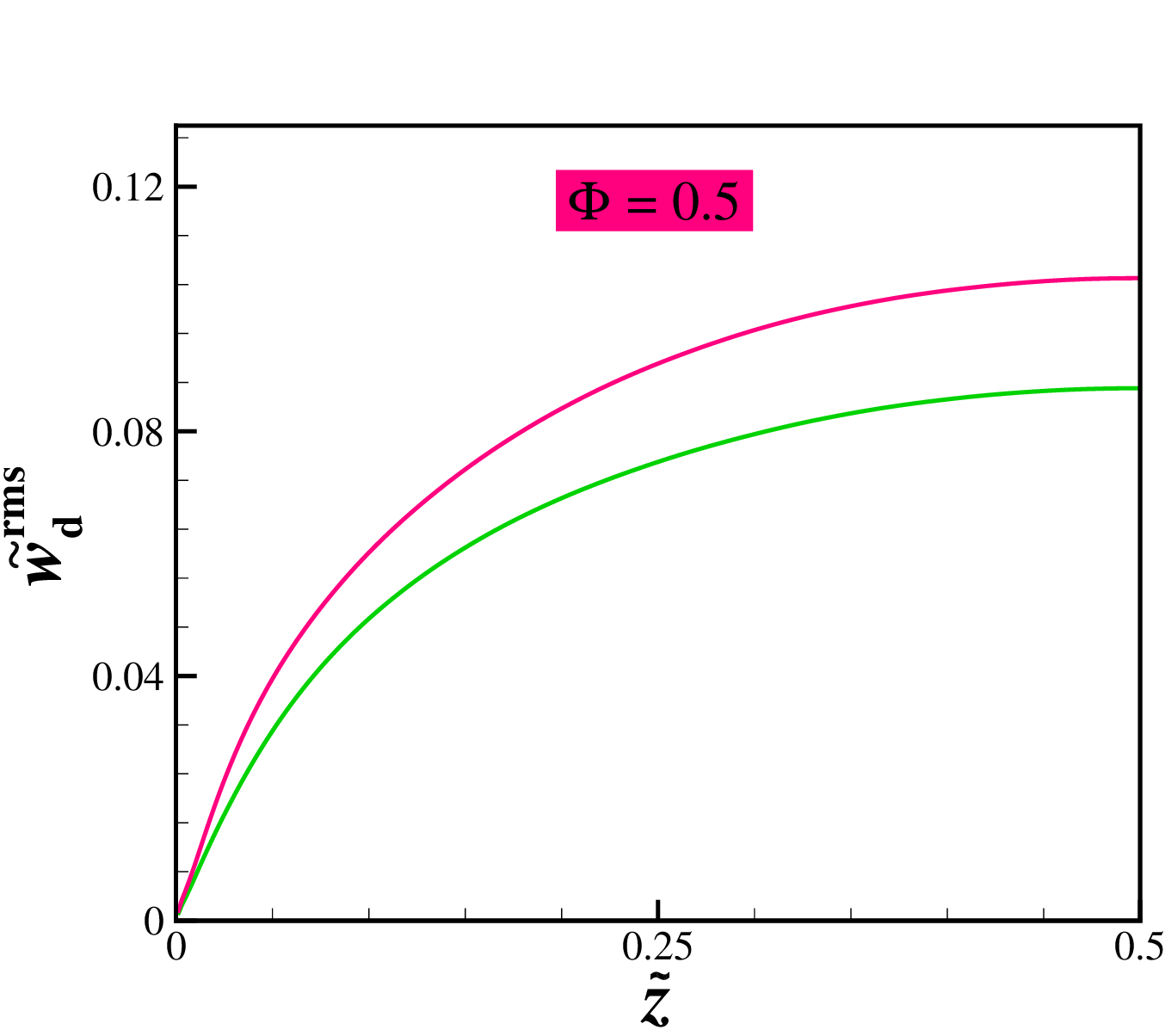}
\phantomsubcaption\label{fig:w rms 50 k dispersed}
\end{subfigure}
\caption{(a-d) Wall-normal profiles of the carrier- and dispersed-phase rms wall-normal velocity for the different thermal diffusivity ratios and droplet volume fractions investigated in this study.}
\label{fig:w_rms_k_c_d}
\end{figure}

The intensity of the wall-normal velocity fluctuations is displayed in figure \ref{fig:w_rms_k_c_d}
for both phases. The figure shows that the fluctuations reduce for $\lambda_{\alpha}=10$ when the dispersed phase more quickly adjusts to the surrounding temperature and the plumes more easily lose their drive. We note also that the cases with $\lambda_{\alpha}=1$ and 0.1 display similar values in Nusselt number and wall-normal velocity fluctuations, despite the dispersed phase is not more likely to be found close to the wall. The weak decrease in near-wall thermal diffusivity appears to be compensated by the weak increase in the number of droplets.

\begin{figure}
  \centering
  \begin{subfigure}[t]{0.03\textwidth}
  \fontsize{6}{9}
    \textbf{(a)}
  \end{subfigure}
  \begin{subfigure}[t]{0.7\textwidth}
    \includegraphics[width=\linewidth, valign=t]{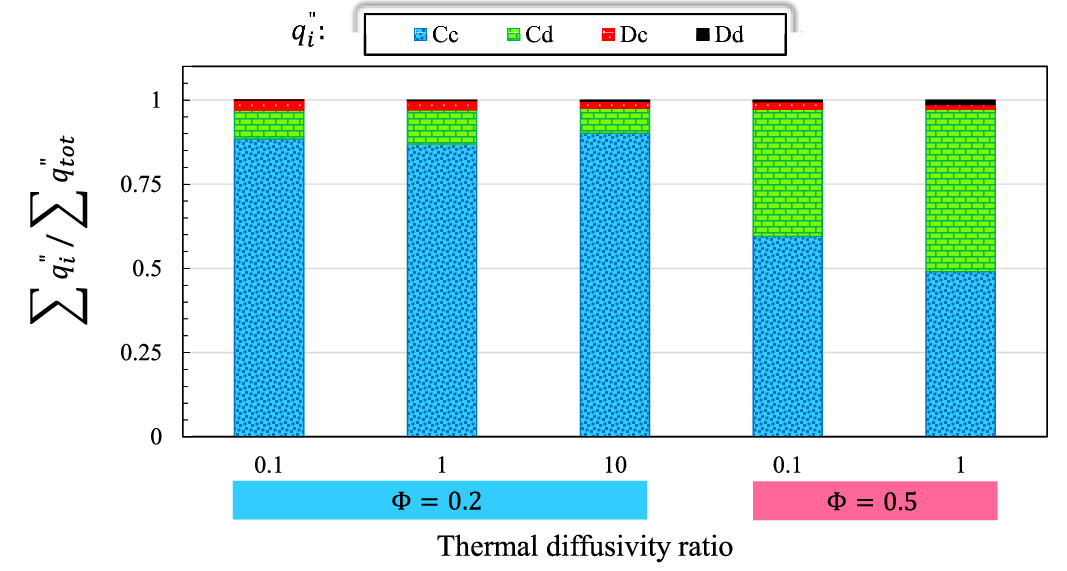}
    \phantomsubcaption\label{fig:heat budget barchart k}
  \end{subfigure}
  \begin{subfigure}[t]{0.03\textwidth}
  \fontsize{6}{9}
    \textbf{(b)} 
  \end{subfigure}
  \begin{subfigure}[t]{0.45\textwidth}
    \includegraphics[width=\linewidth, valign=t]{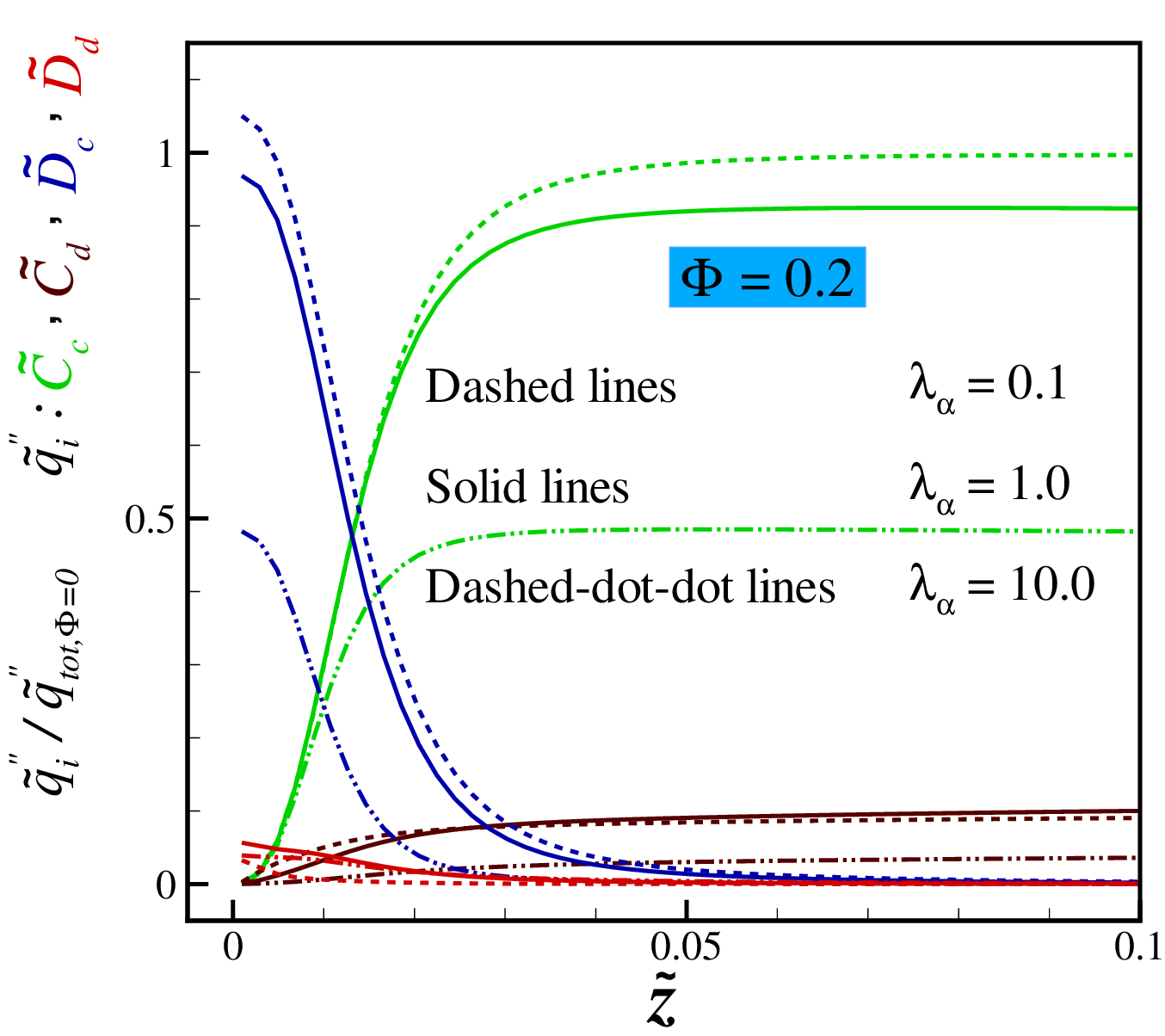}
    \phantomsubcaption\label{fig:heat budget 20 k}
  \end{subfigure}\hfill
  \begin{subfigure}[t]{0.03\textwidth}
  \fontsize{6}{9}
    \textbf{(c)}
  \end{subfigure}
  \begin{subfigure}[t]{0.45\textwidth}
    \includegraphics[width=\linewidth, valign=t]{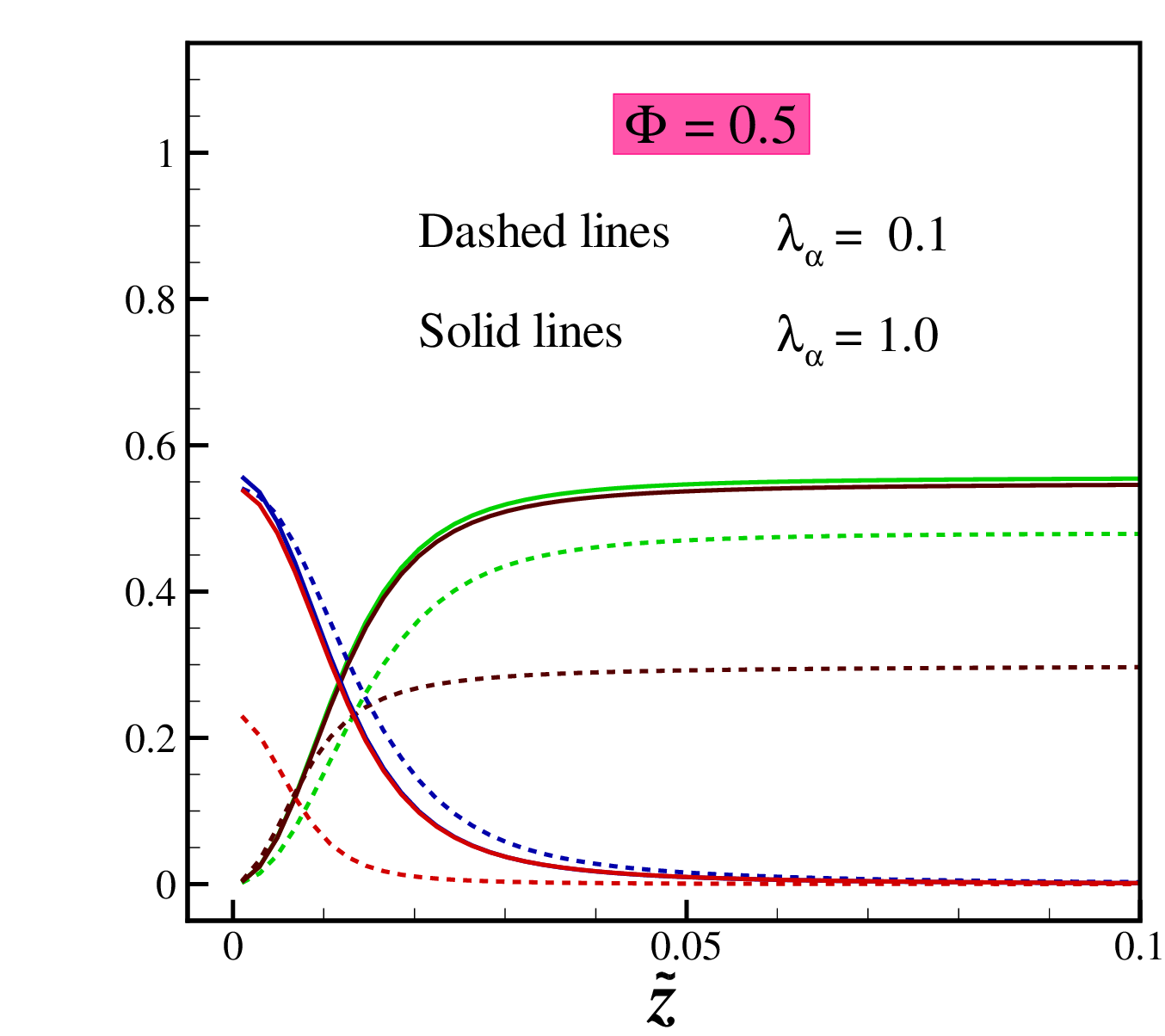}
    \phantomsubcaption\label{fig:heat budget 50 k}
  \end{subfigure}
  \caption{(a) Wall-normal integral of heat fluxes transferred by diffusion and convection, normalized by the total heat flux in $\Upphi = 0.0$; (b,c) wall-normal distributions of convection and diffusion terms of both phases, normalized by the volume averaged total heat flux in $\Upphi=0.0$ and for cases with $\Upphi=0.2$ and $0.5$. }
  \label{fig:heat_budget_k}
\end{figure}
These changes in the droplet distribution patterns and the corresponding values of the local diffusivity lead to the variations in the dispersed-phase convection flux shown in figure \ref{fig:heat_budget_k}.
Note that, since $\alpha_d \neq \alpha_c \neq \alpha$ when $\lambda_{\alpha} \ne 1$, the diffusion terms in the budget equation (\ref{eq:heat_budget_terms}) take the following dimensionless form:
\begin{subequations}
\begin{gather}
 \tilde{D_c} = (1-<\phi>) \frac{\alpha_c}{\alpha} <\frac{d\tilde{\theta}_c}{d\tilde{z}}>,
  \label{eq:Dc_heat_budget_dimless_c_k_ratio}\\
\tilde{D_d} = <\phi>\frac{\alpha_d}{\alpha}<\frac{d\tilde{\theta}_d}{d\tilde{z}}>.
  \label{eq:Dd_heat_budget_dimless_d_k_ratio}
\end{gather}
\end{subequations}
Figure \ref{fig:heat budget barchart k} depicts the wall-normal integral of the convection and diffusion heat fluxes, normalized by the total flux in the single-phase flow. 
Figures \ref{fig:heat budget 20 k} and \ref{fig:heat budget 50 k}, instead, illustrate the distribution of these fluxes along the wall-normal direction. As in figure \ref{fig:heat_budget_visc}, we focus on the region close to the wall ($\Tilde{z}=[0-0.1]$) as it exhibits the most significant variations.
The budget terms in the figure confirm the reduction in near-wall diffusion when $\lambda_{\alpha}=10$, because of the reduced concentration of dispersed-phase in the near-wall region and the local reduction of the average thermal diffusivity. When normalizing the different contributions with the total heat flux, we observe a reduction of the contribution of the dispersed phase despite being more conductive because of the depletion of the near-wall region. 
\section{Conclusions and outlook} \label{sec:conclusion}
We have presented the results of interface-resolved simulations of liquid-liquid emulsions within a turbulent Rayleigh–Bénard convection flow. Our focus lies on examining the effects of three primary properties on heat transfer rates and flow modulations in two-phase systems: (a) dispersed droplet volume fraction within the range of $0 \leq \Upphi \leq 0.5$, (b) dynamic viscosity ratio in the range $0.1 \leq \mu_d / \mu_c \leq 10$, and (c) thermal diffusivity ratio in the range $0.1 \leq \alpha_d / \alpha_c \leq 10$. The remaining dimensionless numbers are kept fixed and equal to $Ra=10^8$, $Pr=1$, $We=6000$, and $Fr=1$. In the following, the key findings are summarized. \par
For fluids characterized by the same thermophysical properties, adding dispersed deformable droplets to the single-phase flow enhances the heat transfer rate in the cavity with respect to the single-phase counterpart. Differently from the case of solid particles, the enhancement is monotonic with the dispersed phase volume fraction, $\Upphi$, with a maximum enhancement rate of $10.2$ $\%$ for the case of a binary mixture. \par
To explain this finding, we recall that the presence of an interface is known to provide an alternative mechanism for energy transfer to small scales, extending the range of active flow structures towards even smaller scales \citep[see][among others]{Perlekar2019,crialesi2022modulation}. Hence, despite the fact that we also report a reduction of the average turbulent kinetic energy, the increase of energy at the smallest scales is found to be responsible for increased mixing and, therefore, at the origin of the increase in the Nusselt number reported here. Note also that the large-scale mixing is almost unaltered, as determined by the cell size and global temperature difference. \par
Our results also show that the dispersed phase is less likely to be found closer to the walls, whereas the carrier phase is most likely to remain within the thermal boundary layers, even if the two fluids have the same thermophysical properties. Hence, the fastest-rising plumes contain more of the carrier than of the dispersed phase.
As a consequence, analysis of the turbulent-kinetic energy and heat-transfer budgets reveals that the turbulence production and heat fluxes are mainly associated with velocity-temperature fluctuations in the carrier phase. \par
As mentioned above, the carrier phase is most likely to stay within the thermal boundary layers, its temperature approaches that of the nearby walls, and its density is therefore more likely to reach low/high values. In other words, the fastest rising plumes are expected to contain more of the carrier than of the dispersed phase, as suggested by the fact that the temperature-velocity fluctuations are larger in the carrier phase. \par
We also consider fluids with different properties, particularly dynamic viscosity and thermal diffusivity, the latter changed by altering the thermal conductivity. The change in thermophysical properties is achieved through two different approaches. The first approach keeps the effective properties constant while changing the volume fraction of the emulsion and the property ratio. This choice allows maintaining a constant Rayleigh number based on the average mixture viscosity. The second approach keeps the properties of the carrier fluid constant, changing only the dispersed-fluid properties. Throughout this work, the first approach is mainly used, but two additional cases are performed using the second approach for a change in dynamic viscosity only and discussed in the Appendix~\ref{sec:appendix_A}. In the first case, unlike the second, we observe an increase of the Nusselt number of about 25\% at $\Upphi=0.2$ when the dispersed phase is more viscous ($\lambda_\mu=10$). This increase is attributed to two concurrent effects. The increase of small-scale mixing is due to the presence of an interface (as for the case $\lambda_\mu=1$) and to an increased level of turbulence in the less viscous carrier phase. In the case $\lambda_\mu=10$, we also observe an almost uniform distribution of the two phases across the cavity and an increased number of droplets despite these being more viscous. Small differences are observed between the emulsions with $\lambda_\mu=1$ and $0.1$, when the near wall region is characterized by a reduction in the concentration of the dispersed phase. In the traditional (second) approach, on the other hand, we observe a significant damping in the turbulence level, leading to $14$ \% reduction in the value of Nusselt number when increasing the viscosity ratio from $0.1$ to $10$. \par
 To investigate the effects of thermal diffusivity, only the first approach is employed throughout the paper. Based on the results, for higher thermal diffusivity of the dispersed phase, we report a lower residence time near the walls, with the dispersed phase almost never reaching the walls. Due to the faster heat diffusion, buoyancy is soon active, and the near-wall region is filled with the fluid with lower diffusivity. This leads to a significant reduction of the Nusselt number, about half that of the corresponding single-phase configuration, which is therefore attributed to a reduction of the local average diffusivity in combination with a reduced number of droplets in the depleted near-wall region, i.e.\ reduced small-scale mixing. 
%
%

Finally, we have examined the droplet-size distribution for the different cases. The results confirm the two scaling laws: $d^{-3/2}$ from \cite{deane2002scale}, and $d^{-10/3}$ from \cite{garrett2000connection} for small and large droplets (Hinze criteria), characterized by the dominance of coalescence or breakup, for cases up to $\Upphi=0.3$. At higher $\Upphi$, a slight deviation from the $-10/3$ scale is observed due to the enhanced coalescence and the large deviation from the spherical shape of the dispersed phase. A better agreement with the scaling law could be observed by accounting for the droplet morphology in the p.d.f. calculation. \par
The configurations examined in this study may serve as a framework for future studies. In addition to direct comparisons with experiments, given the role of the near-wall distribution of the dispersed phase documented here, a potential follow-up investigation could consider thermally patterned walls with distinct structures and shapes on both the top and bottom plates. Varying the wetting properties of these walls could offer valuable insights into the interplay between the wetting properties of the walls and the modulation of heat transfer in the context of multiphase Rayleigh-Bénard convection and provide a strategy to control the system behavior, as shown in \cite{liu2022enhancing}. The role of the droplet-size distribution on the heat transfer should also deserve further attention, as shown by the importance of surfactant for the effective viscosity of emulsions \citep{Yi_Wang_vanVuren_Lohse_Risso_Toschi_Sun_2022}. Moreover, in the present study, we employed a linear relation for calculating the effective reference properties of the emulsion in order to define the non-dimensional numbers. Future work should focus on deriving a non-linear relation for the effective viscosity from the emulsion's rheological curves (under Turbulent Rayleigh-Bénard convective flow) and using this non-linear relation to better estimate the effective viscosity.

\section*{Acknowledgments} \label{sec:acknowledgments}
PM and AMB were supported partially by the National Science Foundation (Award No. 1854376). 
This research used resources from the Argonne Leadership Computing Facility (ALCF), which is a DOE Office of Science User Facility supported under Contract DE-AC02-06CH11357.
\section*{Declaration of interests} \label{sec:declaration_of_interests}
The authors report no conflict of interest.
\appendix
%
\section{Evaluation of the selected reference thermophysical properties.}
\label{sec:appendix_A}
\setcounter{appendixfigure}{1} 
%
In this section, we present and analyze the DNS results from two additional simulations. These simulations were conducted using the properties of the carrier phase as the reference for $\psi_r$, i.e. $\psi_r=\psi_2$. Accordingly, the dimensionless groups $Ra$, $We$, and $Pr$ are defined based on $\rho_c$, $\mu_c$, $k_c$ and $c_{pc}$, with changes in the viscosity or thermal diffusivity ratios achieved by tuning the corresponding properties in the dispersed phase only. Note that hereinafter, we refer to this approach as the traditional one. The two additional cases correspond to cases $12$ and $13$ in \hyperref[table:list_of_simulations]{Table \ref{table:list_of_simulations}} ($\Upphi=0.2$, and $\lambda_{\mu} = 0.1, 10$) and are compared with cases $6$ and $7$ in \hyperref[table:list_of_simulations]{Table \ref{table:list_of_simulations}}. \par
\begin{figure}
\renewcommand{\thefigure}{\theappendixfigure}
\centering
\begin{subfigure}[t]{0.03\textwidth}
\centering
\fontsize{6}{9}
\textbf{(a)}
\end{subfigure}
\begin{subfigure}[t]{0.45\textwidth}
\includegraphics[width=\linewidth, valign=t]{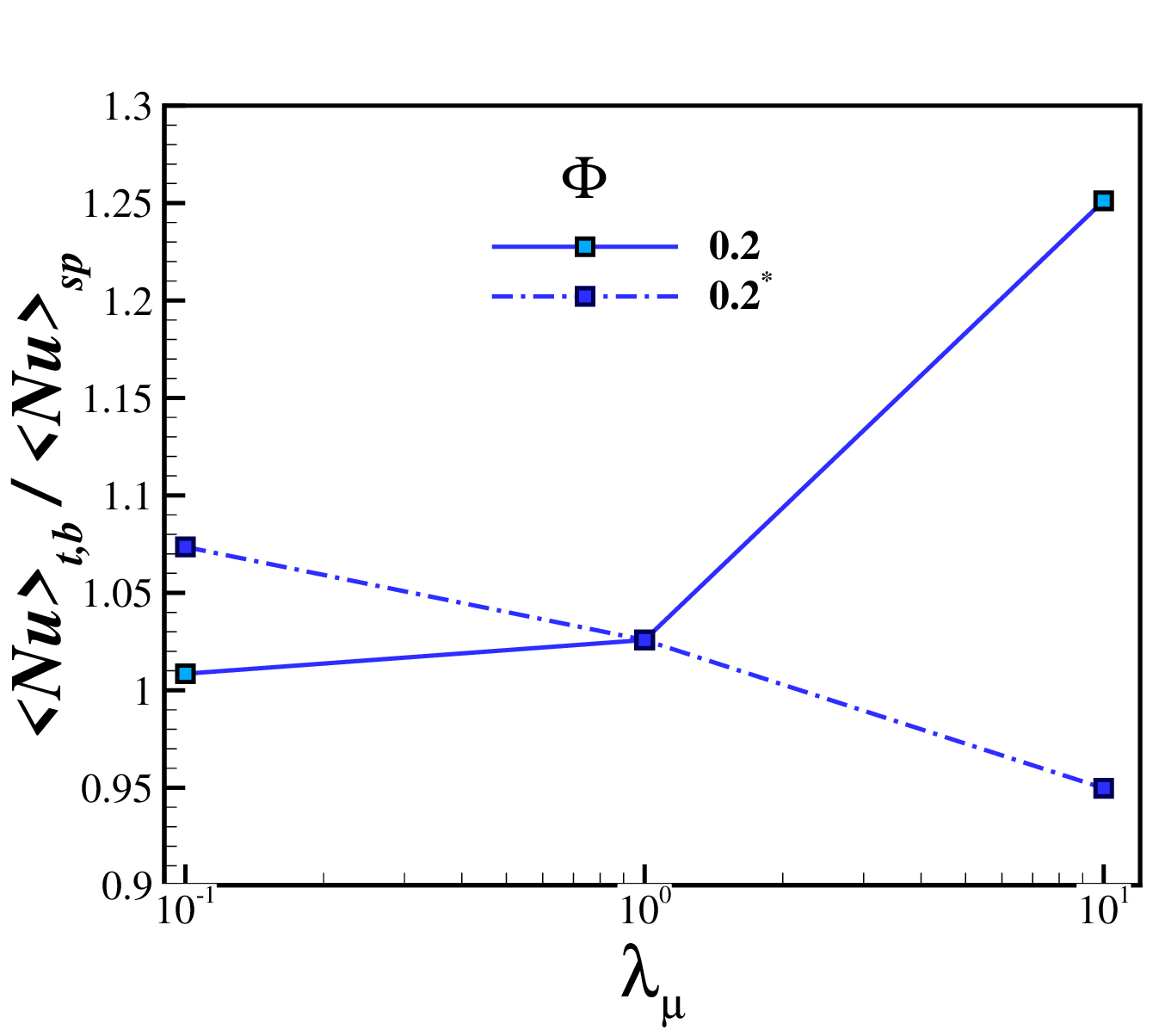}
 \phantomsubcaption\label{fig:nu visc new20}
\end{subfigure}\hfill
\begin{subfigure}[t]{0.03\textwidth}
\centering
\fontsize{6}{9}
\textbf{(b)}
\end{subfigure}
\begin{subfigure}[t]{0.45\textwidth}
\includegraphics[width=\linewidth, valign=t]{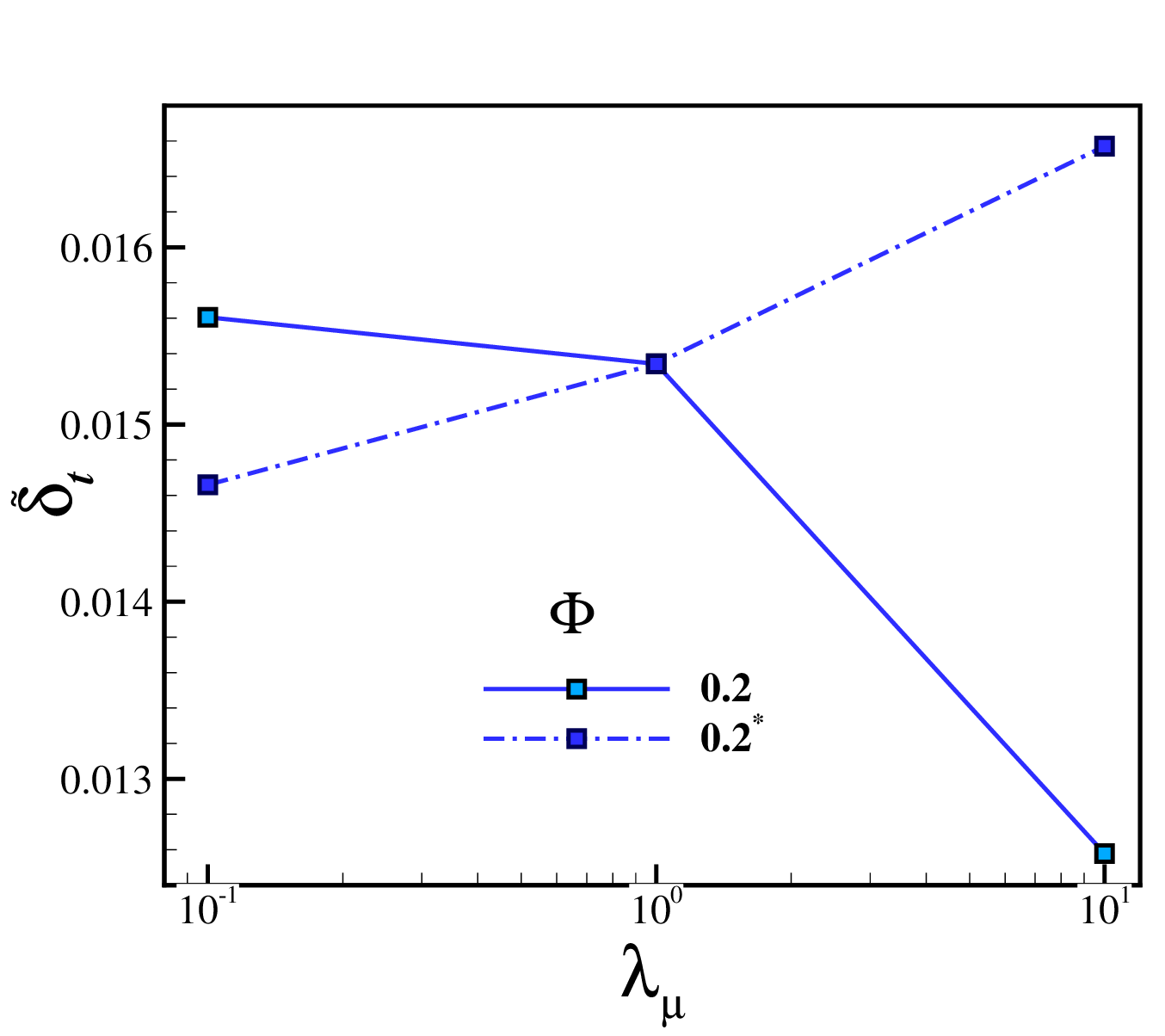}
\phantomsubcaption\label{fig:TBL thickness visc new20}
\end{subfigure}
\begin{subfigure}[t]{0.03\textwidth}
\centering
\fontsize{6}{9}
\textbf{(c)}
\end{subfigure}
\begin{subfigure}[t]{0.7\textwidth}
\centering
\includegraphics[width=\linewidth, valign=t]{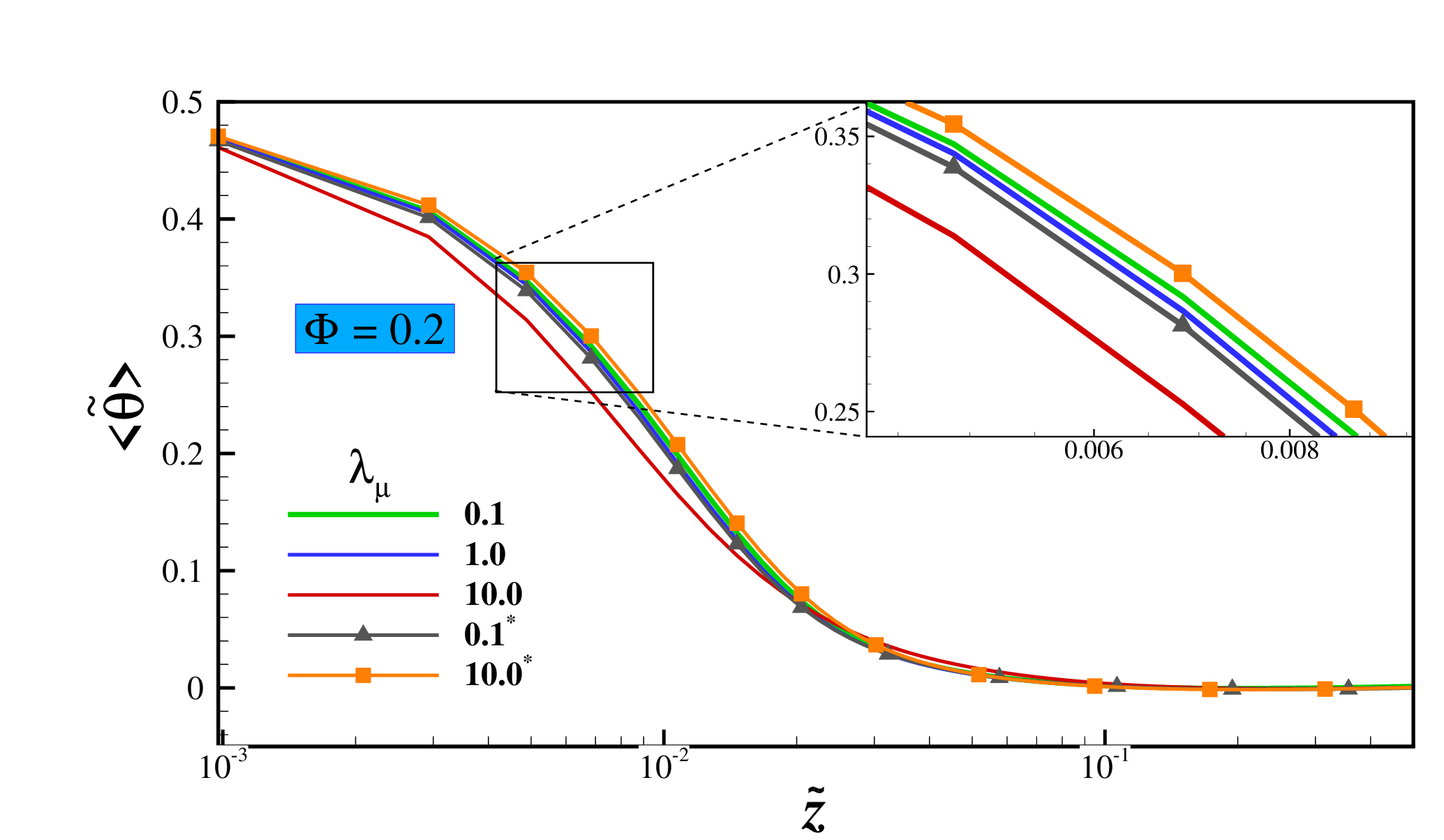}
\phantomsubcaption\label{fig:tmp mean visc new20}
\end{subfigure}
\caption{(a) Nusselt number, (b) thermal boundary layer thickness and (c) mean temperature profiles along the wall-normal direction for the different viscosity ratios and at a moderate droplet volume fraction of $\Upphi=0.2$. Labels shown in legends with $^*$ are the cases obtained based on the traditional way.}
\label{fig:nu_TBL_tmp_mean_visc_new20}
\end{figure}
\begin{figure}
\stepcounter{appendixfigure}
\renewcommand{\thefigure}{\theappendixfigure}
\centering
\begin{subfigure}[t]{0.03\textwidth}
\centering
\fontsize{6}{9}
\textbf{(a)}
\end{subfigure}
\begin{subfigure}[t]{0.45\textwidth}
\includegraphics[width=\linewidth, valign=t]{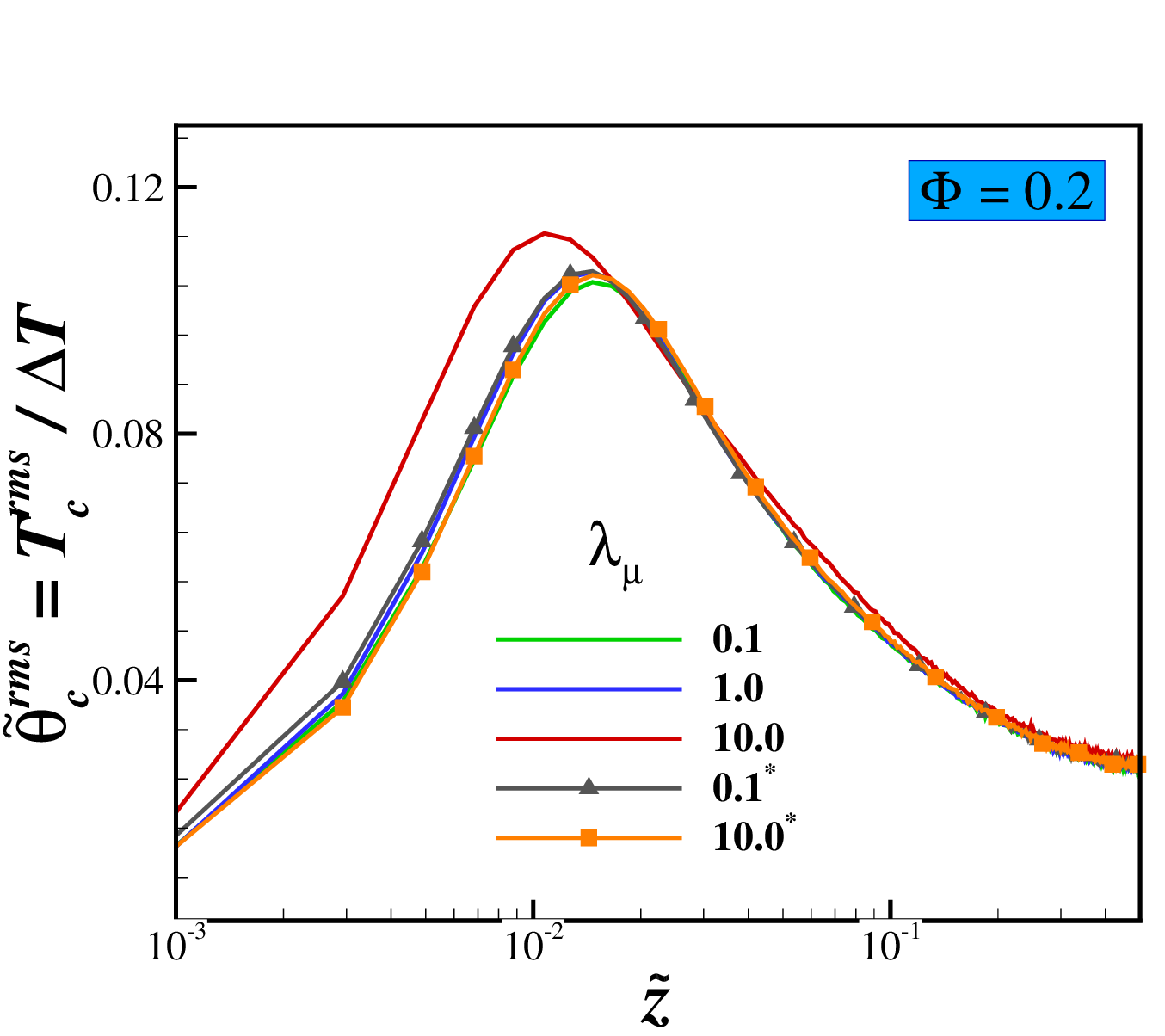}
 \phantomsubcaption\label{fig:tmp rms 20 visc carrier new20}
\end{subfigure}\hfill
\begin{subfigure}[t]{0.03\textwidth}
\centering
\fontsize{6}{9}
\textbf{(b)}
\end{subfigure}
\begin{subfigure}[t]{0.45\textwidth}
\includegraphics[width=\linewidth, valign=t]{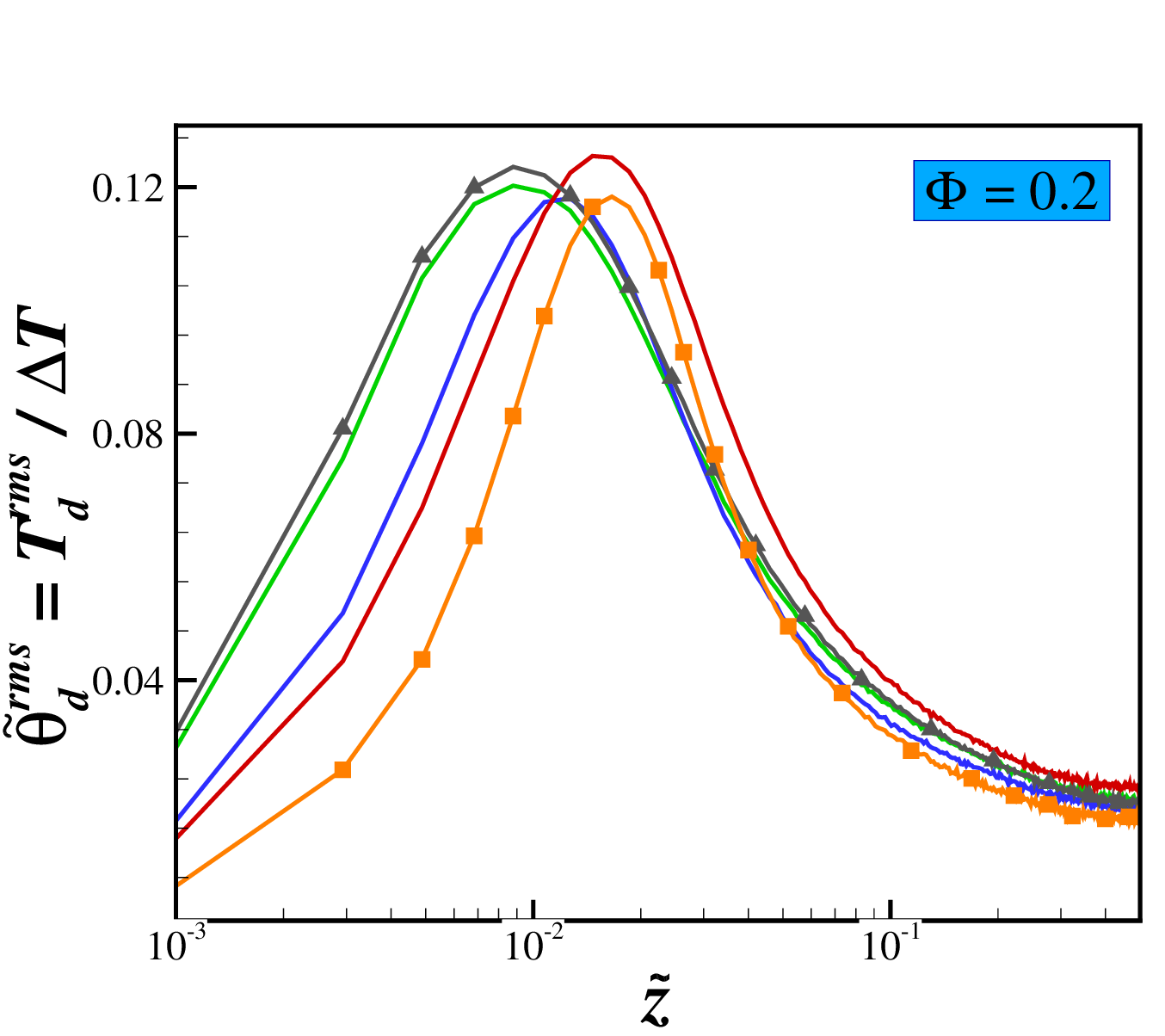}
\phantomsubcaption\label{fig:tmp rms 20 visc dispersed new20}
\end{subfigure}
\caption{(a) Carrier- and (b) dispersed-phase rms temperature profiles along the wall-normal direction at various viscosity ratios and at a droplet volume fraction of $\Upphi=0.2$.}
\label{fig:tmp_rms_visc_c_d_new20}
\end{figure}
\begin{figure}
\stepcounter{appendixfigure}
\renewcommand{\thefigure}{\theappendixfigure}
\centering
\begin{subfigure}[t]{0.03\textwidth}
\centering
\fontsize{6}{9}
\textbf{(a)}
\end{subfigure}
\begin{subfigure}[t]{0.45\textwidth}
\includegraphics[width=\linewidth, valign=t]{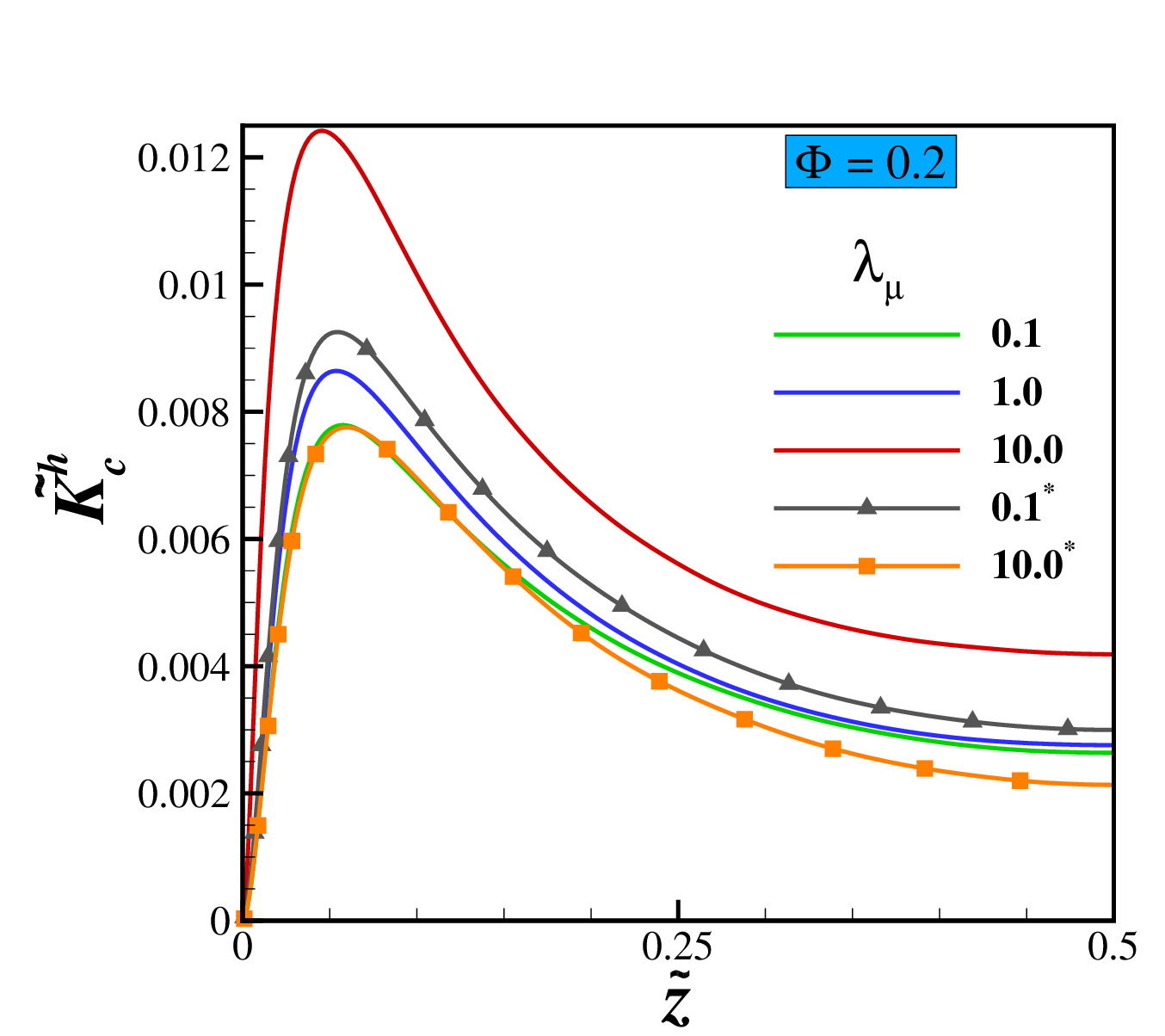}
\phantomsubcaption\label{fig:K h c sym visc new20}
\end{subfigure}\hfill
\begin{subfigure}[t]{0.03\textwidth}
\centering
\fontsize{6}{9}
\textbf{(b)}
\end{subfigure}
\begin{subfigure}[t]{0.45\textwidth}
\includegraphics[width=\linewidth, valign=t]{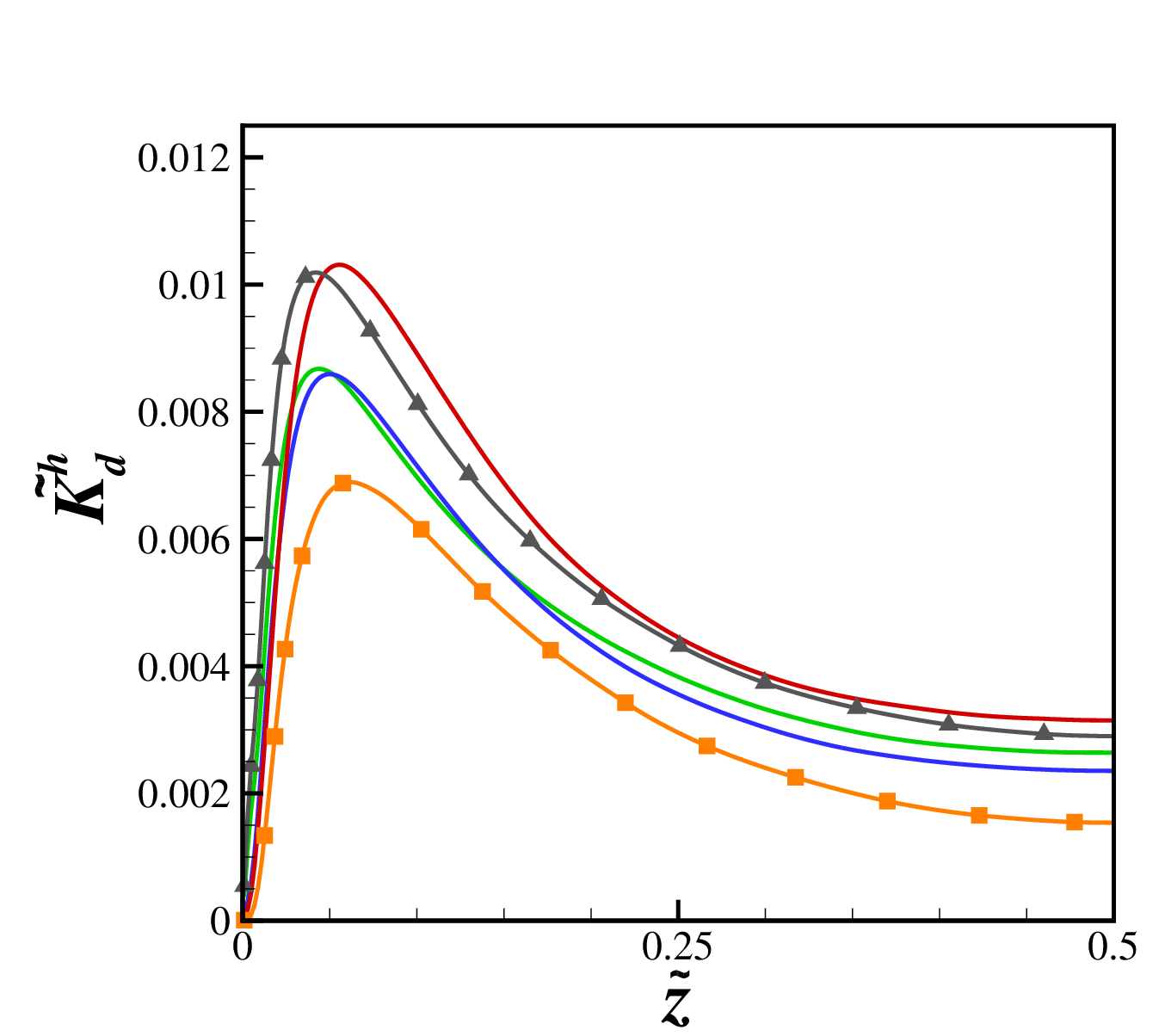}
\phantomsubcaption\label{fig:K h d sym visc new20}
\end{subfigure}
\begin{subfigure}[t]{0.025\textwidth}
\centering
\fontsize{6}{9}
\textbf{(c)}
\end{subfigure}
\begin{subfigure}[t]{0.45\textwidth}
\centering
\includegraphics[width=\linewidth, valign=t]{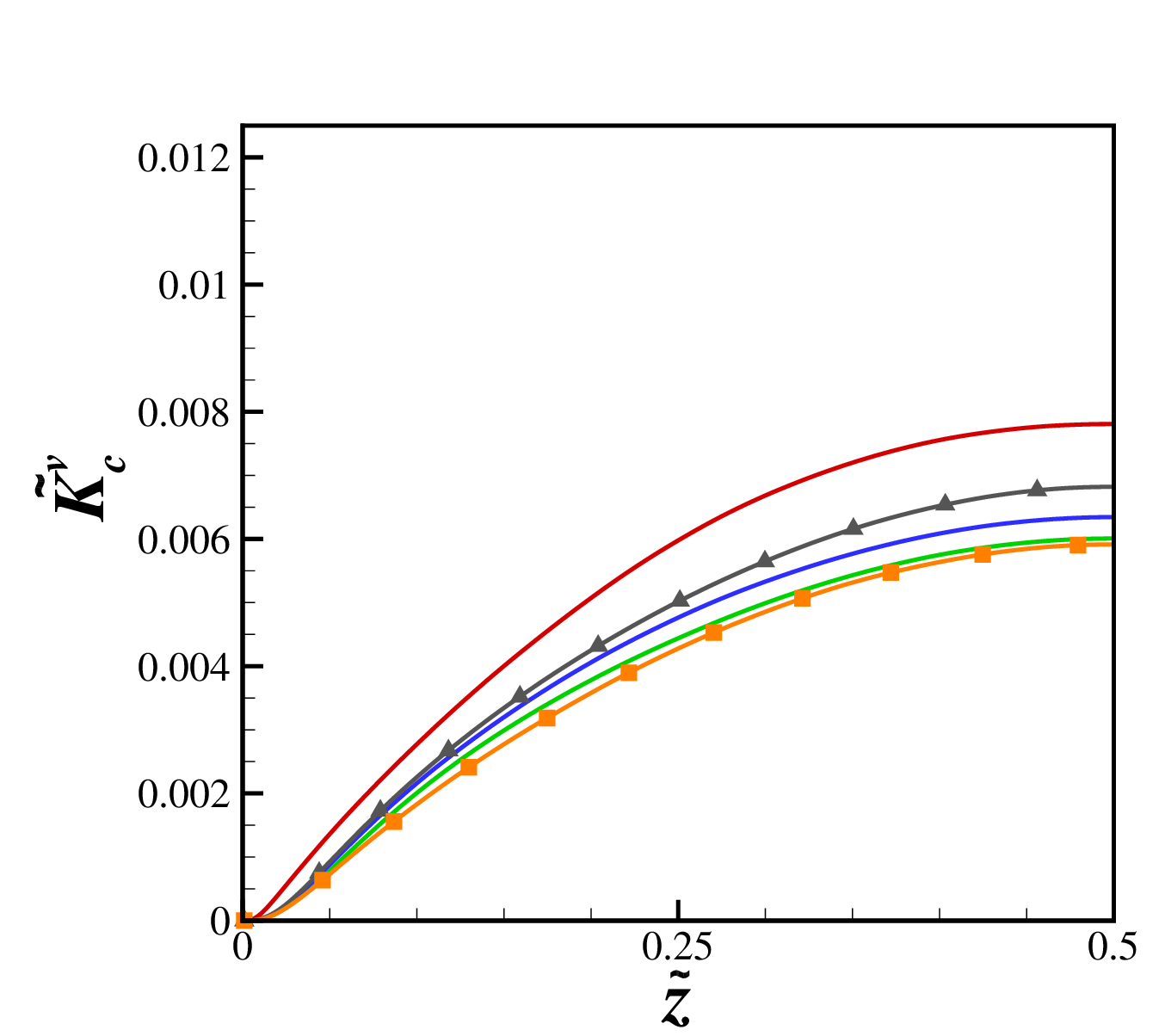}
\phantomsubcaption\label{fig:K v c sym visc new20}
\end{subfigure}
\begin{subfigure}[t]{0.048\textwidth}
\centering
\fontsize{6}{9}
\textbf{(d)}
\end{subfigure}
\begin{subfigure}[t]{0.45\textwidth}
\includegraphics[width=\linewidth, valign=t]{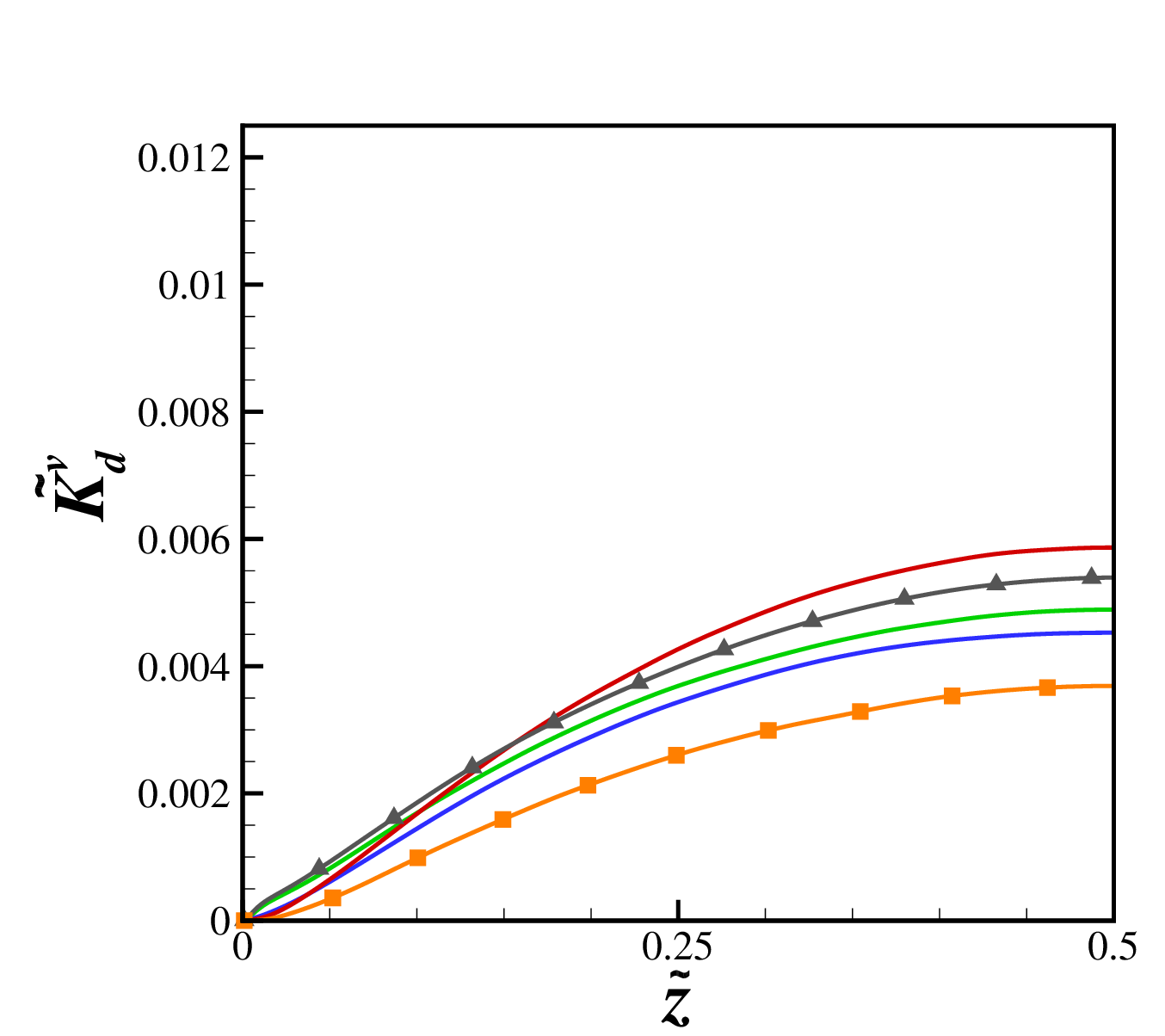}
\phantomsubcaption\label{fig:K v d sym visc new20}
\end{subfigure}
\caption{The horizontal (a,b) and vertical (c,d) components of the average kinetic energy per unit mass, derived from the velocity rms (equations \ref{eq:kinetic_energy_carrier} and \ref{eq:kinetic_energy_droplet}), as a function of the vertical direction for the case of $\Upphi=0.2$ with different viscosity ratios.}
\label{fig:kinetic_energy_visc_new20}
\end{figure}

By comparing the results regarding the cases $2$, $12$, and $13$, it is observed that the Nusselt number (TBL thickness) decreases (becomes thicker) when we introduce a more viscous dispersed fluid ($Nu$ decreases around $14\%$ when changing $\lambda_{\mu}$ from $0.1$ to $10$), which is an opposite behavior compared to the method employed in the paper, which is based on keeping effective viscosity constant. This discrepancy arises because, in the constant-effective-viscosity approach, increasing the viscosity of the dispersed fluid is accompanied by a reduction in the viscosity of the carrier fluid, thereby ultimately enhancing the turbulence intensity of the emulsion, as clearly shown in Figures~\ref{fig:kinetic_energy_visc} and \ref{fig:tmp_rms_visc_c_d}). Conversely, in the traditional method, where the viscosity of the carrier fluid remains unchanged, the introduction of a more viscous dispersed fluid mitigates fluctuations. This damping effect is primarily observed both within the thermal boundary layer, as depicted in Figure \ref{fig:tmp_rms_visc_c_d_new20} and also throughout the entire domain, as shown in Figure \ref{fig:kinetic_energy_visc_new20}. This last picture illustrates the profiles of temperature rms and the kinetic energy (both horizontal and vertical) of both fluids, respectively. Furthermore, in Figure~\ref{fig:tmp mean visc new20}, a higher mean temperature is observed within the TBL at higher $\lambda_{\mu}$ values, indicating lower temperature gradients in close proximity to the wall, corresponding to an increase in the thickness of the thermal boundary layer. \par
In conclusion, it is important to note that in the traditional method, the dimensionless numbers of the carrier-fluid properties remain unchanged when the viscosity ratio varies. However, the same dimensionless numbers of the dispersed-fluid properties do change, which this method overlooks. For instance, the dispersed-fluid Rayleigh number ($Ra_d$) decreases with higher $\lambda_{\mu}$ values. Therefore, if we calculate an effective Rayleigh number using the constant-effective-viscosity approach and define $Ra_\text{eff}$ as done in~\citet{liu2022enhancing}, i.e., $Ra_\text{eff} = \Upphi Ra_d + (1-\Upphi)Ra_c$, we get a lower effective Rayleigh number, $Ra_\text{eff} < 10^8$. Conversely, the approach employed in our work allows us to keep the same $Ra_\text{eff}$ and change the viscosity ratio only. As a result, it offers the possibility to develop correlations for the Nusselt number $Nu$ as a function of the volume fraction $\Upphi$ based on the only parameter that is changed ($\lambda_{\mu}$ in our case).


\newpage
\bibliographystyle{jfm} 
\bibliography{jfm-instructions}

\end{document}